\documentclass[iop,numberedappendix]{emulateapj}

\usepackage{bm}

\usepackage{graphicx}
\usepackage{amssymb}
\usepackage{epstopdf}

\newcommand{\beq}{\begin{equation}}
\newcommand{\eeq}{\end{equation}}
\newcommand{\bea}{\begin{eqnarray}}
\newcommand{\eea}{\end{eqnarray}}
\newcommand{\non}{\nonumber \\}
\newcommand{\trm}[1]{\textrm{#1}}

\newcommand{\gv}[1]{\ensuremath{\mbox{\boldmath$ #1 $}}} 
\renewcommand{\d}[2]{\frac{d #1}{d #2}} % for derivatives
\newcommand{\dd}[2]{\frac{d^2 #1}{d #2^2}} % for double derivatives
\newcommand{\pd}[2]{\frac{\partial #1}{\partial #2}} 
\newcommand{\pdd}[2]{\frac{\partial^2 #1}{\partial #2^2}} 
\newcommand{\grad}[1]{\gv{\nabla} #1} % for gradient
\renewcommand{\div}[1]{\gv{\nabla} \cdot #1} % for divergence
\let\baraccent=\= % rename builtin command \= to \baraccent
\renewcommand{\=}[1]{\stackrel{#1}{=}} % for putting numbers above =

\begin{document}

\shortauthors{WEINBERG}
\shorttitle{GROWTH RATE OF THE TIDAL $p$-$g$ INSTABILITY}

\title{Growth rate of the tidal \lowercase{$p$}-mode \lowercase{$g$}-mode instability in \\ coalescing binary neutron stars}

\author{Nevin N.~Weinberg}
\affil{Department of Physics, and Kavli Institute for Astrophysics and Space Research, Massachusetts Institute of Technology,
 \\Cambridge, MA 02139, USA}
 
\begin{abstract}
We recently described an instability due to the nonlinear coupling of $p$-modes to $g$-modes and, as an application, we studied the stability of the tide in coalescing binary neutron stars.  Although we found that the tide is $p$-$g$ unstable early in the inspiral and rapidly drives modes to large energies, our analysis only accounted for three-mode interactions. Venumadhav et al. showed that four-mode interactions must also be accounted for as they enter into the analysis at the same order.  They found a near-exact cancellation between three- and four-mode interactions and concluded that while the tide in binary neutron stars can be $p$-$g$ unstable, the growth rates are not fast enough to impact the gravitational wave signal. Their analysis assumes that the linear tide is incompressible, which is true of the static linear tide (the $m=0$ harmonic) but not the non-static linear tide ($m=\pm 2$).  Here we account for the compressibility of the linear tide and find that three- and four-mode interactions no longer cancel.  As a result, we find that the instability can rapidly drive modes to significant energies well before the binary merges.  We also show that linear damping interferes with the cancellation and may further enhance the growth rates.  The early onset of the instability (at gravitational wave frequencies  $\approx 50\trm{ Hz}$) and the large number of rapidly growing modes suggest that the instability could impact the  gravitational wave signal.  Assessing its impact will require an understanding of how the instability saturates and is left to future work.
\end{abstract}


\section{Introduction}
\label{sec:intro} 

Coalescing binaries consisting of two neutron stars or a neutron star and a black hole are among the most promising sources for the new generation of advanced ground-based gravitational wave detectors, including Advanced LIGO \citep{Harry:10} in the US, Advanced Virgo \citep{Accadia:12} in Italy, and KAGRA \citep{Somiya:12} in Japan.  Tidal interactions in such binaries imprint a signature on the gravitational waveform that if detected would help constrain the highly uncertain neutron star equation of state.  Quantifying this exciting prospect has been the subject of extensive investigation over the past three decades. 

The strength of the signature depends on both the amplitude of the tidal deformation and its phase lag with respect to the line joining each body's center of mass.  Because the neutron star does not respond instantaneously to the changing tidal potential, the gravitational wave-induced orbital decay causes a phase lag even in the absence of viscous dissipation (see e.g., \citealt{Lai:94}).   Indeed,  numerous studies have shown that if the viscous dissipation is determined entirely by linear processes, its effect on the gravitational waveform is small and the phase lag due to orbital decay dominates.  These include investigations of linear dissipation due to fluid viscosity (\citealt{Kochanek:92, Bildsten:92,  Lai:94}) and due to the linear excitation of resonant, short wavelength internal gravity waves \citep{Reisenegger:94b, Lai:94} and inertial waves  \citep{Ho:99, Lai:06, Flanagan:07}. More recent studies therefore assume that the phase lag due to viscous dissipation can be neglected during the inspiral \citep{Flanagan:08, Hinderer:10,Read:09b,Read:13, Damour:12,Lackey:12, Lackey:15}.

As the binary inspirals and the tidal deformation grows in amplitude, it becomes susceptible to nonlinear instabilities.  These will initially manifest as weakly nonlinear wave interactions in which the long length scale tidal perturbation excites short wavelength fluid waves within the star.  These waves remove energy and angular momentum from the tide and thus act as an additional source of dissipation not accounted for in linear analyses.  Previous studies argued that such nonlinear effects should only become important during the very late stages of the inspiral (gravitational wave frequencies $f_{\rm gw}\ga 400\trm{ Hz}$) because the amplitude of the tidal deformation is too small at larger orbital separations. However, the importance of nonlinear wave-tide interactions depends not only on the amplitude of the tide but also on the strength of the tide's nonlinear coupling to the internal oscillation modes of the star. We must therefore evaluate the nonlinear coupling strengths in order to know when during the inspiral the tidal deformation first becomes unstable and to determine how such an instability affects the gravitational wave signal.

In a previous paper (\citealt{Weinberg:12}, hereafter WAQB), we developed a formalism to study weakly nonlinear wave-tide interactions in close binary systems.  WAQB focused on the well-known parametric instability in which a parent wave (e.g., the tide)  \emph{resonantly} excites a pair of short wavelength daughter waves of approximately half the parent's frequency (for astrophysical applications see also, e.g., \citealt{Kumar:96, Wu:01, Arras:03, Weinberg:08, Barker:10}, \citealt{Essick:16}).  In a follow-up paper, \citeauthor{Weinberg:13} (2013; hereafter WAB) described a different type of instability in which a parent wave \emph{nonresonantly} excites an acoustic wave and a gravity wave (i.e., a $p$-mode and a $g$-mode).  A parent wave that is a $g$-mode or a tidal perturbation is not resonant with a $p$-$g$ daughter pair because its frequency is much lower than the $p$-mode's natural frequency.  Nonetheless, WAB found that high-order $p$-$g$ daughters with similar wavelength couple so strongly that even a highly nonresonant, small amplitude parent can excite them.

The primary application of $p$-$g$ mode coupling that WAB considered was tides in coalescing binary neutron stars. They found that the tide is $p$-$g$ unstable early in the inspiral and rapidly drives modes to large energies.  However,  they only considered three-wave interactions. \citeauthor{Venumadhav:14} (2014; hereafter VZH) showed that four-wave interactions enter the analysis at the same order as the three-wave interactions and found, in particular,  that $\{$tide, tide, $g$-mode, $g$-mode$\}$ interactions cancel significantly with $\{$tide, $p$-mode, $g$-mode$\}$ interactions. They concluded that although the \emph{non-static} tide can be $p$-$g$ unstable, the growth rates are too small to significantly influence the inspiral of binary neutron stars.  

The analysis in VZH assumed that the linear tide does not compress mass elements in the star, i.e., it is incompressible, and thus  $\div{\gv{\chi}^{(1)}}=0$, where $\gv\chi^{(1)}$ is the displacement field of the linear tide. They relied on this assumption in order to carry out a volume-preserving coordinate transformation that relates the energy of a tidally deformed star to that of a radially perturbed spherical star. Using this method, they related the four-mode coupling to the three-mode couplings and found that they almost perfectly cancel.

While the static linear tide (the $m=0$ harmonic) is incompressible, the non-static linear tide ($m=\pm2$) is compressible.  Its compressibility $\div{\gv \chi^{(1)}}$ increases with the ratio of the tidal frequency $\omega=m\Omega$ to the buoyancy frequency $N$ (where $\Omega$ is the orbital frequency). For an inspiraling neutron star $\omega \ga N$, and the non-static tide is highly compressible throughout most of the bandpass of gravitational wave detectors such as LIGO,  Virgo, and KAGRA.  

In this paper, we reanalyze the stability of tidally deformed neutron stars to $p$-$g$ mode coupling.  We include three- and four-wave interactions and account for the compressibility of the non-static tide (and the other finite frequency corrections to the static tide).  We find that these corrections undo the cancellation between three- and four-wave interactions and lead to growth rates that are faster than the incompressible limit by a factor of $\sim \omega_0/\omega \gg 1$, where $\omega_0$ is the dynamical frequency of the star.  As a result, there is a large set of unstable modes that can potentially reach significant energies before the neutron star binary merges.

We also study how linear damping affects the stability of the tide to $p$-$g$ coupling.  Although damping typically reduces the growth rates of instabilities, the cancellation between three- and four-wave interactions requires a balance between the phases of the $p$- and $g$-mode oscillations.  We find that the large $p$-mode damping rates counter the stabilizing effects of the cancellation and significantly increase the growth rate relative to the inviscid, incompressible limit studied in VZH.

The plan of the paper is as follows. In \S~\ref{sec:eom} we derive the nonlinear equations of motion and carry out a stability analysis in the inviscid, incompressible limit that recovers the unstable (but slowly growing) non-static tide solution found in VZH.  In \S~\ref{sec:finite_freq} we derive the functional form of the modified three- and four-wave couplings due to the finite frequency corrections (such as compressibility). In \S~\ref{sec:testing_the_calc} we carry out a series of checks in order to test the accuracy of our analytic and numerical calculations.  In \S~\ref{sec:results_finite_freq} we present the main results of our study, the stability analysis of tidal $p$-$g$ coupling with finite frequency corrections in coalescing binary neutron stars. In \S~\ref{sec:results_damping} we evaluate how linear damping influences the stability of the tide and the $p$-$g$ growth rate. Finally, in \S~\ref{sec:conclusions} we summarize and describe the need for a nonlinear saturation study.

\section{Equations of motion}
\label{sec:eom}

Let $\gv\xi(\gv x,t)$ be the Lagrangian displacement of the stellar fluid at position $\gv x$ and time $t$ relative to the unperturbed spherical background and let the operators $\gv f_i$ represent the restoring forces at order $i$.  The equation of motion for $\gv\xi(\gv x,t)$, including linear forces ($\gv f_1$), three- and four-wave nonlinear interactions ($\gv f_2$ and $\gv f_3$, respectively), and tidal forcing ($\rho \gv a_{\rm tide}$) is 
\beq
\rho \ddot{\gv\xi} = \gv{f}_1[\gv \xi] + \gv{f}_2[\gv \xi, \gv \xi] + \gv{f}_3[\gv \xi, \gv \xi, \gv \xi] + \rho \gv{a}_{\rm tide},
\eeq
where $\rho$ is the background density,
\beq
\gv{a}_{\rm tide} = -\grad U - \left(\gv \xi \cdot\grad\right)\grad U,
\eeq
and the $\ell=2$ tidal potential due to a secondary of mass $M'$ in a circular orbit at separation $a$ is
\beq
\label{eq:tidal_potential}
U(\gv{x},t)=-\epsilon \omega_0^2 r^2 \sum_{m=-2}^2 W_{2m} Y_{2m}(\theta,\phi) e^{-im\Omega t}.
\eeq
Here $\epsilon = (M'/M)(R/a)^3$, $M$, $R$,  and $\omega_0=(GM/R^3)^{1/2}$ are the mass, radius, and dynamical frequency of the primary, $\Omega$ is the orbital frequency, $Y_{\ell m}$ is the spherical harmonic function, and $W_{20}=-\sqrt{\pi/5}$, $W_{2\pm2}=\sqrt{3\pi/10}$, $W_{2\pm1}=0$. Since we restrict the analysis to $\ell=2$, all third and higher derivatives of $U$ in $\gv{a}_{\rm tide}$ vanish.  We express the perturbation $\gv \xi$ to the spherical background as the sum of the linear tide $\gv{\chi}^{(1)}\propto \epsilon$, the second-order nonlinear tide $\gv{\chi}^{(2)}\propto \epsilon^2$, and a perturbation to the tidal flow $\gv{\eta}$:
\beq
\gv \xi = \gv \chi^{(1)} + \gv \chi^{(2)} + \gv \eta.
\eeq
Keeping terms up to order $\epsilon^2$ and linear in $\gv \eta$ (because we are interested in studying the linear stability of the tidal flow to infinitesimal perturbations $\gv\eta$), we have
\bea
&\rho & \left(\ddot{\gv \chi}^{(1)} + \ddot{\gv \chi}^{(2)} + \ddot{\gv \eta}\right)
=\gv{f}_1[\gv \chi^{(1)}] +\gv{f}_1[\gv \chi^{(2)}] +\gv{f}_1[\gv \eta] 
\non &&
+\gv{f}_2[\gv \chi^{(1)}, \gv \chi^{(1)}] + 
2\gv{f}_2[\gv \chi^{(1)}, \gv \eta] + 
2\gv{f}_2[\gv \chi^{(2)}, \gv \eta]  
\non &&
+3\gv{f}_3[\gv \chi^{(1)}, \gv \chi^{(1)}, \gv \eta] 
\non &&
-\rho \grad U - \rho \left(\gv \chi^{(1)}\cdot \grad\right) \grad U - \rho \left(\gv{\eta}\cdot \grad\right) \grad U.
\eea
Since the equation of motion of the linear tide is
\beq
\label{eq:chi1}
\rho \ddot{\gv \chi}^{(1)}
=\gv{f}_1[\gv \chi^{(1)}] -\rho \grad U, 
\eeq
and that of the second-order nonlinear tide is
\beq
\label{eq:chi2}
\rho \ddot{\gv \chi}^{(2)}
=\gv{f}_1[\gv \chi^{(2)}]
+\gv{f}_2[\gv \chi^{(1)}, \gv \chi^{(1)}]  - \rho \left(\gv{\chi}^{(1)}\cdot \grad\right) \grad U,
\eeq
the equation of motion for the perturbation $\gv \eta$ is
\bea
\rho  \ddot{\gv \eta}
&=&\gv{f}_1[\gv \eta] + 
2\gv{f}_2[\gv \chi^{(1)}, \gv \eta] + 
2\gv{f}_2[\gv \chi^{(2)}, \gv \eta] 
\non &&+ 
3\gv{f}_3[\gv \chi^{(1)}, \gv \chi^{(1)}, \gv \eta] 
- \rho \left(\gv{\eta}\cdot \grad\right) \grad U.
\eea
Expanding in the basis of the star's linear modes
\bea
\gv \eta (\gv{x},t)&=& \sum_a \eta_a(t) \gv \xi_a(\gv{x}),\\
\gv{\chi}^{(1)}(\gv{x},t)&=&\sum_a \chi_a^{(1)}(t) \gv{\xi}_a(\gv{x}),\\
\gv{\chi}^{(2)}(\gv{x},t)&=&\sum_a \chi_a^{(2)}(t)  \gv{\xi}_a(\gv{x}),
\eea
using the fact that the displacements are real, normalizing the modes according to
\beq
\label{eq:mode_norm}
\omega_a^2 \int d^3x \, \rho \gv{\xi}_a^\ast \cdot\gv{\xi}_b = E_0 \delta_{ab},
\eeq
and noting that $\gv{f}_1[\gv \xi_a]=-\rho \omega_a^2 \gv{\xi}_a$, we obtain the amplitude equation for mode $a$
\bea
&\ddot{\eta}_a &+\omega_a^2 \eta_a 
=\omega_a^2 \sum_b \left( U_{ab}^\ast+2\sum_c\kappa_{abc}^\ast \chi_c^{(1)\ast}
\right.\non &&\left.
 + 2\sum_c \kappa_{abc}^\ast\chi_c^{(2)\ast}
+3\sum_{cd}\kappa_{abcd}^\ast \chi_c^{(1)\ast}\chi_d^{(1)\ast}\right)\eta_b^\ast,
\eea
where 
\bea
U_{ab}&=&-\frac{1}{E_0}\int d^3x  \rho \, \gv{\xi}_a\cdot \left(\gv{\xi}_b\cdot \grad\right) \grad U,
\\
\kappa_{abc}&=& \frac{1}{E_0} \int d^3x \, \gv{\xi}_a \cdot\gv{f}_2\left[\gv \xi_b ,\gv{\xi}_c\right],
\\
\kappa_{abcd}&=& \frac{1}{E_0} \int d^3x  \, \gv{\xi}_a \cdot\gv{f}_3\left[\gv \xi_b ,\gv \xi_c ,\gv{\xi}_d\right].
\eea
The coefficients  are symmetric in all their indices and satisfy the usual selection rules (see, e.g., WAQB) including angular momentum conservation in the azimuthal direction
\bea
\label{eq:selection_rule1}
m+m_a +m_b &=& 0,\\
\label{eq:selection_rule2}
m_a +m_b +m_c&=&0,\\
\label{eq:selection_rule3}
m_a +m_b +m_c+m_d&=&  0,
\eea
for $U_{ab}$, $\kappa_{abc}$, and $\kappa_{abcd}$, respectively. We can therefore write the amplitude equation as
\bea
\label{eq:amp_eqn_inertial}
\ddot{\eta}_a+\gamma_a \dot{\eta}_a+\omega_a^2 \eta_a &=&
\omega_a^2 \sum_b \left(K_{3ab}^\ast+K_{4ab}^\ast\right)\eta_b^\ast e^{-i(m_a+m_b)\Omega t}
\non&=&
\omega_a^2 \sum_b \left(K_{3\bar{a}b}+K_{4\bar{a}b}\right)\eta_b e^{-i(m_a-m_b)\Omega t}
\non
\eea
where the overbar denotes a mode's complex conjugate, the second equality follows because the displacements are real, 
we added a term $\gamma_a \dot{\eta}_a$ to model linear damping, and we defined the time independent three-mode and four-mode coefficients
\bea
\label{eq:K3ab}
K_{3ab} &=& U_{ab} +2\sum_c \kappa_{abc} \chi_c^{(1)} 
\non &=& 
U_{ab}+2\kappa_{\chi^{(1)}ab},\\
\label{eq:K4ab}
K_{4ab}&=&  2\sum_c\kappa_{abc}\chi_c^{(2)}+3\sum_{cd}\kappa_{abcd} \chi_c^{(1)}\chi_d^{(1)}
\non &=&
2\kappa_{\chi^{(2)}ab}+3\kappa_{\chi^{(1)}\chi^{(1)}ab}.
\eea
For analytic purposes, it proves convenient to change variables to
\beq
\label{eq:change_of_variables}
q_a = \frac{\eta_a}{\omega_a}e^{im_a\Omega t},
\eeq
in which case the amplitude equation  becomes
\bea
\label{eq:ddotq}
&\ddot{q}_a&+\left[\gamma_a-2im_a \Omega \right]\dot{q}_a+\left[\omega_a^2 - (m_a \Omega)^2 -im_a \Omega \gamma_a\right]q_a
\non &&=\omega_a \sum_b\left(K_{3\bar{a}b} + K_{4\bar{a}b}\right)\omega_b q_b.
\eea
This change of variables effectively transforms the amplitude equation from the inertial frame (Equation \ref{eq:amp_eqn_inertial}) to a frame that is co-rotating with the binary, and thereby introduces Coriolis and centrifugal terms on the left-hand side of Equation (\ref{eq:ddotq}).

\subsection{Characteristic equation of tidal $p$-$g$ coupling}

In order to determine the stability of the tidal flow to an infinitesimal perturbation $\gv\eta$, assume that the amplitude of the latter varies as $e^{ist}$. By Equation (\ref{eq:ddotq}), we then have the characteristic equation 
\bea
\label{eq:stab1}
\big[&-s^2&+\left(2m_a\Omega + i\gamma_a\right)s+\omega_a^2-(m_a\Omega)^2-im_a\Omega\gamma_a\big]q_a
\non &&
=\omega_a \sum_b\left(K_{3\bar{a}b} + K_{4\bar{a}b}\right)\omega_b q_b,
\eea 
or in matrix notation
\bea
\label{eq:eig2form}
\Big[&-s^2& \gv{I}+\left(2\Omega \gv{L}+i\gv{\gamma}\right)s-\Omega^2 \gv{L}^2-i\Omega\gamma \gv{L}
\non &&+\gv{M}^{(0)}+\gv{M}^{(1)}\Big]\gv{q}=0,
\eea
where 
\bea
\left(\gv{L}, \gv{\gamma}, \gv{M}^{(0)}\right)&=&\left(m_a, \gamma_a,\omega_a^2\right) \delta_{ab},\\
\gv{M}^{(1)}&=& -\omega_a \omega_b \left(K_{3\bar{a}b}+K_{4\bar{a}b}\right).
\eea
This is similar to Equation (E6) in VZH but with two key differences.  First, they do not include a linear damping term. Second, the nonlinear matrix term in VZH, which they write as $\gv{R}^\dagger(0)\delta \mathcal{M}\gv{R}(0)$, assumes that the coupling coefficients are the same as those of the static ($m=0$) tide, modulo a rotation $\gv{R}(0)$ by $\pi/2$ around the $y$-axis.  It therefore does not account for finite frequency corrections to $K_{3ab}$ and $K_{4ab}$ (due to, e.g., the compressibility of the linear tide).

Equation (\ref{eq:eig2form}) is a quadratic eigenvalue problem.  To reduce it to a standard eigenvalue problem, define
\bea
\gv{A}(s)&=&
-s\left[\begin{array}{cc}\gv{I} & 0 \\0 & \gv{I}\end{array}\right]
+\left[\begin{array}{cc}\gv{C} & \gv{B} \\\gv{I} & 0\end{array}\right],\\
\gv{B}&=&-\Omega^2 \gv{L}^2-i\Omega\gamma \gv{L}+\gv{M}^{(0)}+\gv{M}^{(1)},\\
\gv{C}&=&2\Omega \gv{L}+i\gv{\gamma}.
\eea
Equation (\ref{eq:eig2form}) can then be expressed as a standard eigenvalue problem $\gv{A}(s)\gv{z}=0$, i.e., 
\beq
\left[\begin{array}{cc}\gv{C} & \gv{B} \\\gv{I} & 0\end{array}\right]\gv{z}=s\gv{z},
\eeq
where $\gv{z}=[s\gv q,\gv q]$.
In our calculations below, we solve this equation numerically using the arpack++ library.

For analytic study, it will be convenient to rewrite the characteristic Equation (\ref{eq:stab1}) for the specific case of $p$-$g$ coupling.  Letting $p$ subscripts denote $p$-modes and $g$ subscripts denote $g$-modes, we have the coupled equations
\bea
\label{eq:qp_eom}
\Big[&&-\left(s-m_{p_j}\Omega\right)^2+i\gamma_{p_j}\left(s-m_{p_j}\Omega\right)+\omega_{p_j}^2\Big]q_{p_j}
\non &&=
 \omega_{p_j} \sum_{g_i}K_{3 \bar{p}_j g_i} \omega_{g_i}q_{g_i}
 \\
 \label{eq:qg_eom}
\Big[&&-\left(s-m_{g_j}\Omega\right)^2+i\gamma_{g_j}\left(s-m_{g_j}\Omega\right)+\omega_{g_j}^2\Big]q_{g_j}
\non&&=
\omega_{g_j} \left[\sum_{p_i}K_{3 p_i \bar{g}_j } \omega_{p_i} q_{p_i}+\sum_{g_i}K_{4 \bar{g}_j g_i} \omega_{g_i} q_{g_i} \right].
\eea
The coefficients $K_{3 p_j p_i}$ and $K_{3 g_j g_i}$ have magnitude $\sim \epsilon$ and the coefficient $K_{4 p_j p_i}$ has magnitude $\sim \epsilon^2$ (WAQB, WAB, VZH). Furthermore, as VZH showed (see their Section 2.4),  $K_{4 p_j g_i}$ does not enter the stability analysis at order $\epsilon^2$.  Since these coefficients do not alter the stability at order $\epsilon^2$, we do not include them in the analysis.
Substituting Equation (\ref{eq:qp_eom}) into Equation (\ref{eq:qg_eom}) gives
\bea
\label{eq:char_pg}
&&\left[-\left(s-m_{g_j}\Omega\right)^2+i\gamma_{g_j}\left(s-m_{g_j}\Omega\right)+\omega_{g_j}^2\right]q_{g_j}
\non &&=\omega_{g_j}\sum_{g_i} \Bigg[K_{4 \bar{g}_j g_i} 
\non &&
- \sum_{p_i}\frac{K_{3 p_i \bar{g}_j} K_{3\bar{p}_i g_i} \omega_{p_i}^2}{\left(s-m_{p_i}\Omega\right)^2-i\gamma_{p_i}\left(s-m_{p_i}\Omega\right)-\omega_{p_i}^2}\Bigg]\omega_{g_i} q_{g_i}. \hspace{0.5cm}
\eea
This characteristic equation determines the stability and growth rate of $p$-$g$ coupling to order $\epsilon^2$.

\subsection{Stability analysis in the absence of finite frequency corrections and linear damping}
\label{sec:stability_no_finite_freq_no_damping}

The coupling coefficients $K_{4gg}$ and $\sum K_{3pg}K_{3\bar{p}g}$ that enter Equation (\ref{eq:char_pg}) can each individually be of order unity or larger for high-order modes.  However, if we ignore finite frequency corrections, then as VZH showed 
\beq
\label{eq:K4_K3sq_cancellation}
K_{4 g_j g_i}+\sum_{p_i} K_{3p_i g_j} K_{3 \bar{p}_i g_i} = \mathcal{O}(\epsilon^2),
\eeq
i.e., there is significant cancellation and the residual is merely of order $\epsilon^2\ll 1$. Thus, if we ignore finite frequency corrections and linear damping, equation  (\ref{eq:char_pg})  gives
\bea
&&\left[\left(s-m_{g_j}\Omega\right)^2-\omega_{g_j}^2\right]q_{g_j}
\non && =
\omega_{g_j}\sum_{g_i}\sum_{p_i}\frac{K_{3 p_i \bar{g}_j} K_{3\bar{p}_i g_i}\left(s-m_{p_i}\Omega\right)^2}{\left(s-m_{p_i}\Omega\right)^2-\omega_{p_i}^2}\omega_{g_i} q_{g_i}, 
\eea
where we assumed that the cancellation of Equation (\ref{eq:K4_K3sq_cancellation}) is exact since $\mathcal{O}(\epsilon^2)$ terms cannot anyways alter the stability.  As in VZH, we can estimate the stability of this potentially large system of modes by considering a two mode system consisting of a single $p$-$g$ pair with frequencies $\omega_g\ll\Omega\ll\omega_p$. Since $K_{3pg}\simeq \epsilon \omega_p/\omega_g$ (WAB, VZH), if we define $r=s-m_p\Omega$ and $\alpha=(m_p-m_g)\Omega$, the characteristic equation of the system becomes
\beq
\label{eq:r_char_no_finite_freq_no_damping}
\left[\left(r+\alpha\right)^2-\omega_g^2\right]\left[r^2-\omega_p^2\right]-\epsilon^2 r^2\omega_p^2\simeq0.
\eeq
There are two stable high frequency solutions near $r\simeq \pm\omega_p$ and two low frequency solutions near
\beq
\label{eq:r_low_freq_no_finite_freq_no_damping} 
r\simeq -\alpha \pm \sqrt{\omega_g^2-\epsilon^2\alpha^2}.
\eeq
If $\omega_g \la \epsilon|\alpha|$ then the low frequency solutions are unstable and the $p$-$g$ pair grows at a rate
\beq
\label{eq:growth_rate_no_finite_freq_no_damping}
\Gamma \approx \epsilon\left|m_p-m_g\right|\Omega.
\eeq
By angular momentum conservation (i.e., the coupling coefficient selection rules Equations \ref{eq:selection_rule1}--\ref{eq:selection_rule3}), only $p$-$g$ pairs that satisfy $|m_p-m_g|=m$ couple to harmonic $m$ of the $\ell=2$ linear tide.  Therefore, for the static tide ($m=0$) there is no unstable solution and for the non-static tide ($m=\pm 2$) the growth rate of the instability is $\Gamma\approx 2\epsilon \Omega$ if $\omega_g \la 2\epsilon \Omega$.  This agrees with the results found in VZH (see their Appendix E).  For a coalescing neutron star binary, a growth rate of $2\epsilon \Omega$ is too small to drive $p$-$g$ pairs to large amplitudes before the merger (the modes only have enough time to grow by a factor of $\sim10^5$ in energy; see \S~\ref{sec:efoldings_finite_freq}).

In Appendix \ref{app:static_tidal_field} we present an argument based on the energy principle that shows that the \emph{static} tide must be absolutely stable, consistent with VZH's result and extending it to all orders in $\epsilon$.  In the remainder of the paper, we therefore focus on the stability of the non-static tide.

 \section{Finite frequency corrections}
 \label{sec:finite_freq}
 
In their analysis, VZH assume that the shape of the non-static linear tide $\gv{\chi}^{(1)}$ that enters the calculation of the three- and four-mode coupling coefficients is exactly given by the static (i.e., equilibrium) tide solution
\bea
\div{\gv{\chi}^{(1)}}&=&0, \\
\chi^{(1)}_{r} &=& -\frac{U}{g},
\eea 
where $g$ is the local gravitational acceleration (as in VZH, for simplicity we make the Cowling approximation throughout our analysis.) However, this solution is obtained only in the limit $\omega^2 \ll N^2$ (see, e.g., \citealt{Terquem:98} and \citealt{Ogilvie:14}; here $\omega=m\Omega$ is the tidal frequency and $N$ is the Brunt-V\"ais\"al\"a buoyancy frequency).  Otherwise, the linear tide is compressible ($\div{\gv{\chi}^{(1)}}\neq 0$). In the opposite limit, $\omega^2\gg N^2$, the linear equations of motion are (see \S~\ref{sec:divxi_finite_freq})
\beq
-\omega^2 \gv{\chi}^{(1)}= -\grad \left(\frac{\delta p}{\rho} + U\right)
\eeq
and $\gv\chi^{(1)}$ is compressible but irrotational ($\grad\times\gv{\chi}^{(1)}=0$; here $\delta p$ is the Eulerian pressure perturbation).

In a neutron star $N\approx 0.1 \omega_0 r/R$ and for a coalescing binary neutron star within LIGO's bandpass $\omega^2 \ga N^2$.  Finite frequency corrections to $\gv{\chi}^{(1)}$ can therefore be significant throughout the inspiral and, importantly, not only when the orbit sweeps through resonances with individual modes of the star (i.e., not only when the \citealt{Lai:94} ``dynamical tide" is excited\footnote{In \citet{Lai:94}, the dynamical tide is defined to be the set of resonantly excited modes. In the case of coalescing binary neutron stars, these modes have little energy and cannot significantly influence the inspiral waveform even if they are $p$-$g$ unstable (WAB). However, it is also common to define the dynamical tide as the difference in shape between the non-static and static linear tides,  $\gv{\chi}_d^{(1)}\equiv\gv{\chi}^{(1)}-\gv{\chi}^{(1)}_e$ (this is how it is defined by, e.g., \citealt{Zahn:70, Terquem:98, Goodman:98}; WAQB).  If the finite frequency corrections are large, then $\gv{\chi}^{(1)}$ and $\gv{\chi}^{(1)}_e$ are not similar and the energy in $\gv{\chi}_d^{(1)}$ can be significant (even between resonances).}).

\begin{figure}
\centering
\includegraphics[width=3.4in]{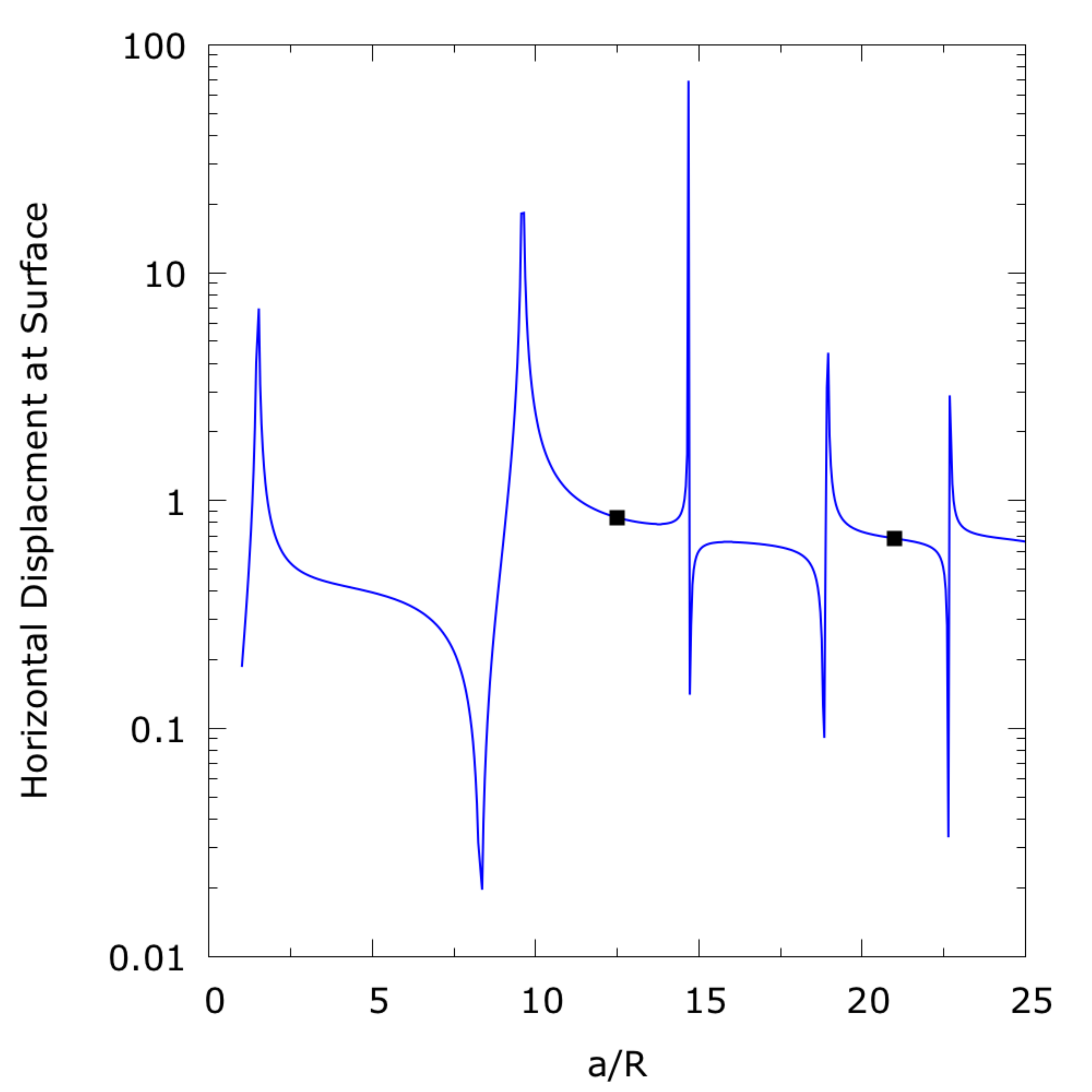} 
\caption{Horizontal displacement of the linear tide at the stellar surface $\chi^{(1)}_h(R)$ (divided by $\epsilon W_{22} R$; see Equation (\ref{eq:tidal_potential})) as a function of orbital separation $a/R$.  The displacement is found by solving the linear inhomogeneous Equation (\ref{eq:chi1}) for $m=\pm2$; the peaks occur when the driving is resonant with individual $\ell=2$ modes of the star.
In Figure \ref{fig:div_xi_and_xir} we plot $\div\gv{\chi}^{(1)}$ and $\chi_r^{(1)}$ at $a/R=12.5$ and $21$, indicated here by the two black squares located between resonances. 
\label{fig:linear_resonances}}
\end{figure}

\begin{figure*}
\centering
\includegraphics[width=3.4in]{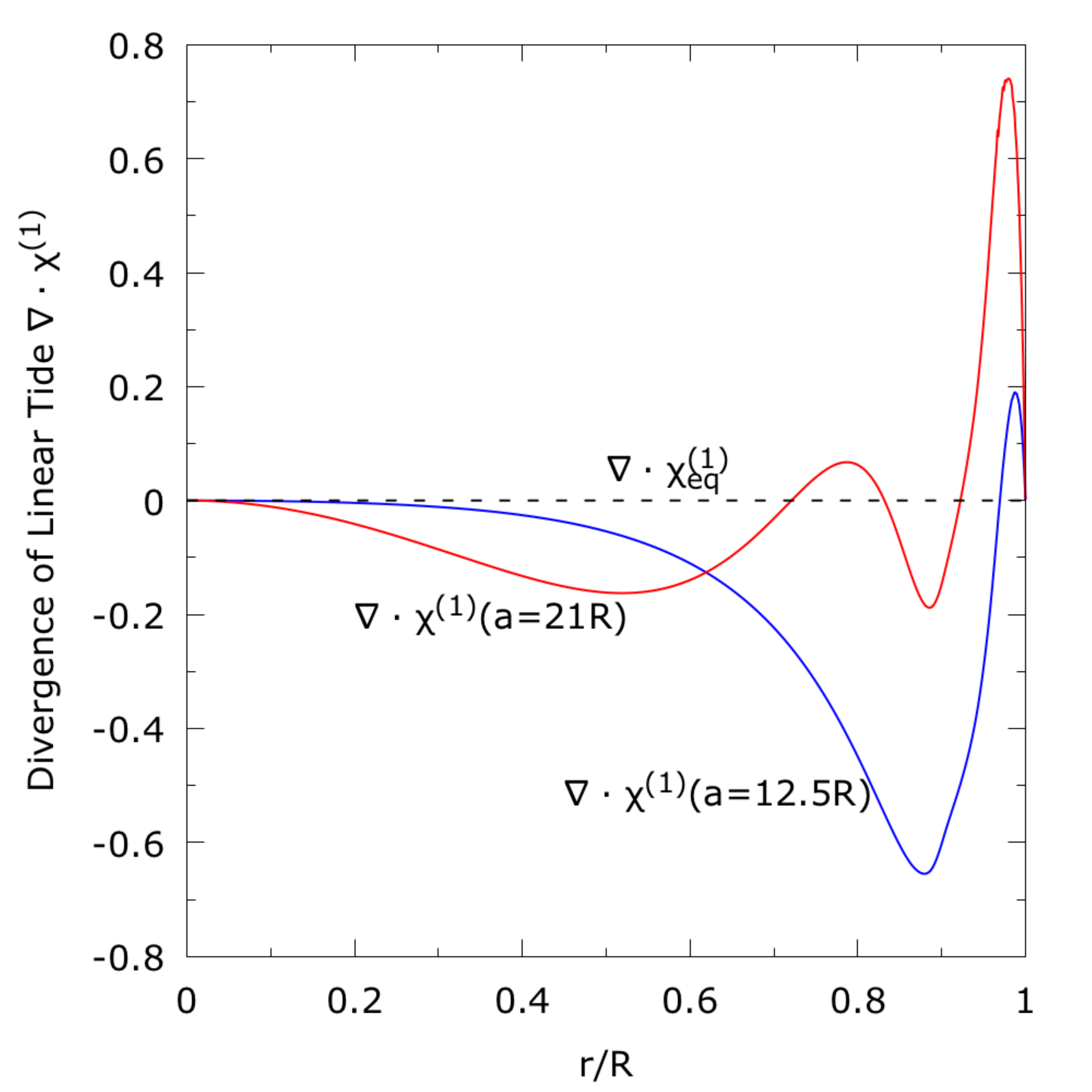} 
\includegraphics[width=3.4in]{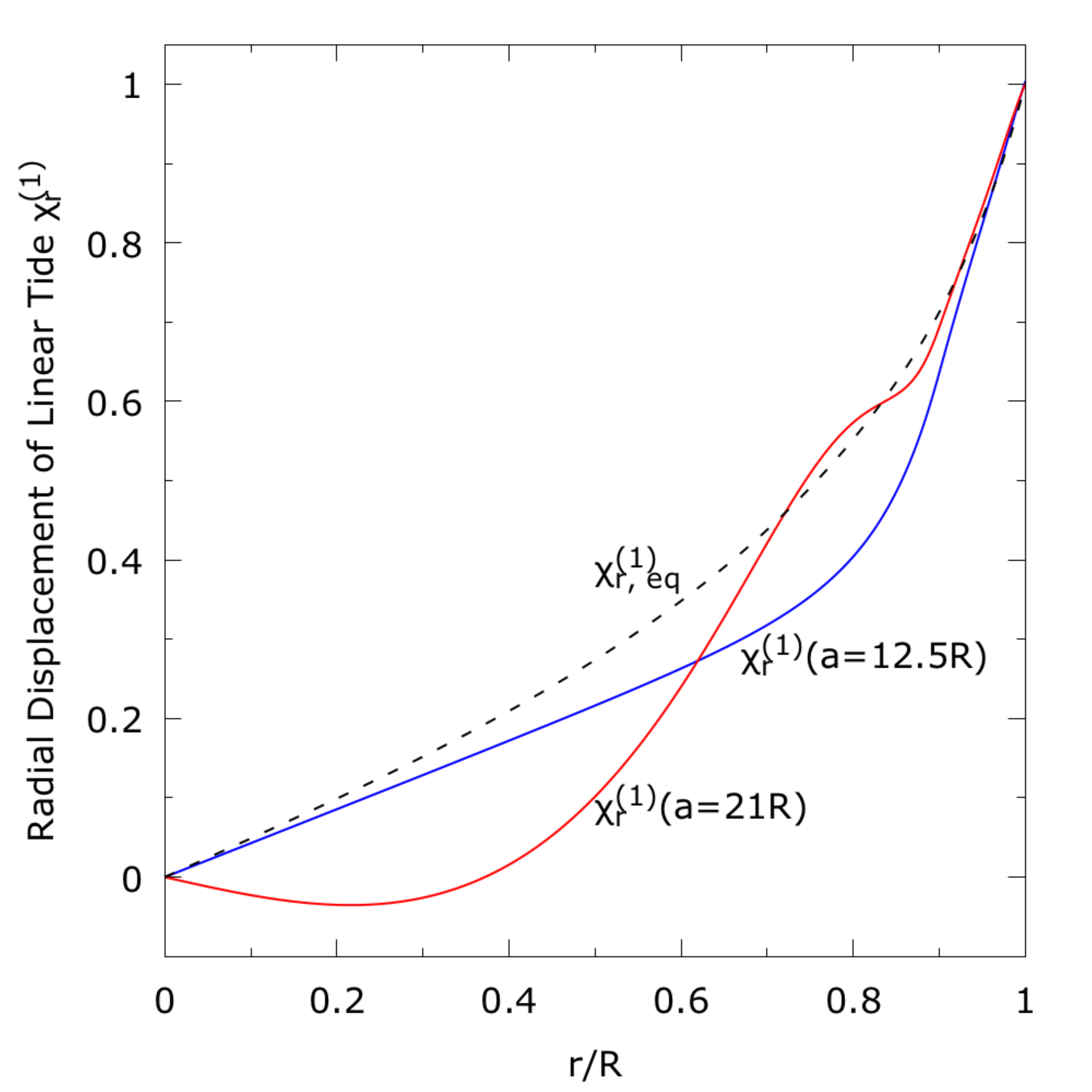} 
\caption{Divergence of the non-static linear tide $\div{\gv{\chi}}^{(1)}$ (divided by $\epsilon W_{22}$; left panel) and the radial displacement of the non-static linear tide $\chi^{(1)}_r$ (divided by $\epsilon W_{22}R$; right panel)  as a function of radius $r$ at two orbital separations that are not near resonances: the blue solid lines are for $a/R=12.5$ ($f_{\rm gw}\simeq100\trm{ Hz}$) and the red solid lines are for $a/R=21$ ($f_{\rm gw}\simeq 45\trm{ Hz}$; see black squares in Figure \ref{fig:linear_resonances}). The black dashed lines show the linear static (i.e., equilibrium) tide divergence and radial displacement (divided by $\epsilon W_{20}R$).
\label{fig:div_xi_and_xir}}
\end{figure*}

We illustrate this in Figure \ref{fig:linear_resonances} and Figure \ref{fig:div_xi_and_xir}. Here and throughout the paper we use the same neutron star model as in WAB;  it has a mass $M=1.4M_\odot$ and radius $R\simeq 11.7\trm{ km}$ and assumes the Skyrme Lyon (SLy4) equation of state \citep{Chabanat:98} and a non-buoyant crust.  In Figure  \ref{fig:linear_resonances} we show the horizontal displacement of the linear tide at the stellar surface $\chi^{(1)}_h(R)$ as a function of orbital separation $a/R$. We compute $\gv \chi^{(1)}$ by solving the linear inhomogeneous Equation (\ref{eq:chi1}) using the method described in WAQB.  The peaks in $\chi^{(1)}_h(R)$ occur where the linear driving is resonant with an $\ell=2$ mode of the star. In Figure \ref{fig:div_xi_and_xir} we show the radial profiles of $\div\gv\chi^{(1)}$ and $\chi^{(1)}_r$ for the non-static and static tide at two separations not near linear resonances ($a/R=12.5$ and $21$, which for our neutron star model correspond to gravitational wave frequencies $f_{\rm gw}\simeq 100\trm{ Hz}$ and $45\trm{ Hz}$; see black squares in Figure \ref{fig:linear_resonances}). Despite not being near resonances,  $\div\gv\chi^{(1)}$ and $\chi^{(1)}_r$ differ significantly from the static tide solution due to finite frequency corrections.

The remainder of this section is organized as follows.  We first derive (\S~\ref{sec:divxi_finite_freq}) an analytic estimate of the finite frequency corrections to $\div{\gv{\chi}^{(1)}}$ and $\gv\chi^{(1)}$  in order to better understand the numerical values of Figure \ref{fig:div_xi_and_xir}. Then, over several subsections (\S\S~\ref{sec:k3_finite_freq}-\ref{sec:corrections_wc2}), we  describe our procedure for calculating the three and four mode coupling coefficients.  Recall that VZH use a volume preserving coordinate transformation in order to map the tidally deformed star back into a spherically symmetric configuration and from there calculate the  coupling coefficients.  However, because the linear tide is compressible and thus not volume preserving, this procedure cannot be used to calculate the finite frequency corrections.   Our strategy instead is to adopt the more general approach  taken by WAQB, who calculate the coupling coefficients to order $\epsilon$, and to now extend the calculation to order $\epsilon^2$. Specifically,  in \S~\ref{sec:k3_finite_freq} we consider the finite frequency correction to $K_{3pg}$ by making use of the exact expressions for $U_{ab}$ and $\kappa_{abc}$ derived in WAQB.  In \S~\ref{sec:k3sq_sum_rule} we derive a sum rule for $\sum K_{3pg} K_{3\bar{p}g}$ that allows us to calculate it accurately without doing an explicit numerical sum over $p$-modes. In \S~\ref{sec:kap4} we describe the four-mode coupling coefficient $\kappa_{abcd}$ that we use to calculate $K_{4gg}$.  In \S~\ref{sec:corrections_wc3&4} we show analytically that the finite frequency corrections to the terms that enter at $\mathcal{O}([\omega_0/\omega_g]^{4})$ and $\mathcal{O}([\omega_0/\omega_g]^{3})$ do not undo the cancellation of the zero frequency (i.e., incompressible) limit.  In \S~\ref{sec:corrections_wc2} we derive expressions for the terms that enter $\mathcal{O}([\omega_0/\omega_g]^{2})$. When we numerically evaluate these terms in \S~\ref{sec:results_finite_freq}, we find that they do undo the cancellation.

\subsection{Estimate of the finite frequency corrections to $\gv{\chi}^{(1)}$}
\label{sec:divxi_finite_freq}

In order to obtain an analytic estimate of the finite frequency corrections to the non-static linear tide $\gv{\chi}^{(1)}$ and its  compressibility $\div\gv\chi^{(1)}$, consider the equation of motion for the linear tide (in the Cowling approximation) 
\beq
\label{eq:lin_tide_inhomog_eqn}
\rho \frac{d^2\gv{\chi}}{dt^2}=-\grad \delta p +\gv{g}\delta \rho - \rho \grad U,
\eeq
where $\delta p$ and $\delta \rho$ are the Eulerian pressure and density perturbations (here and for the remainder of this section we drop the ``1'' superscript; note that in this section $p$ refers to  pressure and not the $p$-mode).  Since $U\sim e^{-im\Phi(t)}$, assume a 
solution of the form
\beq
\gv{\chi}(\gv{x},t)=\gv{\alpha}(\gv{x},t) e^{-im\Phi(t)},
\eeq
where $\dot{\Phi}=\Omega$. Then
\bea
\dd{\gv{\chi}}{t}&=&\left[\ddot{\gv{\alpha}}-2im\Omega \dot{\gv{\alpha}}-\left(m^2\Omega^2+im\dot{\Omega}\right)\gv{\alpha}\right]e^{-im\Phi}.
\non
\eea
Focusing on coalescing binary neutron stars, define  the orbital decay time due to gravitational wave emission
\beq
\label{eq:ta}
t_a =\frac{a}{\left|\dot{a}\right|}=
8.9\left(\frac{\mathcal{M}}{1.2M_\odot}\right)^{-5/3}\left(\frac{f_{\rm gw}}{100\trm{ Hz}}\right)^{-8/3}\trm{ s}
\eeq
(\citealt{Peters:63}), where $f_{\rm gw}=\Omega/\pi$ is the gravitational wave frequency and $\mathcal{M}=\left[(MM')^3/(M+M')\right]^{1/5}$ is the chirp mass ($\mathcal{M}\simeq1.2M_\odot$ for $M=M'=1.4M_\odot$).
When the driving is not near a resonance with a normal mode of the star,  the amplitude of the non-static tide varies on a timescale $t_a$ and  (\citealt{Lai:94})
\beq
\dot{\gv{\alpha}}\sim \gv{\alpha}/t_a, \hspace{1.0cm} \ddot{\gv{\alpha}}\sim \gv{\alpha}/t_a^2.
\eeq
The duration of a resonance is $\delta a \sim 0.5 R (a/10R)^{-1/4}$ (\citealt{Lai:94} Equation 3.24, after correcting the typo in the exponent for $a$) and the above scalings should apply when outside these narrow resonances (cf. Fig. \ref{fig:linear_resonances}).   Since $t_a\gg \Omega^{-1}$, during non-resonant driving 
\beq
\dd{\gv{\chi}}{t}\simeq -\omega^2 \gv{\chi}
\eeq
and we can make the usual separation of variables
\beq
\label{eq:chi1_sep_var}
\gv{\chi}(\gv{x},t)=\left[\chi_r(r), r\chi_h(r) \grad_\perp \right] Y_{\ell m}(\theta,\phi) e^{-i\omega t},
\eeq
where $\omega=m\Omega$ and we treat the amplitudes $\chi_r$ and $\chi_h$ as constant in time.  

We can estimate the second order (i.e., $\omega^2$) finite frequency corrections to $\gv\chi^{(1)}$ with an analysis similar to \citet{Arras:10}. By Equation (\ref{eq:lin_tide_inhomog_eqn}), 
\bea
\label{eq:horiz_mom}
\omega^2 \chi_h &=&\frac{\delta p/\rho + U}{r} \\
\label{eq:radial_mom}
\omega^2 \chi_r &=& \frac{1}{\rho}\frac{d\delta p}{d r} + g\frac{\delta \rho}{\rho} + \frac{dU}{dr},
\eea
where we expand perturbed scalar quantities, such as pressure, as $\delta p(r,\theta,\phi) = \delta p(r) Y_{\ell m}(\theta,\phi)$. The $\omega^2$ correction to the $\omega=0$ equilibrium tide Eulerian pressure perturbation is therefore
\bea
\delta p
&=&  \delta p^{({\rm eq})} + \omega^2 r \rho \chi_h 
\non&\simeq&  \delta p^{({\rm eq})} + \omega^2 r \rho \chi_h^{({\rm eq})},
\label{eq:deltap}
\eea
where $\delta p^{({\rm eq})}  = -\rho U.$
Plugging (\ref{eq:deltap}) into (\ref{eq:radial_mom}) gives
\bea
\delta \rho \simeq \delta \rho^{({\rm eq})} + \frac{\omega^2}{g}\left[\rho \chi_r^{({\rm eq})} -\frac{d}{dr}\left(\rho r \chi_h^{({\rm eq})}\right)\right],
\eea
where $\delta \rho^{({\rm eq})}=(U/g)d\rho/dr.$
Assuming the perturbations are adiabatic, the Lagrangian pressure and density perturbations are related by $\Delta p/p=\Gamma_1 \Delta \rho/\rho$ (where $\Gamma_1$ is the adiabatic index), implying
\bea
\label{eq:mass_cons}
\frac{\delta \rho}{\rho}=\frac{\delta p}{\Gamma_1 p}+\frac{N^2}{g}\chi_r.
\eea
In the $\omega^2=0$ limit of the equilibrium tide this gives $\chi_r^{({\rm eq})}=-U/g$
and by mass conservation $\delta \rho =-\grad\cdot\left(\rho \gv{\chi}\right)$ we have
\beq
\div{\gv{\chi}^{(\rm eq)}}=0,
\eeq
i.e.,
\beq
\chi_h^{({\rm eq})}=\frac{1}{r \Lambda^2}\frac{d}{dr}\left(r^2 \chi_r^{({\rm eq})}\right).
\eeq
We can now solve for the second order correction to $\chi_r$ by plugging (\ref{eq:deltap}) into (\ref{eq:mass_cons}) to get
\beq
\label{eq:xir}
\chi_r \simeq
  \chi_r^{({\rm eq})}\left[1+\left(\frac{\omega}{N}\right)^2 f(r)\right],
 \eeq
 where
\bea
f(r) &=& \left[1-\frac{\chi_h^{({\rm eq})}}{\chi_r^{({\rm eq})}}\left(1-\frac{rN^2}{g}\right)-\frac{r}{\chi_r^{({\rm eq})}}\frac{d \chi_h^{({\rm eq})}}{dr}\right]
\\&=&
\left[1-\frac{E+2}{\Lambda^2}\left(1+E-\frac{rN^2}{g}\right)+\frac{1}{\Lambda^2}\frac{d^2\ln g}{d\ln r^2} \right]\hspace{0.5cm}
\eea
(here $E=\ell-d\ln g/d\ln r$).
Similarly, we find
\bea
\chi_h &\simeq&  \chi_h^{({\rm eq})}+\frac{r\omega^2}{\Lambda^2 g}\left\{\chi_r^{({\rm eq})}-\frac{1}{\rho}\frac{d}{dr}\left(\rho r \chi_h^{({\rm eq})}\right)
\right. \non &&\left.
+\frac{g}{\rho r^2}\frac{d}{dr}\left[\frac{\rho r^2}{N^2}f\chi_r^{({\rm eq})}\right]\right\}.
\eea
Finally, the second order correction to the divergence is
\bea
\div{\gv{\chi}}
&=& -\frac{\delta \rho}{\rho}-\chi_r\d{\ln \rho}{r}
\non &\simeq& \left(\frac{\omega}{N}\right)^2 \! \frac{g}{c_s^2}\frac{\chi_r^{(\rm eq)}}{\Lambda^2}
\! \left[\Lambda^2- (E+2)(E+1)+\dd{\ln g}{\ln r}\right]
\non &\simeq&\frac{1}{2}\left(\frac{\omega}{N}\right)^2\frac{g}{c_s^2}\d{\ln \rho}{\ln r}\chi_r^{(\rm eq)},
\eea
where the last equality is for the specific case of $\ell=2$ in the neutron star core, where $g\simeq 4\pi G\rho r/3$.
Since $N\simeq 0.1 \omega_0 r/R$ and $c_s\simeq 1.5\omega_0 R$ in the core,
\bea
\d{\ln \rho}{\ln r}=\frac{rN^2}{g}-\frac{gr}{c_s^2}\simeq -\frac{gr}{c_s^2}
\eea
and we find
\bea
\label{eq:divxi_finite_freq}
\div{\gv{\chi}}&\simeq& -\frac{1}{2}\left(\frac{\omega}{N}\right)^2\left(\frac{gr}{c_s^2}\right)^2\frac{\chi_r^{(\rm eq)}}{r}
\non&\simeq &0.3 \,\epsilon W_{22} \left(\frac{\omega}{N}\right)^2\left(\frac{r}{R}\right)^4
\non&\simeq& 30\,\epsilon W_{22} \left(\frac{\omega}{\omega_0}\right)^2\left(\frac{r}{R}\right)^2.
\eea
We thus see that $\div{\gv{\chi}}\propto \left(\omega/N\right)^2$ and that for $\omega\simeq 0.1\omega_0$ ($f_{\rm gw}\simeq 100\trm{ Hz}$), the finite frequency correction to the divergence is $\div{\gv{\chi}}\sim 0.1 \epsilon$ in the core, consistent with the numerical results shown in Figure \ref{fig:div_xi_and_xir}.
 
\subsection{Finite frequency corrections to $K_{3pg}$}
 \label{sec:k3_finite_freq}

As VZH showed and as we found in \S~\ref{sec:eom}, the stability of $p$-$g$ coupling  depends on the degree of cancellation among the terms in 
\beq
\label{eq:K4_K3sq}
K_{4gg}+\sum_{p} K_{3p g}K_{3\bar{p}g}.
\eeq
In this section we calculate $K_{3pg}$ accounting for the finite frequency $\omega$ of the linear tide $\gv{\chi}$ (here we again drop the ``1" superscript).  
WAQB give the full expression for $K_{3pg}$,
including finite frequency corrections.  The coupling is strongest for high order $p$-$g$ pairs whose wavelengths match (i.e., whose wavenumbers satisfy $k_p \simeq k_g$), which implies $\omega_p/\omega_0 \sim \omega_0/\omega_g$.   The largest individual terms in $K_{3pg}$ are of magnitude $\sim \epsilon (\omega_0/\omega_g)^2$, which we write as $\epsilon\mathcal{O}(\omega_g^{-2})$. Here we are interested in accounting for all terms bigger than or equal to $\epsilon\mathcal{O}(1)$; since the dynamics depend on $\sum_p K_{3pg}K_{3\bar{p}g}$, such terms contribute  at least at $\epsilon^2\mathcal{O}(\omega_g^{-2})$  and for small enough $\omega_g$ can lead to instability if they are not cancelled by $K_{4gg}$. From Equations (A55)--(A62) and (A71) in WAQB,
\bea
&&K_{3pg}= 
-\sum_m \epsilon W_{2m} \frac{T\left(\ell+2\right)}{MR^\ell }\int dr \rho r^{\ell+1}g_r \div{\gv{p}}
\non &&
+\frac{1}{E_0}\int dr \rho r \Bigg\{\omega_p^2F_p\left(\chi_r-\chi_h\right) p_r g_h 
-\omega_p^2G_p \chi_h p_h g_h 
\non &&
+\omega_p^2\left(3F_\chi+2F_p\right) \chi_r p_h g_h+\omega_p^2F_p \chi_h p_r g_r 
\non &&
+\Lambda_g^2 T c_s^2 \div{\gv{\chi}}\div{\gv{p}}g_h
\non &&
+Tc_s^2 \div{\gv{p}}
\bigg[\left(\Gamma_1+1 +\left(\pd{\ln \Gamma_1}{\ln \rho}\right)_S\right) r\div{\gv{\chi}}\div{\gv{g}}
\non &&
+ \left(\chi_h \Lambda_a^2 -4\chi_r\right)\div{\gv{g}}-4g_r\div{\gv{\chi}}\bigg]
 \non &&
+T\left(4g + r\d{g}{r}\right)\div{\gv{p}}\chi_r g_r
\non &&
+ \omega^2 \left[\left(3F_p+2F_\chi\right) \chi_h +F_\chi \chi_r\right]p_r g_h
\non &&
+ \omega_p^2 \left[\left(3F_g+2F_p\right) \chi_h -6T\chi_r\right]p_h g_r
\Bigg\}+\epsilon\mathcal{O}(\omega_g).
\label{eq:K3pg_monster}
\eea
Here $\gv{g}$ is the $g$-mode eigenvector (not gravity), $T$, $F_a$, and $G_a$ are three-mode angular integrals, and for consistency with the convention used in previous sections we write the frequency of the tide as $\omega$ (rather than $\omega_\chi$). For high-order $p$- and $g$-modes 
(\citealt{Unno:89}; WAB)
\bea
\left(p_r, p_h\right)&\simeq& \frac{A_p}{\omega_p}\left(\cos k_p r,\, \frac{c_s\sin k_p r}{\omega_p r}\right)\propto \left(\omega_p^{-1}, \omega_p^{-2}\right),
\hspace{0.5cm}
\\
\left(g_r, g_h\right)&\simeq& \frac{A_g}{\omega_g}\left(\frac{\omega_g \sin k_g r}{N}, \,\frac{\cos k_g r}{\Lambda_g}\right)\propto \left(1, \omega_g^{-1}\right),
\eea
where $A_{p,g}=\sqrt{E_0 \alpha_{p,g}/\rho r^2}$, $\alpha_p = c_s^{-1}/\int c_s^{-1} dr$, and  $\alpha_p = (N/r)/\int N d\ln r$. 
The $\epsilon\mathcal{O}(\omega_g)$ corrections consist of terms like, e.g., 
\beq
\frac{T}{E_0}\int dr \rho \left(4g +r \d{g}{r}\right)\left(r\div{\gv{\chi}} + \d{\ln\rho}{\ln r}\chi_r \right) p_r g_r.  
\eeq  
A straightforward but lengthy calculation (that makes extensive use of integration by parts and the linear equations of motion) reveals that for high-order $p$-$g$ coupling,  Equation (\ref{eq:K3pg_monster}) reduces to
\bea
\label{eq:K3pg_finite_freq}
K_{3pg}&=&\frac{1}{E_0}\int d^3x \rho \, \gv{p}\cdot\left[\left(\omega_p^2-\omega^2\right) \gv{\psi}_{\chi g}-\omega_0^2 \gv{\zeta}_{\chi g}\right]
\non &&
+\epsilon\mathcal{O}(\omega_g)
\eea
where 
\bea
\label{eq:psi1}
\gv{\psi}_{\chi g}&\equiv& \left(\gv{g}\cdot\grad\right)\gv{\chi},\\
\label{eq:zeta1}
\gv{\zeta}_{\chi g}&\equiv& \frac{g}{\omega_0^2}\left[\Gamma_1\d{\ln \Gamma_1}{\ln p}-\left(\pd{\ln \Gamma_1}{\ln \rho}\right)_S\right]\div{\gv{\chi}}
\d{g_r}{r}\gv{\hat{r}}.\hspace{0.4cm}
\eea
For the static tide, $\omega=0$ and $\div{\gv{\chi}}=0$ (i.e., $\gv{\zeta}_{\chi g}=0$), and our expression for $K_{3pg}$ matches the expression given in VZH (see their Equations 54, 62, and 69).  The last two terms in the integrand of Equation (\ref{eq:K3pg_finite_freq}) arise due to finite frequency corrections to $K_{3pg}$ and come in at order $\epsilon \mathcal{O}(1)$.

\subsection{A sum rule for $\sum K_{3p g}K_{3\bar{p} g}$} 
 \label{sec:k3sq_sum_rule}
 
We need to evaluate the finite frequency corrections to $\sum_p K_{3p g}K_{3\bar{p} g}$, where the sum is over all $p$-modes. In this section we derive a sum rule for $\sum_p K_{3p g}K_{3\bar{p} g}$ that reduces the sum to a spatial integral that involves only the displacements of the tide and the $g$-modes (i.e., the $p$-modes do not explicitly enter the calculation).  This allows us to evaluate  $\sum_p K_{3p g}K_{3\bar{p} g}$  much more accurately than if we had to numerically sum over $p$-modes. For simplicity, in this section index $a$ labels the linear tide (whose frequency we continue to write as $\omega$) and indices $c$ and $d$ label the $g$-modes.

To derive the sum rule, we adopt the procedure \citet{Reisenegger:94a} used to find a sum rule for the linear tide overlap integral. First, we expand the vectors $\gv{\psi}_{ac}$ and $\gv{\zeta}_{ac}$ (Equations \ref{eq:psi1} and \ref{eq:zeta1}) as a sum over modes
\bea
\gv{\psi}_{ac}&=&\sum_b M_{abc}\gv{\xi}_b,\\
\gv{\zeta}_{ac}&=&\sum_b N_{abc}\gv{\xi}_b,
\eea
where $M_{abc}$ and $N_{abc}$ are the coefficients of the expansion
which, from the orthogonality of the modes (Equation \ref{eq:mode_norm}), are given by
\bea
M_{abc}&=& \frac{\omega_b^2}{E_0}\int d^3x \rho \, \gv{\xi}_b^\ast \cdot  \gv{\psi}_{ac},\\
N_{abc}&=& \frac{\omega_b^2}{E_0}\int d^3x \rho \, \gv{\xi}_b^\ast \cdot  \gv{\zeta}_{ac}.
\eea
Comparing with  Equation (\ref{eq:K3pg_finite_freq}),  we have
\beq
\label{eq:k3_MN}
K_{3\bar{b}c}= 
\frac{\left(\omega_b^2-\omega^2\right)M_{abc}-\omega_0^2N_{abc}}{\omega_b^2}+\epsilon\mathcal{O}(\omega_c)
\eeq
for a $p$-mode $\gv \xi_b$. Let $\gv{f}_1[\gv{\eta}]$ be the standard linear oscillation operator acting on some arbitrary vector $\gv{\eta}$.  For a linear eigenmode $\gv{\xi}_a$, $\gv{f}_1[\gv{\xi}_a]= -\rho \omega^2 \gv{\xi}_a$. Therefore
\bea
\int d^3x \,  \gv{\psi}_{ac}^\ast \cdot \gv{f}_1\left[\gv{\psi}_{ad}\right] 
 &=& \sum_{be} M_{abc}^\ast M_{aed}  \int d^3x \, \gv{\xi}_b^\ast  \cdot \gv{f}_1\left[\gv{\xi}_e\right]
\non &=& -E_0 \sum_b M_{abc}^\ast M_{abd},
\eea
and we find the sum rule
\beq
 \sum_b M_{abc}^\ast M_{abd} =-\frac{1}{E_0}  \int d^3x \,  \gv{\psi}_{ac}^\ast \cdot \gv{f}_1\left[\gv{\psi}_{ad}\right]. 
\eeq
Using the same procedure, we find the sum rules
\bea
 \sum_b \frac{M_{abc}^\ast M_{abd}}{\omega_b^2} &=&
 \frac{1}{E_0}\int d^3x \rho \,  \gv{\psi}_{ac}^\ast \cdot \gv{\psi}_{ad} \\
 \sum_b \frac{M_{abc}^\ast N_{abd}}{\omega_b^2} &=&
 \frac{1}{E_0}\int d^3x \rho \,  \gv{\psi}_{ac}^\ast \cdot \gv{\zeta}_{ad}. 
\eea

Note that these are sums over all modes ($p$-modes, $g$-modes, and the fundamental ($f$-) modes). However, based on the stability analysis of  \S~\ref{sec:eom} (see Equation \ref{eq:char_pg}), we are interested in the restricted sum
\beq
K_{4\bar{c}d}\;+\sum_{b\, \in\, \{p\}}K_{3b\bar{c}} K_{3\bar{b}d},
\eeq
where the sum is only over $p$-modes and
\bea
\label{eq:sum_k3pg_v1}
&&\sum_{b\, \in\, \{p\}}K_{3b\bar{c}} K_{3\bar{b}d}=
\sum_{b\, \in\, \{p\}}K_{3\bar{b}c}^\ast K_{3\bar{b}d}
\non &&=
\sum_{b\, \in\, \{p\}}  \left[\left(1-\frac{\omega^2}{\omega_b^2}\right)M_{abc}^\ast- \frac{\omega_0^2}{\omega_b^2}N_{abc}^\ast\right]
\non &&
\hspace{2.0cm}
\times
 \left[\left(1-\frac{\omega^2}{\omega_b^2}\right)M_{abd}  - \frac{\omega_0^2}{\omega_b^2}N_{abd}\right]
\non &&=
\sum_{b\, \in\, \{p\}} \bigg[M_{abc}^\ast M_{abd} - 2\frac{\omega^2}{\omega_b^2}M_{abc}^\ast M_{abd} 
\non &&
\hspace{0.4cm}
- \frac{\omega_0^2}{\omega_b^2}\left(M_{abc}^\ast N_{abd}+M_{abd}N_{abc}^\ast\right)\bigg]
 +\epsilon^2\mathcal{O}\left(\omega_c^{-1}\right)
\eea
(the error is $\epsilon^2\mathcal{O}(\omega_c^{-1})$ because $K_{3bc}\sim \epsilon\mathcal{O}(\omega_c^{-2})$ and we accounted for all terms bigger than $\epsilon\mathcal{O}(\omega_c)$ in $K_{3bc}$).  If $b$ is a $p$-mode then $M_{abc}\sim \mathcal{O}(\omega_c^{-2})$ and if $b$ is a $g$-mode then $M_{abc}\sim \mathcal{O}(1)$.  The sum over all modes $\sum M_{abc}^\ast M_{abd}$ is therefore dominated by the $p$-modes, which contribute at $\mathcal{O}(\omega_c^{-4})$; the $g$-modes contribute at only $\mathcal{O}(1)$. I.e.,
\beq
 \sum_{b\,\in\, \{p, f, g\}}M_{abc}^\ast M_{abd}
 =\sum_{b\,\in\, \{p\}}M_{abc}^\ast M_{abd}+\epsilon^2\mathcal{O}(1).
\eeq
 However, for the sums over all modes $\sum M_{abc}^\ast M_{abd}/\omega_b^2$ and $\sum M_{abc}^\ast N_{abd}/\omega_b^2$, the $p$-modes and $g$-modes both contribute at $\mathcal{O}(\omega_c^{-2})$. Since $K_{3pg}$ is large only for $k_b\simeq k_c$, we show in \S~\ref{sec:check4} that we can nonetheless accurately calculate these sums using the approximate relation
\beq
\sum_{b\,\in\, \{p\}} \frac{M_{abc}^\ast M_{abd}}{\omega_b^2}
\simeq \frac{1}{\langle \omega_b^2\rangle}\sum_{b\,\in\, \{p,f,g\}} M_{abc}^\ast M_{abd}
\eeq
(and similarly for $\sum M_{abc}^\ast N_{abd}/\omega_b^2$),
where 
\beq
\langle\omega_b^2\rangle(r) \equiv \left(\frac{\Lambda_c N c_s}{\omega_c r}\right)^2  
\label{eq:avg_wp2},
\eeq
which follows from the condition $k_b\simeq k_c$.  We can therefore express the sum over $p$-modes as a spatial integral involving only the tide and the $g$-modes
\bea
&&\sum_{b\, \in\, \{p\}} K_{3bc} K_{3\bar{b}d}
\non &&\simeq
- \frac{1}{E_0}\int d^3x \,\bigg[ \left(1- \frac{2\omega^2}{\langle \omega_b^2\rangle}\right)\gv{\psi}_{ac} \cdot \gv{f}_1\left[\gv{\psi}_{ad}\right]
\non &&
- \frac{\omega_0^2}{\langle \omega_b^2\rangle}\left(\gv{\psi}_{ac} \cdot  \gv{f}_1\left[\gv{\zeta}_{ad}\right]+\gv{\psi}_{ad}\cdot \gv{f}_1\left[\gv{\zeta}_{ac}\right]\right)\bigg]
 +\epsilon^2\mathcal{O}(\omega_c^{-1}),
 \non
\label{eq:K3sq_sum_rule}
\eea
where we used the fact that $\gv a = \gv a^\ast$ since the tidal displacement is real.  In Appendix \ref{app:sum_rule_useful_form} we show that this result can be expressed in the following form (which will prove useful for comparing with $K_{4cd}$ in \S~\ref{sec:corrections_wc3&4})
\bea
&&\sum_{b\, \in\, \{p\}} K_{3bc} K_{3\bar{b}d}
\non &&\simeq
\frac{1}{E_0}\int d^3x
\bigg\{\Gamma_1 p \Big[\big(a^i_{;j}c^j_{;i}\big)\left(a^k_{;s}d^s_{;k}\right)+  \big(a^i_{;j}c^j_{;i}\big)\left(d^k a^s_{;ks}\right)
\non &&
+ \big(c^i a^j_{;ij}\big)\left(a^k_{;s}d^s_{;k}\right)
 + \big(c^i a^j_{;ij}\big)\left(d^k a^s_{;ks}\right)\Big]
 \non &&
-\frac{\rho g}{r} \left[ \psi_{ac,r} z_{ad}+\psi_{ad,r} z_{ac} -\left(2+\d{\ln g}{\ln r}\right)\psi_{ac,r} \psi_{ad,r} \right] 
\non &&
 -\frac{\Gamma_1 p}{\langle \omega_b^2\rangle}
  \bigg[2\omega^2\d{\psi_{ac,r}}{r}\d{\psi_{ad,r}}{r}
  \non &&
  + \omega_0^2\left(\d{\psi_{ac,r}}{r}\d{\zeta_{ad,r}}{r}
+\d{\zeta_{ac,r}}{r}\d{\psi_{ad,r}}{r}\right)\bigg]\bigg\}
+\epsilon^2\mathcal{O}(\omega_c^{-1}),
\non
\label{eq:k3pg_sq_final}
\eea
where the subscript semicolon denotes covariant derivative and
\beq
 z_{ac}(r,\theta,\phi)\equiv
r\left[ \div{\gv{\psi}_{ac}} -\d{\psi_{ac,r}}{r}\right].
\eeq
 The first term in Equation (\ref{eq:k3pg_sq_final}) is $\mathcal{O}(\omega_c^{-4})$, the next two terms are $\mathcal{O}(\omega_c^{-3})$, and the remaining terms are each $\mathcal{O}(\omega_c^{-2})$.
In \S~\ref{sec:testing_the_calc} we compare calculations of $\sum K_{3\bar{b}c} K_{3bd}$ using both methods --- explicitly summing over $p$-modes versus using the sum rule (Equation \ref{eq:K3sq_sum_rule}) --- and show that they are in good agreement.

\subsection{Four-mode coupling coefficient $\kappa_{abcd}$}
\label{sec:kap4}

In order to calculate $K_{4cd}$ (Equation \ref{eq:K4ab}) we need to evaluate the four-mode coupling coefficient $\kappa_{abcd}$.  Using a Hamiltonian formalism, \citet{VanHoolst:94} derives an expression for $\kappa_{abcd}$ (see  \citealt{VanHoolst:93} for an alternative derivation).  In terms of a displacement $\gv{\xi}$, he shows that the fourth-order coupling density is (Equation 49 in Van Hoolst with our normalization and in the Cowling approximation) 
\bea
&&\d{\kappa_4}{V}=-\frac{1}{6E_0}
\bigg[
G_1p\left(\div{\gv{\xi}}\right)^4+
6G_2p\left(\div{\gv{\xi}}\right)^2\xi^i_{;j}\xi^j_{;i}
\non &&
+3\Gamma_1p\big(\xi^i_{;j}\xi^j_{;i}\big)\left(\xi^k_{;s}\xi^s_{;k}\right)
+8\Gamma_1p\div{\gv{\xi}}\xi^i_{;j}\xi^j_{;k}\xi^k_{;i}
\non &&
+\rho\xi^i\xi^j\xi^k\xi^s\phi_{;ijks}
\bigg]
\non &&=
-\frac{\Gamma_1 p}{6E_0}
\bigg[
\left(\frac{G_1}{\Gamma_1}-4\right)\left(\div{\gv{\xi}}\right)^4
+3\big(\xi^i_{;j}\xi^j_{;i}\big)\left(\xi^k_{;s}\xi^s_{;k}\right)
\non &&
+6\left(\frac{G_2}{\Gamma_1}+2\right)\left(\div{\gv{\xi}}\right)^2\xi^i_{;j}\xi^j_{;i}
+24\div{\gv{\xi}}\det\left|\xi^i_{;j}\right|
\non &&
+c_s^{-2}\xi^i\xi^j\xi^k\xi^s\phi_{;ijks}
\bigg],
\label{eq:dk4_dV}
\eea
where
\bea
\label{eq:G1}
G_1 &\equiv& \Gamma_1\left(\Gamma_1^2-3\Gamma_1+3\right)+3\left(\Gamma_1-1\right)\left(\pd{\Gamma_1}{\ln \rho}\right)_S
\non &&
+\left(\pdd{\Gamma_1}{\ln\rho}\right)_S,\\
\label{eq:G2}
G_2 &\equiv& \Gamma_1\left(\Gamma_1-2\right)+\left(\pd{\Gamma_1}{\ln \rho}\right)_S
\eea
(subscript $S$ indicates derivatives taken at constant entropy).
The second equality in the equation for $d\kappa_4/dV$ follows from the relation (\citealt{Wu:01, Schenk:02})
\beq
\det \left|\xi^i_{;j}\right| = \frac{1}{6}\left(\div{\gv{\xi}}\right)^3-\frac{1}{2}\div{\gv{\xi}}\xi^i_{;j}\xi^j_{;i}+\frac{1}{3}\xi^i_{;j}\xi^j_{;k}\xi^k_{;i}.
\eeq

Consider a perturbation $\gv\xi=\gv{a}+\gv{c}+\gv{d}$, where $a$ is the linear tide and $c$ and $d$ are high-order $g$-modes.  The ordering of terms in $d\kappa_4/dV$ is then
\bea
\left(a^i_{;j}c^j_{;i}\right)\left(a^i_{;j}d^j_{;i}\right)&\sim& \mathcal{O}(\omega_c^{-4}),\non
\div{\gv{a}}\det\left|(a+c+d)^i_{;j}\right| &\sim& \mathcal{O}(\omega_c^{-3}),\non
\left(\div{\gv{a}}\right)^2 \,c^i_{;j} d^j_{;i}&\sim& \mathcal{O}(\omega_c^{-2}),\non
\left(a^i_{;j}a^j_{;i}\right)\left(c^k_{;s}d^s_{;k}\right)&\sim& \mathcal{O}(\omega_c^{-2}),\non
\div{\gv{a}}\div{\gv{c}} \,a^i_{;j} d^j_{;i}+\left(c\leftrightarrow d\right)&\sim& \mathcal{O}(\omega_c^{-2}),\non
\div{\gv{c}}\det\left|(a+a+d)^i_{;j}\right|+\left(c\leftrightarrow d\right)&\sim& \mathcal{O}(\omega_c^{-2}),\non
a^i a^j c^k d^s\phi_{;ijks}&\sim& \mathcal{O}(\omega_c^{-2}),\non
\left(\div{\gv{a}}\right)^2\left(\div{\gv{c}}\right)\left(\div{\gv{d}}\right) &\sim& \mathcal{O}(1),\non
\left(\div{\gv{c}}\right)\left(\div{\gv{d}}\right) \, a^i_{;j} a^j_{;i}&\sim& \mathcal{O}(1),
\label{eq:k4_ordering_of_terms}
\eea
where we assume the strong coupling condition for the $g$-modes $k_c\simeq k_d$ and thus $\omega_c\approx \omega_d$.
The four-mode coupling coefficient, accurate to $\mathcal{O}(\omega_c^{-2})$, is therefore
\bea
&&
\kappa_{aacd}=
\non && -\frac{1}{6E_0}\int d^3x \Gamma_1 p\bigg[
2\big(a^i_{;j}c^j_{;i}\big)\left(a^k_{;s}d^s_{;k}\right)+\big(a^i_{;j}a^j_{;i}\big)\left(c^k_{;s}d^s_{;k}\right)
\non &&
+c_s^{-2} a^i a^j c^k d^s \phi_{;ijks}
+2\div{\gv{a}}\det\left|\left(a+c+d\right)^i_{;j}\right|
\non &&
+\div{\gv{c}}\det\left|\left(a+a+d\right)^i_{;j}\right|
+\div{\gv{d}}\det\left|\left(a+a+c\right)^i_{;j}\right|
\non &&
+\left(\frac{G_2}{\Gamma_1} +2\right)
\Big(\left(\div{\gv{a}}\right)^2 c^i_{;j}d^j_{;i} + 2\left(\div{\gv{a}}\right) \left(\div{\gv{c}}\right) a^i_{;j}d^j_{;i} 
\non && 
+2\left(\div{\gv{a}}\right) \left(\div{\gv{d}}\right) a^i_{;j}c^j_{;i}\bigg) \bigg] +\epsilon^2\mathcal{O}(\omega_c^{-1}).
\label{eq:kaacd}
\eea
This form for $\kappa_{aacd}$ will be useful for carrying out the analytic calculations in \S~\ref{sec:corrections_wc3&4} at $\mathcal{O}(\omega_c^{-4})$ and $\mathcal{O}(\omega_c^{-3})$.  In Appendix~\ref{app:friendly_four_mode_coup_coef} we give an alternative form for $\kappa_{aacd}$ that is less compact but more suitable for carrying out the $\mathcal{O}(\omega_c^{-2})$ numerical calculations in \S~\ref{sec:corrections_wc2} and \S~\ref{sec:results_finite_freq}.

\subsection{Finite frequency corrections at $\mathcal{O}(\omega_g^{-4})$ and $\mathcal{O}(\omega_g^{-3})$}
\label{sec:corrections_wc3&4}

Having obtained expressions that account for finite frequency corrections to the coupling coefficients  to $\mathcal{O}(\omega_g^{-2})$, we now evaluate 
$K_{4gg}+\sum K_{3pg}K_{3\bar{p}g}$ and determine whether the cancellation (Equation \ref{eq:K4_K3sq_cancellation}) still holds. In this section we consider the finite frequency corrections that enter at $\mathcal{O}(\omega_g^{-4})$ and $\mathcal{O}(\omega_g^{-3})$ and show analytically that they do not undo the cancellation. In \S~\ref{sec:corrections_wc2} we consider the finite frequency corrections that enter at $\mathcal{O}(\omega_g^{-2})$.  In the numerical calculations of \S~\ref{sec:results_finite_freq} we find that these corrections do undo the cancellation.

If we again let index $a$ label the linear tide $\gv{\chi}^{(1)}$ and indices $c$ and $d$ label the $g$-modes, then the  $\mathcal{O}(\omega_c^{-4})$ and $\mathcal{O}(\omega_c^{-3})$ contributions to the coupling coefficient (Equation \ref{eq:K4ab}) 
\beq
K_{4cd} =  2\kappa_{\chi^{(2)}cd} + 3\kappa_{aacd}
\eeq
come entirely from the $\mathcal{O}(\omega_c^{-4})$ and $\mathcal{O}(\omega_c^{-3})$ contributions to $\kappa_{aacd}$. From the expressions for three-mode coupling in WAQB, we know that $\kappa_{\chi^{(2)}cd}$ will at most contribute at $\mathcal{O}(\omega_c^{-2})$.  Comparing $\kappa_{aacd}$ (Equation \ref{eq:kaacd}) to our sum rule result (Equation \ref{eq:k3pg_sq_final}) we immediately see that the $\mathcal{O}(\omega_c^{-4})$ terms $\left(a^i_{;j}c^j_{;i}\right)\left(a^k_{;s}d^s_{;k}\right)$ cancel upon calculating $3\kappa_{aacd} + \sum K_{3bc} K_{3\bar{b}d}$.

We now show that the $\mathcal{O}(\omega_c^{-3})$ terms cancel too.
The remaining terms in $K_{4cd}+\sum K_{3bc} K_{3\bar{b}d}$ that contribute at $\mathcal{O}(\omega_c^{-3})$ are, by Equations (\ref{eq:k3pg_sq_final}) and  (\ref{eq:kaacd}),
\bea
&& 
3\kappa_{aacd}+\sum_b K_{3bc} K_{3\bar{b}d}
\non&&=\!\frac{1}{E_0}\int \! d^3x 
\Gamma_1 p  \bigg\{\left[
 \big(a^i_{;j}c^j_{;i}\big) \left(d^k a^s_{;sk}\right)+\big(c^i a^j_{;ij}\big) \left(a^k_{;s} d^s_{;k}\right)
\right]
\non &&
-\div{\gv{a}}\det\left|(a+c +d)^i_{;j}\right|\bigg\}
+ \epsilon^2\mathcal{O}(\omega_c^{-2})
\non &&=\!
\frac{1}{E_0}\!\int \! dr \Gamma_1 p \div{\gv{a}}\left(a_r-a_h\right)\! 
\bigg[c_h\pd{d_h}{r}\!\left(\!E_{ac,ad}^{(22)}-S_{a,ad, c}\!\right)
\non && 
+d_h\pd{c_h}{r}\left(\!E_{ac,a d}^{(22)}-S_{a,ac,d}\!\right)\bigg]
+ \epsilon^2\mathcal{O}(\omega_c^{-2}),
\eea
where we used results from Appendix \ref{app:friendly_four_mode_coup_coef} (specifically, Equations \ref{eq:hab1}--\ref{eq:hab3} and the four-mode angular integrals defined in Equations \ref{eq:Tabcd}--\ref{eq:Sabcd}). In Appendix \ref{app:Sabcd_Eabcd_relation} we show that the angular integrals satisfy the relation
\beq
\label{eq:ESsubtract}
E_{ac,bd}^{(22)}-S_{a,bd, c} = 
E_{bc,a d}^{(22)}-S_{a,bc,d}.
\eeq
The bracketed term in the previous equation is therefore proportional to $\partial (c_h d_h)/\partial r$ and the expression is seen to contribute only at $\mathcal{O}(\omega_c^{-2})$ after integration by parts. 

We have thus shown that the finite frequency corrections that enter Equation (\ref{eq:K4_K3sq}) at $\mathcal{O}(\omega_c^{-4})$ and $\mathcal{O}(\omega_c^{-3})$ do not undo the cancellation between three- and four-mode coupling.  

\subsection{Finite frequency corrections at $\mathcal{O}(\omega_g^{-2})$}
\label{sec:corrections_wc2}
 
There are a large number of terms that contribute at $\mathcal{O}(\omega_c^{-2})$ to $K_{4gg}+\sum K_{3pg}K_{3\bar{p}g}$ and they must be evaluated numerically.  In Appendix \ref{app:Komc2} we describe how we carry out the calculation.  First, we give numerically useful expressions for the terms in $3\kappa_{\chi^{(1)}\chi^{(1)}gg} + \sum K_{3pg} K_{3\bar{p}g}$. We then consider two methods for calculating the nonlinear tide $\gv{\chi}^{(2)}$, which we need in order to compute $2\kappa_{\chi^{(2)}gg}$.  The first method involves expanding $\gv\chi^{(2)}$ as a sum over modes and solving for the coefficients of the expansion.  The second method involves directly solving the inhomogeneous equation of motion of the nonlinear tide (Equation \ref{eq:chi2}).  We find that the second method is significantly more accurate.
We will present the results of the $\mathcal{O}(\omega_c^{-2})$ calculation in \S~\ref{sec:results_finite_freq}.

\section{Testing the calculation}
\label{sec:testing_the_calc}
 
In order to test the accuracy of our analytic and numerical calculations, in this section we perform the following four checks:
 \begin{enumerate}
 \item In  \S~\ref{sec:check1} we evaluate whether our numerical calculations can recover the expected cancellation to $\mathcal{O}(\epsilon^2)$ in the case of the static tide.
 \item In \S~\ref{sec:check2} we compute  $3\kappa_{\chi^{(1)}\chi^{(1)}gg} +\sum K_{3pg} K_{3\bar{p}g}$ by explicitly summing over $p$-modes and integrating the full four-mode coupling coefficient.  We then compare this with the analytic method described in \S\S~\ref{sec:corrections_wc3&4} and \ref{sec:corrections_wc2} in which the cancellation at $\mathcal{O}(\omega_c^{-4})$ and $\mathcal{O}(\omega_c^{-3})$ is first carried out by hand and then the remaining $\mathcal{O}(\omega_c^{-2})$ contributions computed numerically.
 \item In \S~\ref{sec:check3} we check our solution of the nonlinear tide  $\gv\chi^{(2)}$ by comparing results from our two calculational methods (solving for the coefficients of the mode expansion versus directly solving the inhomogeneous equation).
   \item In \S~\ref{sec:check4} we compute the sum over $1/\omega_b^{2}$ terms in Equation (\ref{eq:sum_k3pg_v1}) by directly summing over $p$-modes and compare this result with our approximate analytic expression for the sum (the $1/\langle \omega_b^2\rangle$ terms in Equation \ref{eq:k3pg_sq_final}).
 \end{enumerate}

\subsection{Testing the static tide cancellation}
\label{sec:check1}

If we only include the $m=0$ component of the linear tide $\gv\chi^{(1)}$ in our calculation, then there are no finite frequency corrections and based on the analysis in VZH and the argument from energy principles given in Appendix \ref{app:static_tidal_field}, we expect near-exact cancellation (i.e., cancellation to $\mathcal{O}(\epsilon^2)$; see Equation \ref{eq:K4_K3sq_cancellation}). We have already shown analytically that the $\mathcal{O}(\omega_g^{-4})$ and $\mathcal{O}(\omega_g^{-3})$ contributions cancel.  We now want to test whether in the case of the $m=0$ static tide our numerical calculation of the $\mathcal{O}(\omega_g^{-2})$ terms cancel as expected.  To provide a measure for the degree of cancellation, define the residual
\bea
\label{eq:Rgg}
\mathcal{R}_{g g}
&=&
K_{4gg}+\sum_{p} K_{3pg} K_{3\bar{p} g}
\non &=&
2\kappa_{\chi^{(2)} g g}+3\kappa_{\chi^{(1)}\chi^{(1)}g g} + \sum_{p} K_{3pg} K_{3\bar{p} g}
\eea
and the fractional residual $\sigma_{gg}$ 
\beq
\label{eq:sigma}
\sigma_{gg}=\left|\frac{\mathcal{R}_{gg}}{2\kappa_{\chi^{(2)} g g}} \right|.
\eeq
For the case of the static tide, we expect $\mathcal{R}_{g g}\sim\mathcal{O}(\epsilon^2)$ and  $\sigma_{gg} \sim (\omega_g / \omega_0)^2$ since $\kappa_{\chi^{(2)}gg}\sim \epsilon^2 (\omega_0/\omega_g)^2$.  

To carry out this calculation, we only include the $m=0$ linear tide and compute the coupling coefficients using Equations (\ref{eq:t1})--(\ref{eq:t8}) and the direct $\gv{\chi}^{(2)}$ solution (Equation \ref{eq:chi2_direct}).  We find the $g$-modes (and the $p$-modes in the calculations below) using the Aarhus adiabatic oscillation package ADIPLS \citep{Dalsgaard:08}. At the maximum resolution, our background model and eigenfunctions have $\simeq 5\times10^5$ grid points over the star.

\begin{figure}
\centering
\includegraphics[width=3.3in]{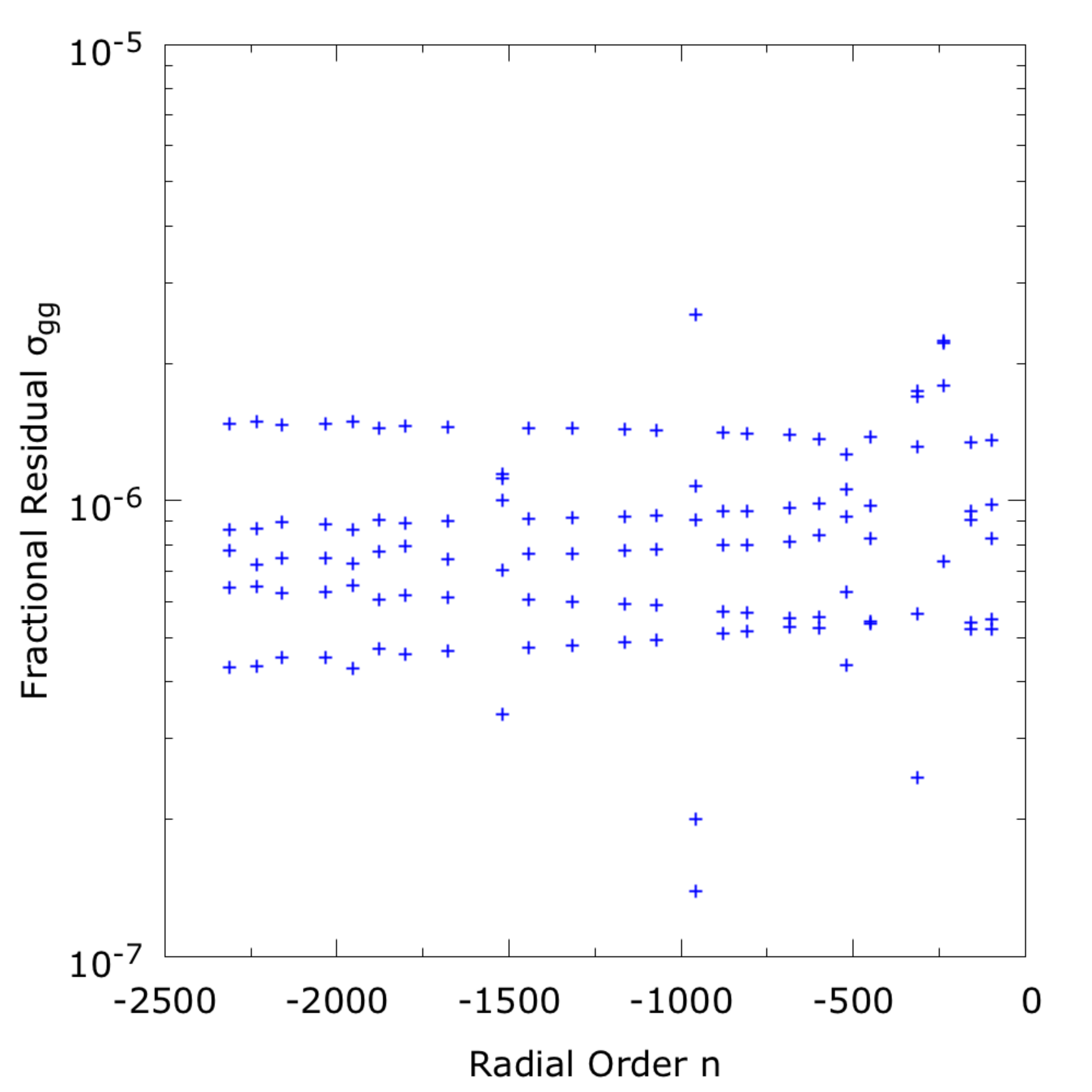} 
\caption{Fractional residual $\sigma_{gg}$ as a function of $g$-mode radial order $n$ for the case when only the static ($m=0$) tide is included.  Each point is for an $\ell=4$ daughter pair of the neutron star model with $n_1=n_2$ and azimuthal orders that must satisfy $m_1+ m_2=0$. 
\label{fig:static_tide_frac_diff}}
\end{figure}

In Figure \ref{fig:static_tide_frac_diff} we show  $\sigma_{gg}$ for the static tide coupled to high-order $\ell=4$ $g$-modes with $100\la n \la 1000$. 
We find $\sigma_{gg} \sim 10^{-6}$ over the entire range in $n$, implying that the individual terms in $\mathcal{R}_{gg}$ cancel to a part in $\sim 10^6$ in our calculation.  For these modes, $\omega_g \approx 0.2\omega_0/n$ and we therefore expect $10^{-8} \la \sigma_{gg} \la 10^{-6}$.   However, given the resolution set by the $5\times10^5$ background grid points,  a value of  $\sigma_{gg} \sim 10^{-6}$ is  consistent with near-exact cancellation (we checked that $\sigma_{gg}$ increases in proportion to decreasing background resolution).  We therefore conclude that our calculation successfully recovers the correct result in the static tide (i.e., incompressible) limit.

\subsection{Testing the calculation of  $3\kappa_{\chi\chi gg} +\sum K_{3pg}K_{3\bar{p}g}$}
\label{sec:check2}

In order to accurately calculate 
\beq
\label{eq:k4_plus_k3sq}
3\kappa_{\chi^{(1)}\chi^{(1)}gg} + \sum_{p} K_{3pg} K_{3\bar{p} g}
\eeq
we carried out a series of analytic steps, which we described in detail in \S~\ref{sec:finite_freq}.  Briefly summarizing these steps, we showed in \S~\ref{sec:k3sq_sum_rule} that the sum
over $p$-modes can be expressed as an integral that involves only the tide and the $g$-modes.  After simplifying that integral (Appendix \ref{app:sum_rule_useful_form}), we showed in \S~\ref{sec:corrections_wc3&4} that the two cancel to $\mathcal{O}(\omega_g^{-4})$ and $\mathcal{O}(\omega_g^{-3})$ even with finite frequency corrections.  We then rewrote the $\mathcal{O}(\omega_g^{-2})$ contribution in a more numerically accurate form (Equations \ref{eq:t1}--\ref{eq:t8}).

\begin{figure}
\centering
\includegraphics[width=3.3in]{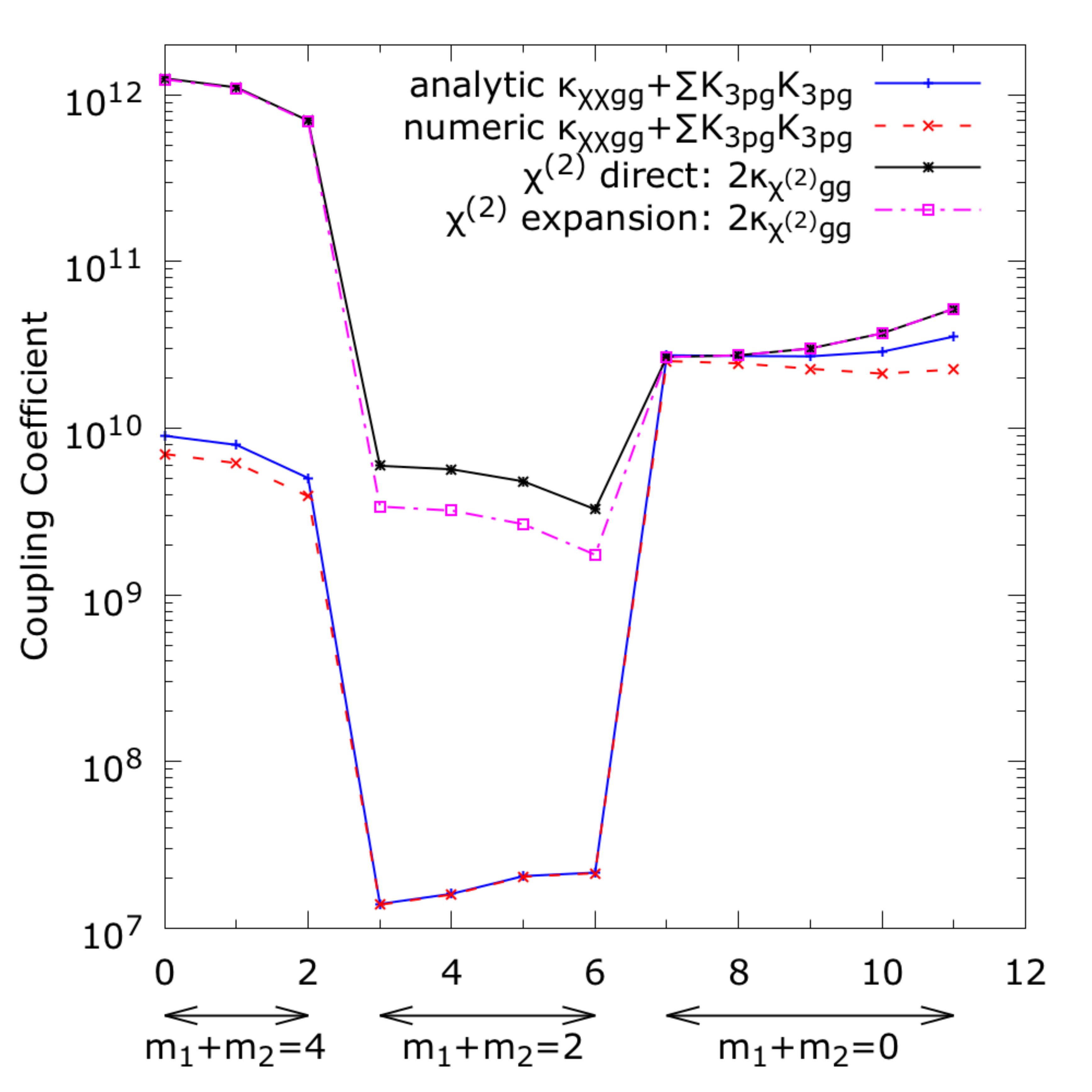} 
\caption{Comparison of coupling coefficients calculated using the analytic approach and the direct numerical approach. We show results for the $\ell=2$, $n=1000$ $g$-mode pairs of the $n_{\rm poly}=2$, $A=0.1\omega_0^2$ polytrope at $a/R=1.23$ (corresponding to the $f$-mode resonance). The first three points along the $x$-axis correspond to the $g$-mode pairs with $m_1+m_2=4$, the next four points are those with $m_1+m_2=2$, and the final five points are those with $m_1+m_2=0$. The blue solid line shows the ``analytic" calculation of  Equation (\ref{eq:k4_plus_k3sq}) as given by the sum of the eight terms discussed in Appendix \ref{app:3kaagg_sum_k3pg_sq}. The red dashed line shows the numerical calculation of  Equation (\ref{eq:k4_plus_k3sq}) as given by explicitly summing over $p$-modes. The black solid line shows the calculation of $2\kappa_{\chi^{(2)}g_1 g_2}$ with $\gv{\chi}^{(2)}$ found by directly solving the inhomogeneous Equation (\ref{eq:chi2_direct}).   The magenta dash-dotted line shows the calculation of $2\kappa_{\chi^{(2)}g_1 g_2}$ with  $\gv{\chi}^{(2)}$ found by using the mode expansion solution (Equation \ref{eq:chi2_expansion}). 
\label{fig:numeric_vs_analytic}}
\end{figure}

As a check of these analytic steps and their numerical implementation, we compare our results to a more direct, though less numerically accurate, calculation of Equation (\ref{eq:k4_plus_k3sq}).   We carry out the comparison for the full tide, including the non-static components, in order to test the calculation with the finite frequency corrections. In Figure \ref{fig:numeric_vs_analytic}, the blue solid line shows the results given by the analytic approach summarized  in the preceding paragraph.  The red dashed line shows the results given by the explicit sum over $p$-modes and relying entirely on the numerics to handle cancellations between terms (including the very large $\mathcal{O}(\omega_g^{-4})$ and $\mathcal{O}(\omega_g^{-3})$ contributions). We describe the latter calculation in the next paragraph.  We see that the red and blue lines nearly overlap over the full range in $m$.  While the agreement is very good, it is not perfect.  Given that the analytic approach passes the  static tide test to $\sim10^{-6}$ precision (\S~\ref{sec:check1}), the most likely explanation for the small but $>10^{-6}$ differences between the blue and red lines is numerical imprecision on the part of the red line, which must rely entirely on the numerics to cancel the large $\mathcal{O}(\omega_g^{-4})$ and $\mathcal{O}(\omega_g^{-3})$ terms.   This test should therefore give us confidence in our analytic approach to calculating Equation (\ref{eq:k4_plus_k3sq}) and its numerical implementation.

\begin{figure}
\centering
\includegraphics[width=3.3in]{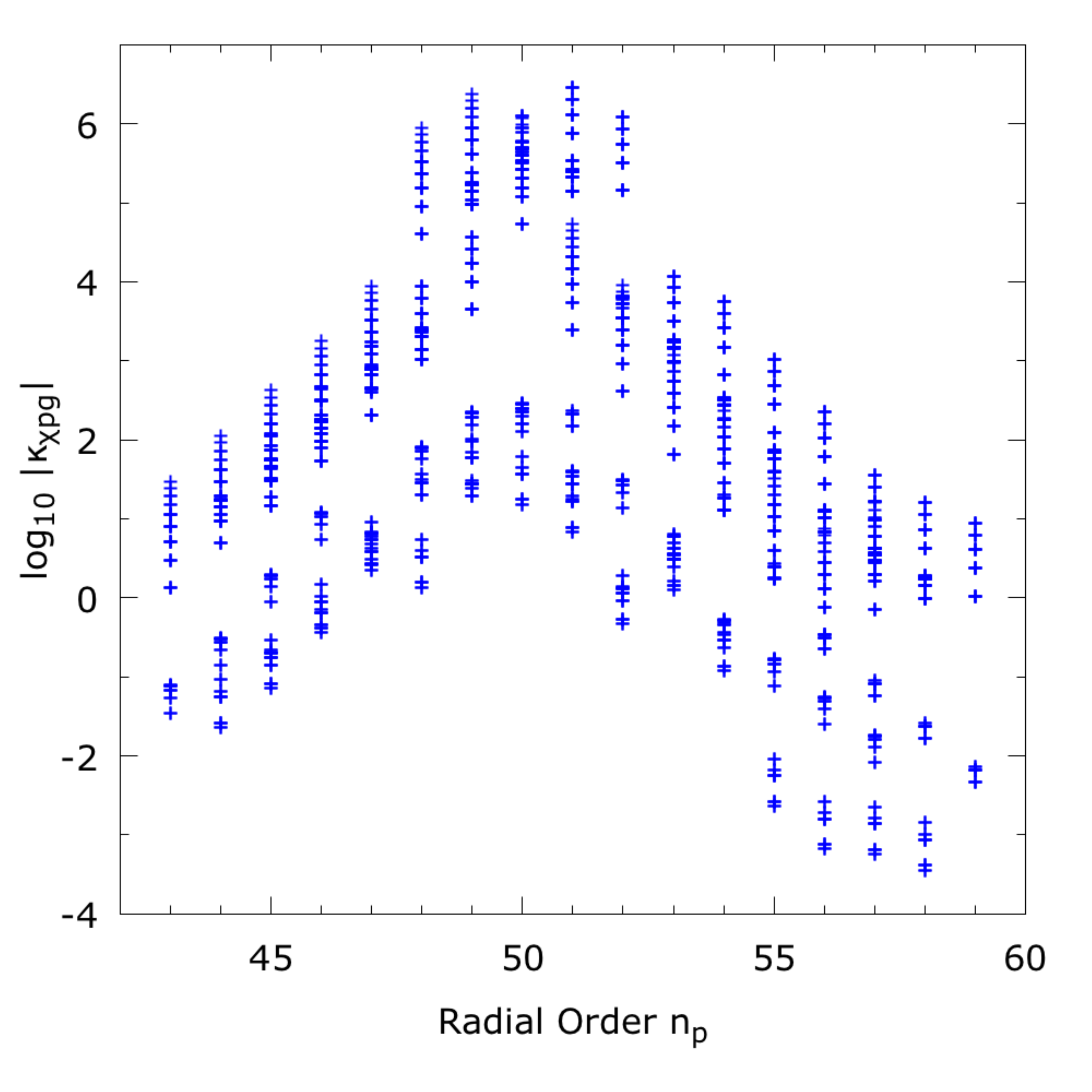} 
\caption{Three-mode coupling coefficient $\kappa_{\chi^{(1)}pg}$ as a function of the $p$-mode radial order $n_p$ for the $p$-modes that dominate the sum given by the blue line in Figure \ref{fig:numeric_vs_analytic}. 
\label{fig:kap3pg_sum_over_pmodes}}
\end{figure}

The $p$-modes that contribute to the sum are those for which $k_p\simeq k_g$, which corresponds to $\omega_p \simeq \Lambda_g N c_s/\omega_g r$.
For a typical stellar model, including our neutron star model, the quantity $Nc_s/r$ varies somewhat with radius, especially outside the stellar core.  As a result, there are many $p$-modes that contribute to the sum.
 In order to minimize the error that arises from summing over many modes, rather than use a neutron star model to carry out the test, we use a polytropic model and set the adiabatic index $\Gamma_1(r)$ such that 
\beq
\frac{Nc_s}{r}=A,
\eeq
where $A$ is a constant.  Specifically, when calculating the polytrope's buoyancy 
\beq
N^2=-g\left[\d{\ln \rho}{r}-\frac{1}{\Gamma_1(r)}\d{\ln p}{r}\right]
\eeq
we set
\beq
\Gamma_1(r) = \gamma \left[1+\left(\frac{A }{g/r}\right)^2\right],
\eeq
where $\gamma=1+1/n_{\rm poly}$ and $n_{\rm poly}$ is the (constant) polytropic index.  The results in Figure \ref{fig:numeric_vs_analytic} are for an $\ell=4$, $n=50$ $g$-mode of an $n_{\rm poly}=2$ polytrope with $A=0.1\omega_0^2$, the latter being similar to the value in the core of our neutron star model.  In Figure \ref{fig:kap3pg_sum_over_pmodes} we show $\kappa_{\chi^{(1)}pg}$, which enters the calculation of $\sum K_{3pg}K_{3\bar{p}g}$, as a function of the radial order $n_p$ of the $p$-modes included in the sum.  The magnitude of $\kappa_{\chi^{(1)}pg}$ peaks sharply for those $p$-modes that best satisfy the condition $k_p= k_g$. For modes that are $\Delta n_p\simeq 10$ away from the peak, $\kappa_{\chi^{(1)}pg}$ is $\sim 10^{5}$ times smaller and therefore the $p$-modes in the tail make only a very small contribution to  $\sum K_{3pg}K_{3\bar{p}g}$ (if we had not set $A$ to be a constant the distribution would be much broader and many more $p$-modes would contribute).  

\begin{figure*}
\centering
\includegraphics[width=3.1in]{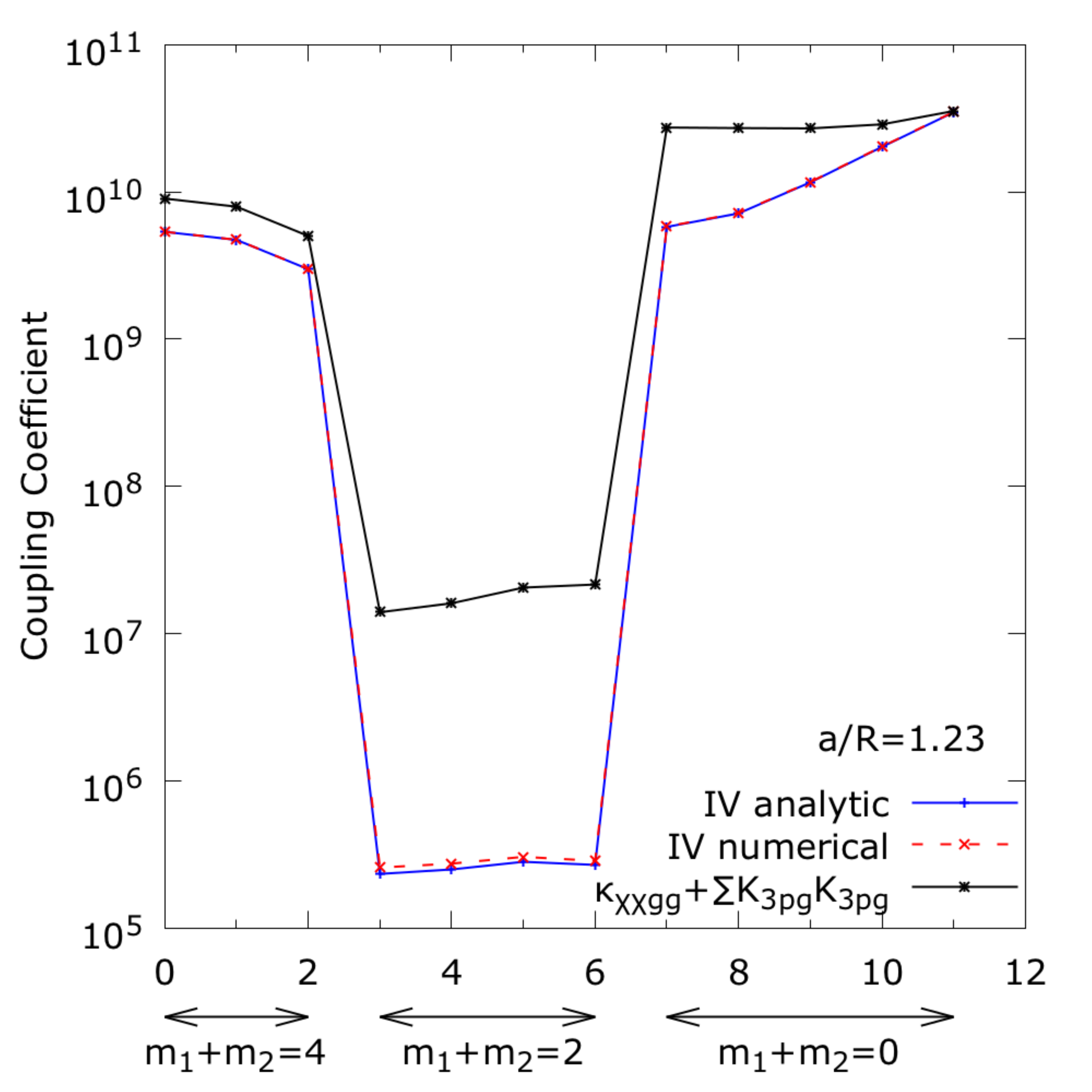} 
\hspace{1.0cm}
\includegraphics[width=3.1in]{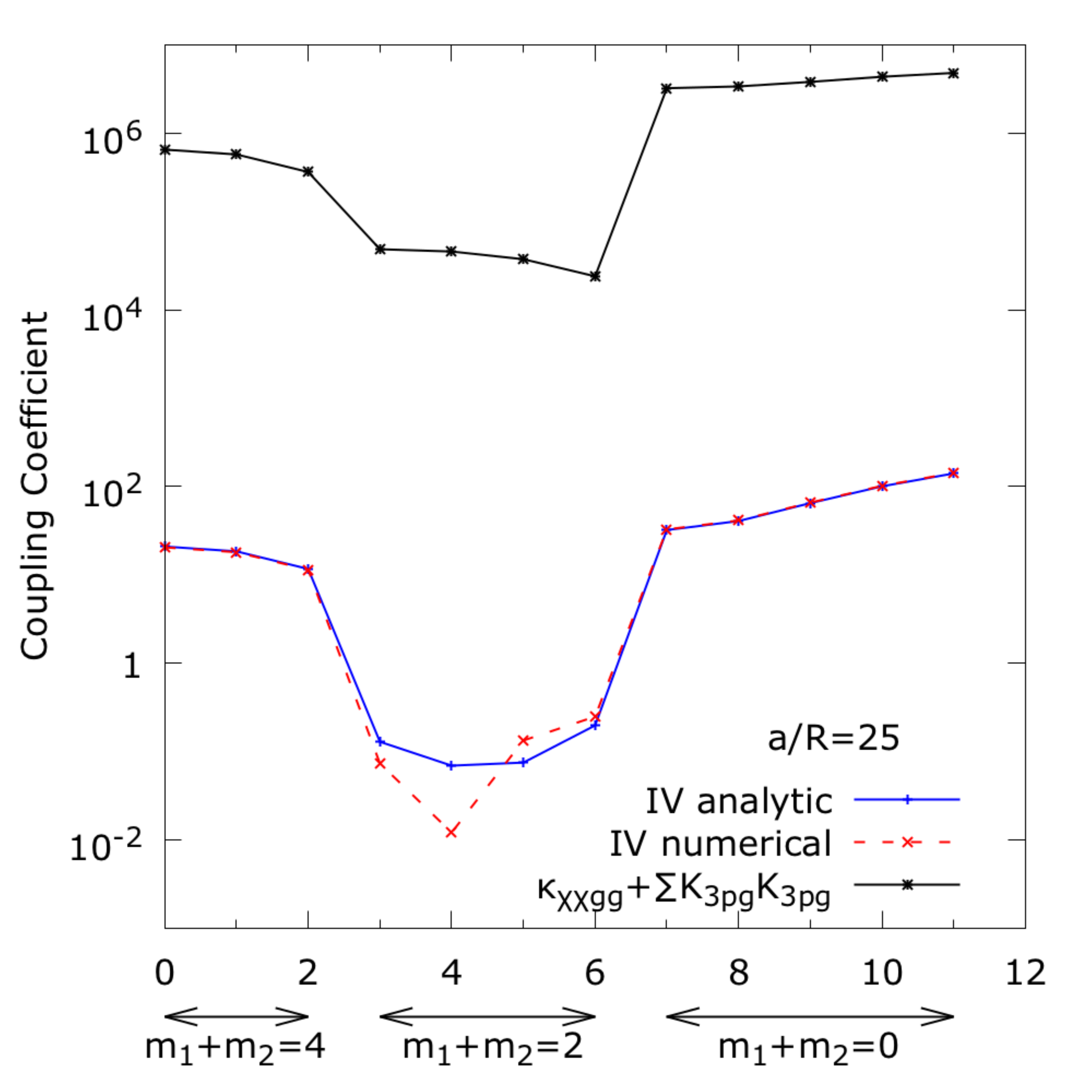} 
\caption{Comparing the two different methods for calculating the $1/\omega_b^2$ coupling terms of $\sum K_{3pg}K_{3\bar{p}g}$ given by Equation (\ref{eq:sum_k3pg_v1}).  As in Figure \ref{fig:numeric_vs_analytic}, we show results for the $\ell=2$, $n=1000$ $g$-mode pairs of the $n_{\rm poly}=2$, $A=0.1\omega_0^2$ polytrope.  The left panel is for $a/R=1.23$ (corresponding to the $f$-mode resonance) and the right panel is for $a/R=25$ (which is not near a linear resonance).  The first three points along the $x$-axis correspond to the $g$-mode pairs with $m_1+m_2=4$, the next four points are those with $m_1+m_2=2$, and the final five points are those with $m_1+m_2=0$.  The blue solid line shows the analytic method for calculating the $1/\omega_b^2$ terms (which makes the approximation described in \S~\ref{sec:k3sq_sum_rule} and is given by term $IV$, Equation \ref{eq:t4}). The red dashed line shows the numerical method for calculating the $1/\omega_b^2$ terms (explicitly summing over $p$-modes). For comparison, the black solid line shows the analytic method for calculating $\kappa_{\chi^{(1)}\chi^{(1)}gg}+\sum K_{3pg}K_{3\bar{p}g}$ given by the eight term sum of Equations (\ref{eq:t1})--(\ref{eq:t8}).
\label{fig:check_Mabc_sq_freq_corr}}
\vspace{0.0cm}
\end{figure*}

\subsection{Testing the calculation of $\kappa_{\chi^{(2)}g g}$}
\label{sec:check3}

Here we compare our direct solution of the nonlinear tide $\gv\chi^{(2)}$ found by solving the inhomogeneous Equation (\ref{eq:chi2_direct}) with the mode expansion solution found by solving Equation (\ref{eq:chi2_expansion}).  While we expect the former to be more numerically accurate, such a check is useful, especially given the complicated forcing terms (Equations \ref{eq:fr_full}, \ref{eq:fh_full}).  The black and magenta curves in Figure \ref{fig:numeric_vs_analytic} compare $\kappa_{\chi^{(2)}g g}$ calculated using the two different methods for the same polytropic model and modes as \S~\ref{sec:check2}.  We see that they are in good agreement. Given that the high-precision static tide cancellation made use of the direct solution of $\gv\chi^{(2)}$,  the small differences for certain values of $m$ are more likely due to some imprecision in calculating $\gv\chi^{(2)}$ using the mode expansion.

We find in \S~\ref{sec:results_finite_freq} that the finite frequency corrections due to the non-static tide do not cancel and yield $\sigma_{gg} \sim 0.01-0.1$ (when not near a linear resonance). While the checks above suggest that the analytic approach is in good agreement with the direct calculation involving sums over modes, the limited accuracy of the latter precludes us from using it to calculate such small $\sigma_{gg}$'s. However, near resonances the residual $\sigma_{gg}$ becomes especially large and even with the direct calculation it is easy to distinguish $2\kappa_{\chi^{(2)}g g}$ from $3\kappa_{\chi^{(1)}\chi^{(1)}g g} + \sum K_{3pg} K_{3\bar{p} g}$ and see that the $\mathcal{O}(\omega_g^{-2})$ terms do not cancel as a result of the finite frequency corrections.  The calculation shown in Figure  \ref{fig:numeric_vs_analytic} is at the resonance with the $f$-mode ($a/R=1.23$) of the $n_{\rm poly}=2$ polytrope. We see that the residuals calculated using the analytic approach and the direct approach are in good agreement, with both yielding $\sigma_{gg} \sim 100$ for the $m=2$ and $m=4$ harmonics of the nonlinear tide. 

\subsection{Testing the $1/\omega_b^2$ sum rule terms}
\label{sec:check4}

As we describe in \S~\ref{sec:k3sq_sum_rule}, when calculating $\sum K_{3pg}K_{3\bar{p}g}$ using the sum rule (Equation \ref{eq:K3sq_sum_rule}), we approximate the $\omega_b^{-2}$ terms as
\bea
\label{eq:sum_rule_wb2_approx}
\sum_{b\, \in\,\{p\} } \frac{M_{abc}^\ast M_{abd}}{\omega_b^2}
&\simeq&
\frac{1}{\langle\omega_b^2\rangle}\sum_{b\, \in\,\{p,f,g\}} M_{abc}^\ast M_{abd}
\non&\simeq&
- \frac{1}{E_0}\int d^3x \, \frac{\gv{\psi}_{ac}^\ast \cdot \gv{f}_1\left[\gv{\psi}_{ad}\right]}{\langle \omega_b^2\rangle}
\eea
(and similarly for the  $\omega_b^{-2}$ terms involving $N_{abc}$), where $\langle \omega_b^2\rangle = \left(\Lambda_g N c_s /\omega_g r\right)^2$.  This approximation enters our calculation of $\kappa_{\chi^{(1)}\chi^{(1)}gg}+\sum K_{3pg}K_{3\bar{p}g}$
 as term $IV$ of Appendix \ref{app:Komc2} (Equation \ref{eq:t4}).  In order to check the accuracy of the approximation, we explicitly solve the sum over $p$-modes using the same polytropic model and $g$-mode pairs as  \S~\ref{sec:check2}.  We show the results in Figure \ref{fig:check_Mabc_sq_freq_corr} for $a/R= 1.23$ (near the $\ell=2$ $f$-mode resonance) and  $a/R=25$ (not near a resonance). We find good agreement between the analytic calculation given by term $IV$ (blue curves) and the numerical sum over modes (red curves).  At large orbital separations, term $IV$ makes only a very small contribution to the total coupling $\kappa_{\chi^{(1)}\chi^{(1)}gg}+\sum K_{3pg}K_{3\bar{p}g}$ (which is shown as the black curve).  At smaller separations, the contribution is more significant but still subdominant and much less than $2\kappa_{\chi^{(2)}gg}$ (cf. black curve in Fig. \ref{fig:numeric_vs_analytic}).  As a result, we find that the fractional errors introduced by the approximations in Equation (\ref{eq:sum_rule_wb2_approx}) are always much smaller than the values of the residual $\sigma_{gg}$ ($\sim 0.01-0.1$ when not near a resonance) and therefore conclude that they do not affect the accuracy of our calculation.
 
\begin{figure*}
\begin{center}
\hspace{0.0cm}
\includegraphics[width=3.2in]{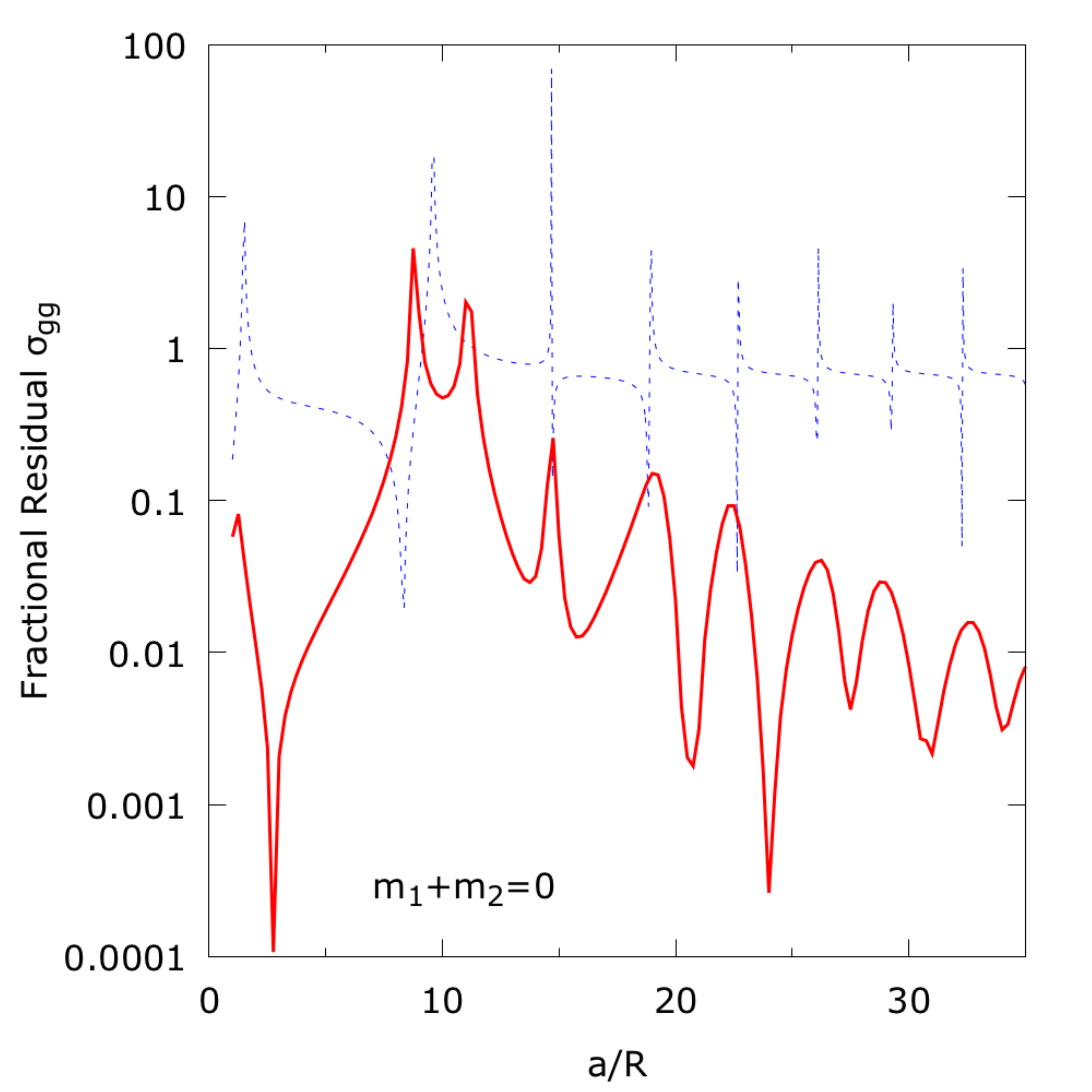}
\includegraphics[width=3.2in]{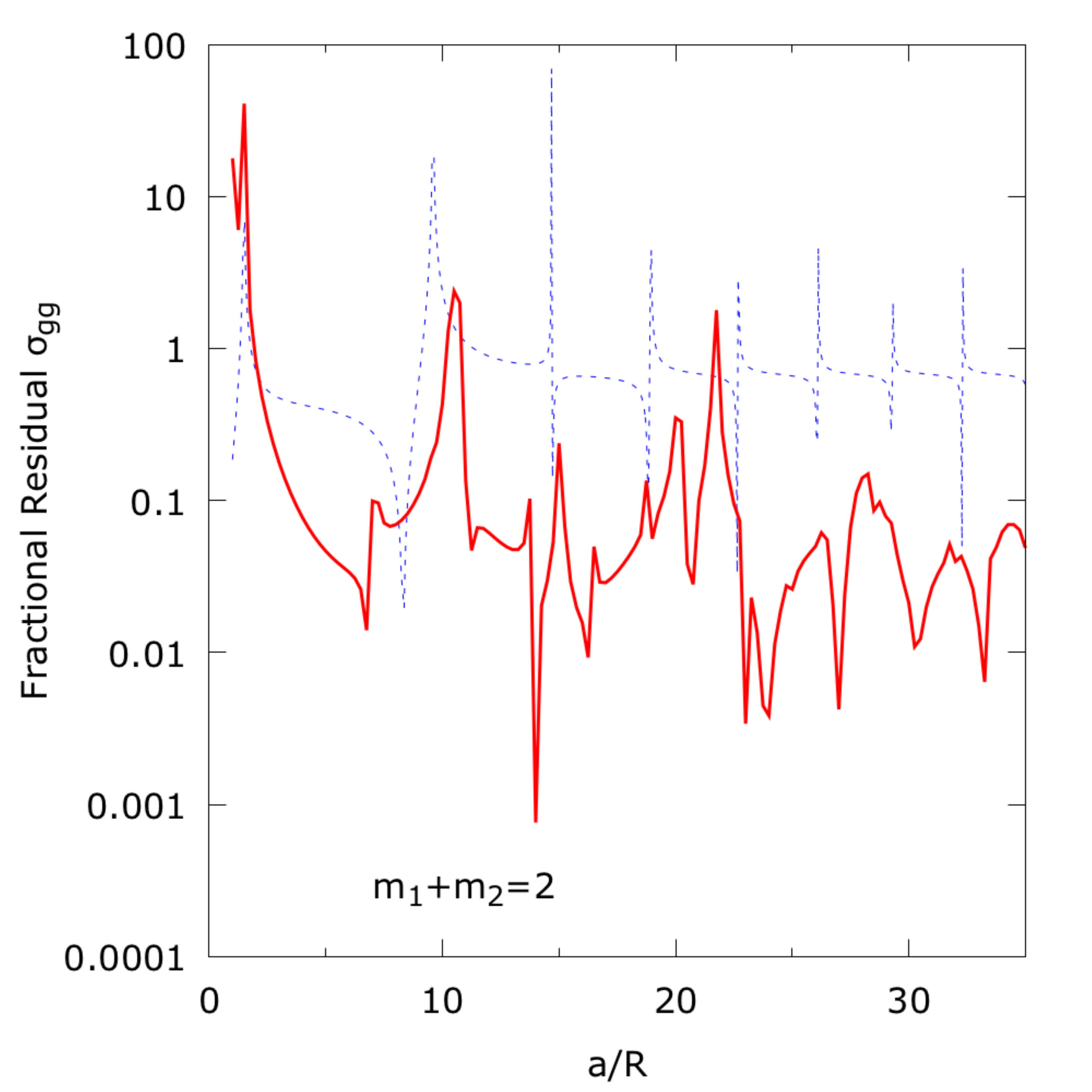} 
\includegraphics[width=3.2in]{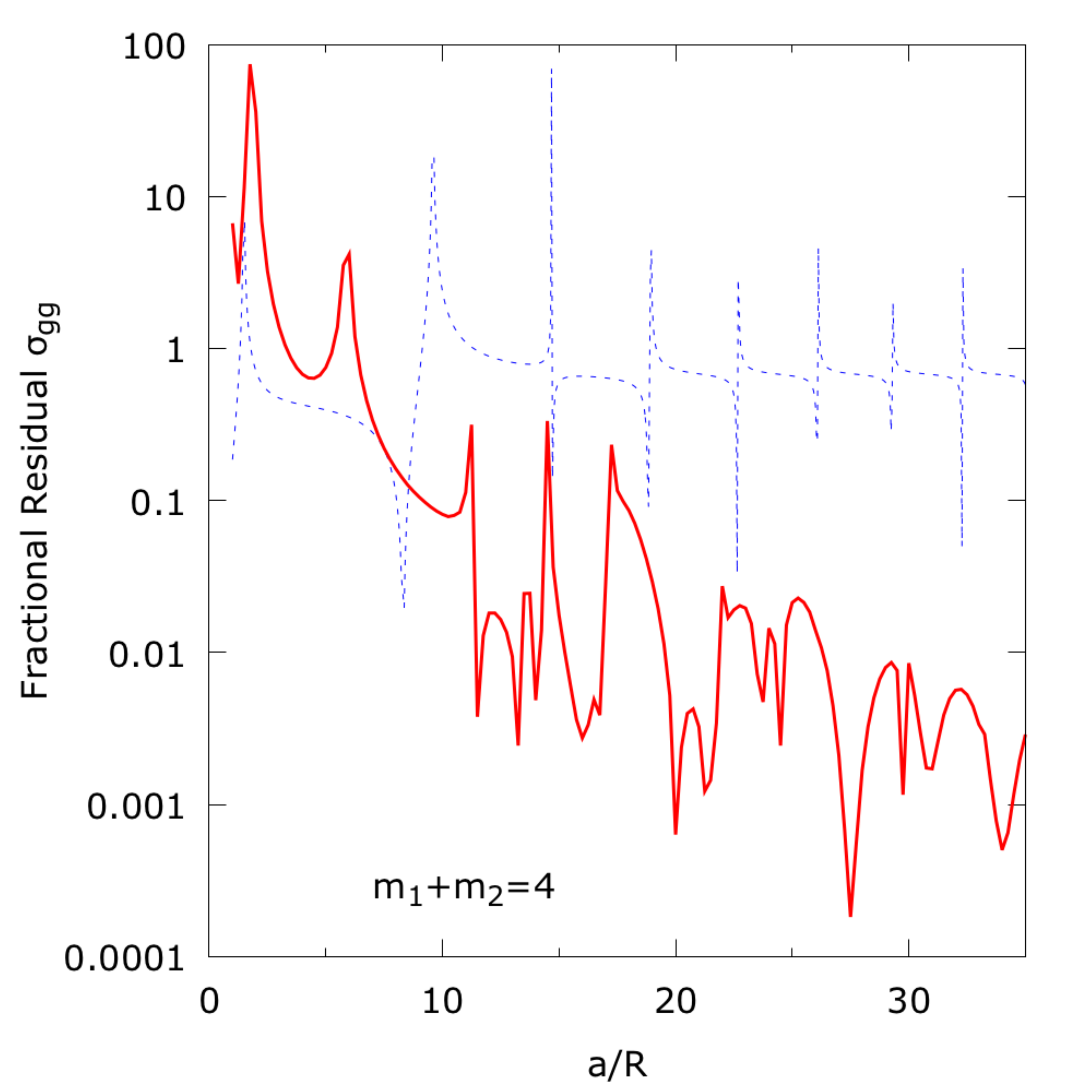} 
\end{center}
\caption{Fractional residual $\sigma_{g_1g_2}$ (red solid line) with finite frequency corrections due to the non-static linear tide, as a function of orbital separation $a/R$ for the neutron star model. The blue dashed line shows $\chi_h^{(1)}(R)/R$ (as plotted in Fig. \ref{fig:linear_resonances}) and indicates the location of linear resonances.  The $g$-mode pair is $\ell=2$, $n=1000$ and  $(m_1,m_2)=(2,-2)$, $(2,0)$, $(2,2)$ clockwise starting from the top left panel.
\label{fig:frac_diff}}
\end{figure*}

\section{Results with finite frequency corrections}
\label{sec:results_finite_freq}

In this section we carry out a stability analysis of $p$-$g$ coupling accounting for the finite frequency corrections due to the non-static tide of a coalescing binary neutron star.  In \S~\ref{sec:sigma_gg} we calculate the degree of cancellation between three- and four-wave coupling as measured by the residual $\mathcal{R}_{gg}$  (Equation \ref{eq:Rgg}).  In \S\S~\ref{sec:growth_rate_finite_freq} and \ref{sec:lambda} we evaluate the stability and  growth rate of $p$-$g$ mode pairs given the computed values of $\mathcal{R}_{gg}$.  We use these results in \S~\ref{sec:efoldings_finite_freq} to estimate the number of $e$-foldings of growth an unstable $p$-$g$ pair can undergo before the binary merges (ignoring the nonlinear saturation of the instability).

\subsection{Residual $\mathcal{R}_{gg}$ due to finite frequency corrections}
\label{sec:sigma_gg}

If we ignore finite frequency corrections then we expect a residual $\mathcal{R}_{g g}\sim\mathcal{O}(\epsilon^2)$.  By Equation (\ref{eq:sigma}), this corresponds to a fractional residual $\sigma_{gg} \sim (\omega_g / \omega_0)^2$ since $\kappa_{\chi^{(2)}gg}\sim \epsilon^2 (\omega_0/\omega_g)^2$.  For illustrative purposes, in this section we show results from our SLy4 neutron star model for $g$-mode pairs $(g_1,g_2)$ with $\ell=2$, $n=1000$, and $|m_1+m_2|=\{0,2,4\}$ (other combinations do not couple to the tide).  The results we show are representative of what we find with other high-order, low $\ell$, $g$-modes. Since this particular $g$-mode has a frequency $\omega_g \simeq 10^{-4}\omega_0$, if finite frequency corrections  to the coupling coefficients are not significant we would expect $\sigma_{gg}\sim 10^{-8}$.  However, based on the results of \S~\ref{sec:check1}, we can only expect our calculation of $\sigma_{gg}$ to be accurate at the $\sim 10^{-6}$ level (as determined by the resolution of our model and eigenfunctions).  Therefore, if the computed $\sigma_{gg} \la 10^{-6}$ it suggests finite frequency corrections are not significant and vice versa.

In Figure \ref{fig:frac_diff} we show $\sigma_{gg}$ accounting for finite frequency corrections due to the non-static tide.  The red curve shows $\sigma_{gg}$  as a function of orbital separation $a/R$.  The blue curve is the same as that of Figure \ref{fig:linear_resonances} and illustrates the variation of $\gv\chi^{(1)}$ with $a/R$ and the location of linear resonances.  The $g$-mode pairs in the three panels are $\ell=2$, $n=1000$, and $(m_1,m_2)=(2,-2)$, $(2,0)$ and $(2,2)$; other possible pair combinations such as $(1,1)$ and $(1,-1)$ yield similar results.  

We find that between resonances, $10^{-3}\la \sigma_{gg}\la 1$ for $10 \la f_{\rm gw}\la 10^3 \trm{ Hz}$.\footnote{\label{footnote:near_resonances} Although $\sigma_{gg}$ can be much larger near resonances, because our calculation does not account for the limited duration of each resonance as the neutron star inspirals, the narrow set of points that fall within the resonance window $\delta a / R \sim 0.5 (a/10R)^{-1/4}$ should be ignored (see \citealt{Lai:94} and \S~\ref{sec:divxi_finite_freq}).} This result suggests that the finite frequency corrections are significant and undo the cancellation of  three- and four-wave coupling found in the static tide (i.e., incompressible) limit. The variations in $\sigma_{gg}$ with $a/R$ are due to variations in $\gv\chi^{(1)}$ and $\gv\chi^{(2)}$ with $a/R$ and, as the figures show, correlate with the locations of linear resonances.  The magnitude of  $\sigma_{gg}$ tends to decrease with increasing $a/R$, as might be expected given that finite frequency corrections, including the compressibility $\div\gv\chi^{(1)}$, decrease with decreasing driving frequency. Note, however, that although  $\div\gv\chi^{(1)}\propto (\omega/\omega_0)^2\propto (R/a)^{3}$  (see Equation \ref{eq:divxi_finite_freq}), the magnitude of $\sigma_{gg}$ falls off more slowly than $(R/a)^3$.
As we explain more fully in \S~\ref{sec:lambda}, this is because even at larger separations, the finite frequency corrections remain significant near the inner turning radius of the tide (where $\omega\approx N(r)$).  Since the nonlinear coupling is strongest near the inner turning radius, the residual approaches the static tide limit more slowly than $(R/a)^3$.

\begin{figure}
\centering
\includegraphics[width=3.4in]{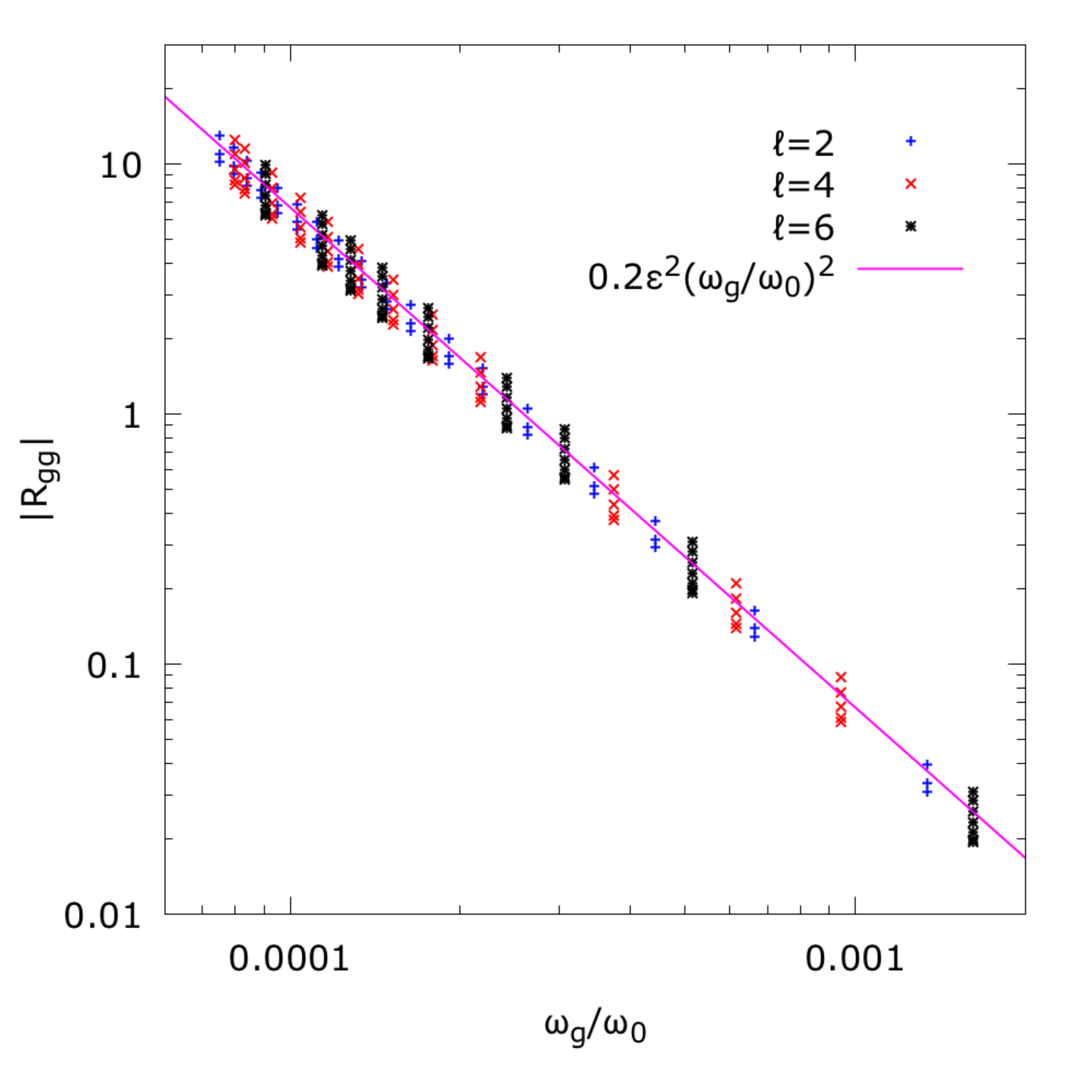} 
\caption{Magnitude of the residual $R_{gg}$ as a function of $g$-mode frequency $\omega_g/\omega_0$ at $a/R=12$ ($f_{\rm gw}\simeq100\trm { Hz}$; this is not near a linear resonance, see Fig. \ref{fig:linear_resonances}).   We show results for $g$-mode pairs with $\ell=2,4,6$ and $m_1+m_2=0$. We find $|R_{gg}|\propto \epsilon^2 (\omega_g/\omega_0)^2$, as shown by the magenta line (different values of $m_1+m_2$ show the same scaling).
\label{fig:Rgg_vs_omg}}
\end{figure}

\subsection{Stability analysis}
\label{sec:growth_rate_finite_freq}

We return to the stability analysis of \S~\ref{sec:eom} but now include the effects of finite frequency corrections.  From the definition of $\mathcal{R}_{gg}$ (Equation \ref{eq:Rgg}) 
\beq
K_{4 \bar{g}_j g_i} =- \sum_{p_i}K_{3 p_i \bar{g}_j} K_{3\bar{p}_i g_i}+\mathcal{R}_{\bar{g}_jg_i}.
\eeq
The characteristic equation for $p$-$g$ coupling (Equation \ref{eq:char_pg})
including finite frequency corrections (but ignoring linear damping) is then 
\bea
\label{eq:char_eqn_finite_freq}
&&\!\left[-\left(s-m_{g_j}\Omega\right)^2+\omega_{g_j}^2\right]q_{g_j}
\non &&
\!=\!\omega_{g_j}\!\sum_{g_i} \!\Bigg[
\mathcal{R}_{\bar{g}_jg_i}
-\sum_{p_i}\frac{K_{3 p_i \bar{g}_j} K_{3\bar{p}_i g_i}\left(s-m_{p_i}\Omega\right)^2}{\left(s-m_{p_i}\Omega\right)^2-\omega_{p_i}^2}\Bigg]\omega_{g_i} q_{g_i}.
\non
\eea
  As in \S~\ref{sec:stability_no_finite_freq_no_damping}, we can obtain an accurate estimate of the stability of this potentially large system of modes by considering a two mode system consisting of a single $p$-$g$ pair (we have verified this by numerically solving the eigenvalue problem given by equation \ref{eq:eig2form} for large sets of strongly coupled modes; see \S~\ref{sec:efoldings_finite_freq}).  
Since $K_{3pg}\simeq \epsilon \omega_p /\omega_g$ (WAB, VZH), we then have 
\bea
&&\left[\left(r+\alpha\right)^2-\omega_g^2\right]\left[r^2-\omega_p^2\right]
-\epsilon^2 r^2 \omega_p^2 
\non &&
+\,\omega_g^2 \mathcal{R}_{gg}\left(r^2-\omega_p^2\right)\simeq 0,
\eea
where $r=s-m_p\Omega$ and $\alpha=(m_p-m_g)\Omega$. 
There are two stable high frequency solutions near $r\simeq \pm\omega_p$ and two low frequency solutions near
\beq
\label{eq:low_freq_soln_finite_freq}
r\simeq -\alpha \pm \sqrt{\omega_g^2(1-\mathcal{R}_{gg})-\epsilon^2\alpha^2}
\eeq
(cf. Equation \ref{eq:r_low_freq_no_finite_freq_no_damping}). In the absence of finite frequency corrections ($\mathcal{R}_{gg}\simeq 0$) we recover the unstable solution of the incompressible limit in which the growth rate is $\Gamma\approx2\epsilon \Omega$ if $\omega_g\la 2\epsilon\Omega$ (see VZH and \S~\ref{sec:stability_no_finite_freq_no_damping}).  If, however, the finite frequency corrections are significant and $\mathcal{R}_{gg}\ga 1$, there is an unstable solution with a growth rate 
\beq
\Gamma \approx \omega_g \sqrt{\mathcal{R}_{gg}}.  
\eeq

In Figure \ref{fig:Rgg_vs_omg} we show $|\mathcal{R}_{gg}|$  as a function of $\omega_g$ at $f_{\rm gw}\simeq100\trm{ Hz}$ (this is not near a linear resonance). We find that $\mathcal{R}_{gg}$ is well-fit by a relation of the form
\beq
\left|\mathcal{R}_{gg}\right|\simeq \lambda^2 \epsilon^2 \left(\frac{\omega_0}{\omega_g}\right)^2,
\eeq  
where the factor $\lambda=\lambda(\omega)$, which we describe in detail in \S~\ref{sec:lambda}, is a function of orbital separation and depends on $\gv\chi^{(1)}$ and $\gv\chi^{(2)}$.  The $\omega_g^{-2}$ scaling is a consequence of the fact that the $\mathcal{O}(\omega_g^{-4})$ and $\mathcal{O}(\omega_g^{-3})$ cancel in Equation (\ref{eq:Rgg}) but the $\mathcal{O}(\omega_g^{-2})$ terms do not.  The instability criterion $\mathcal{R}_{gg}\ga 1$ is then equivalent to $\omega_g \la\lambda \epsilon \omega_0$ and the growth rate can be expressed as
\beq
\label{eq:Gamma_finite_freq}
\Gamma\approx 
\lambda \epsilon \omega_0.
\eeq
The lack of cancellation due to finite frequency corrections therefore leads to growth rates that are potentially a factor of $\approx \lambda \omega_0/2\Omega$ larger than the incompressible limit growth rate.

The sign of $R_{gg}$ depends on the quantum numbers of the particular $g$-mode pair and in practice we find that there is no preference for either sign.  In our numerical experiments with more realistic multi-mode systems, we find that the residual drives an instability whenever there are mode pairs with $|R_{gg}|\ga1$.

\begin{figure*}
\centering
\includegraphics[width=3.2in]{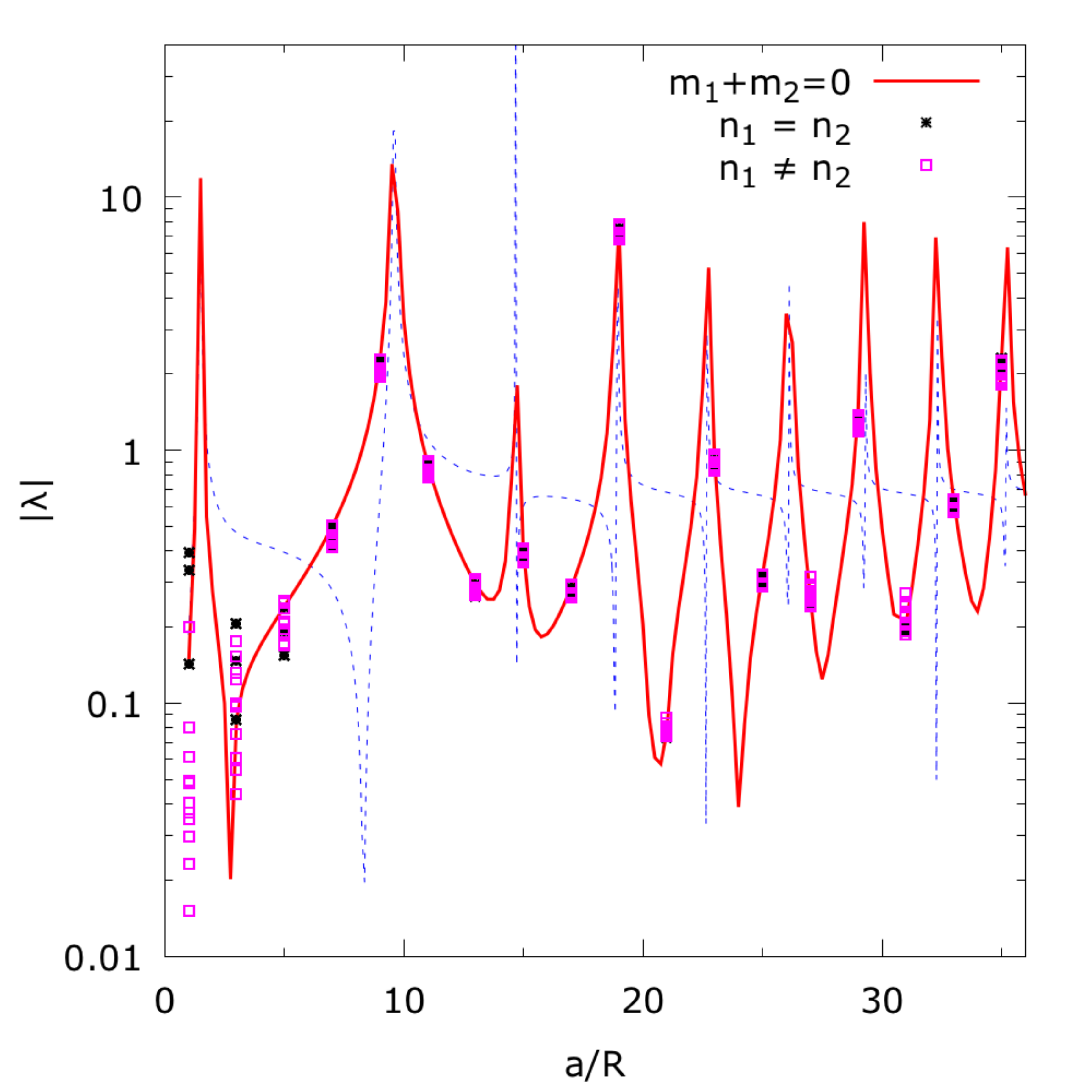} 
\includegraphics[width=3.2in]{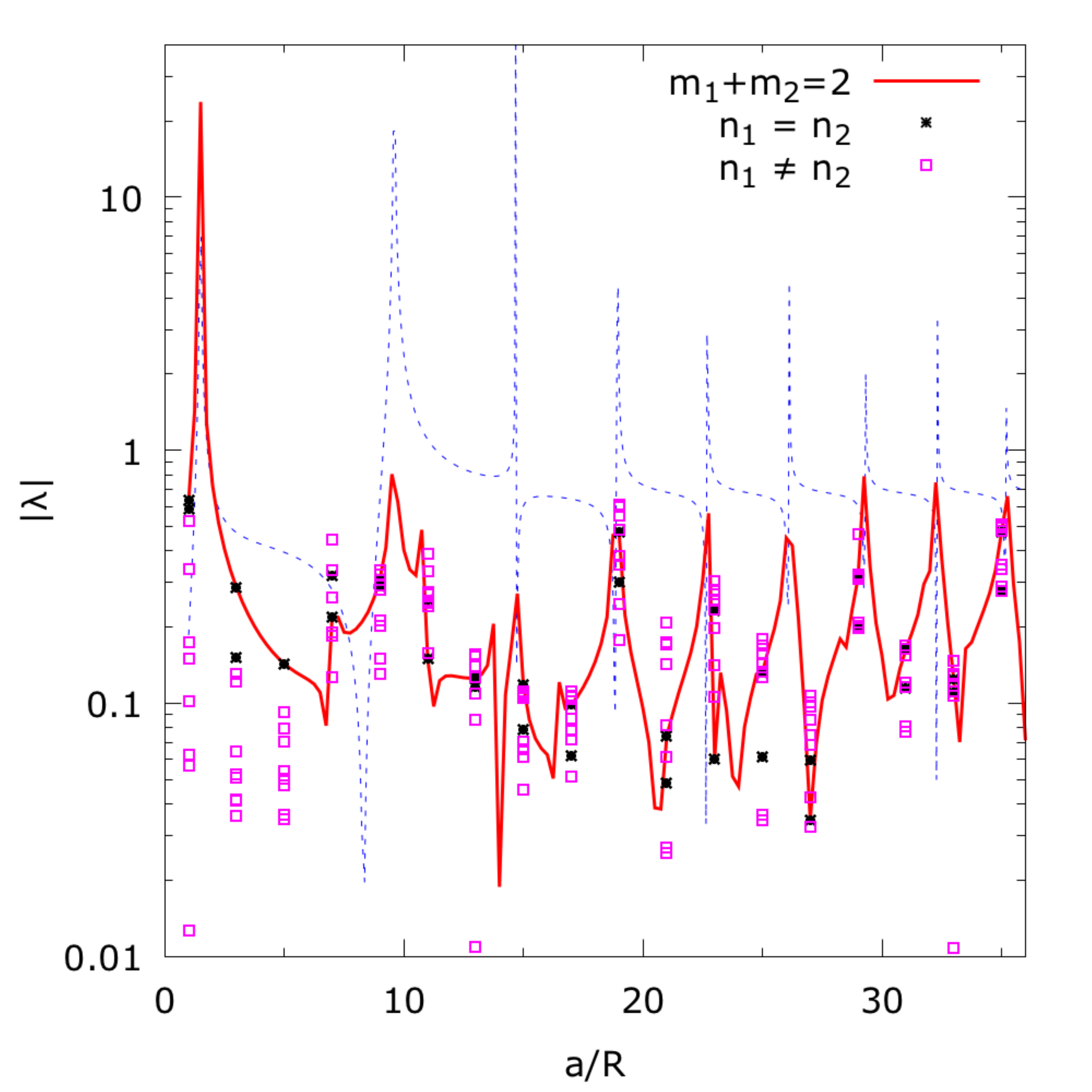} 
\includegraphics[width=3.2in]{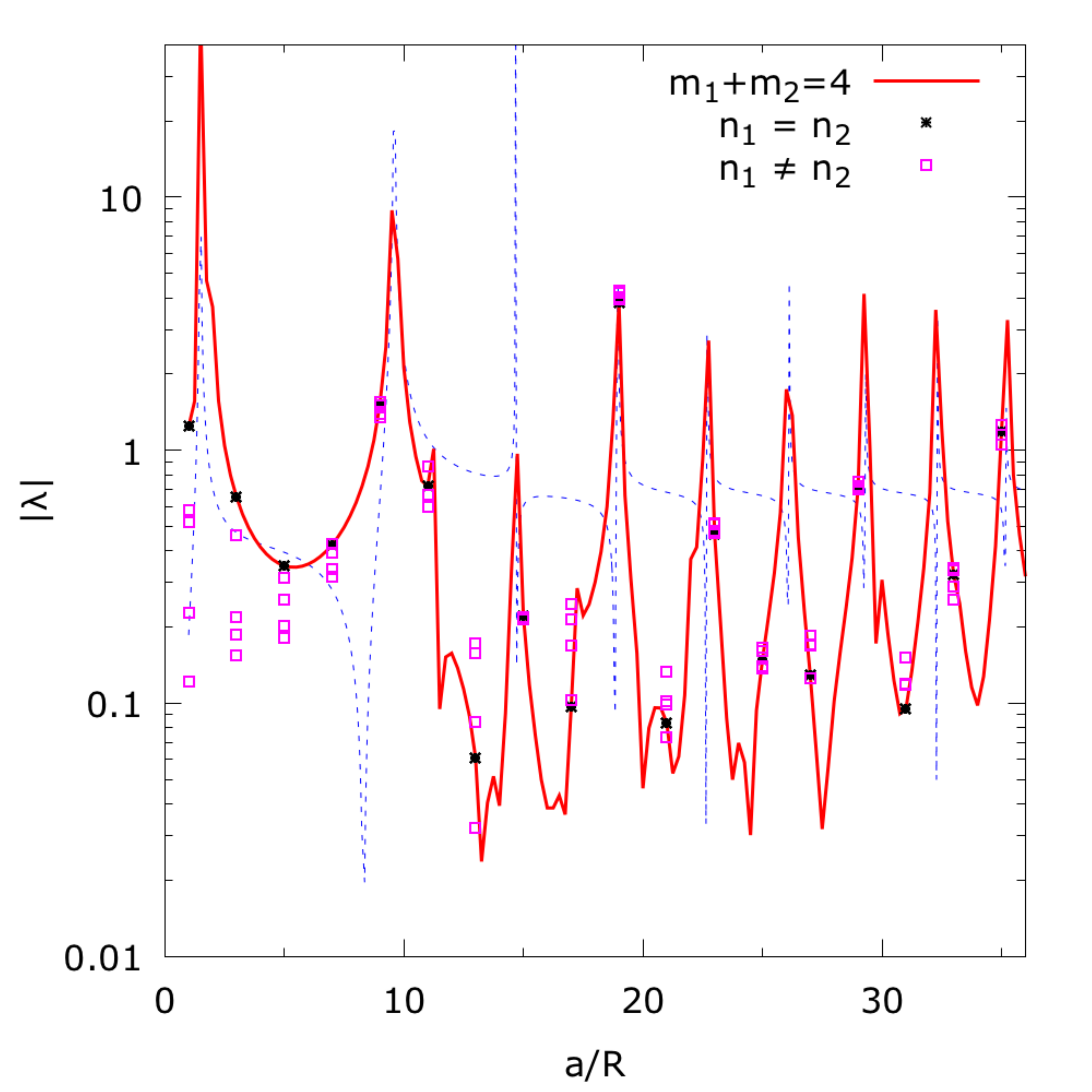} 
\caption{Magnitude of the residual factor $\lambda$ (red solid line) as a function of $a/R$.  The blue dashed line shows $\chi_h^{(1)}(R)/R$ (as plotted in Fig. \ref{fig:linear_resonances}) and indicates the location of linear resonances. The $g$-mode pair is $\ell=2$, $n=1000$ and $(m_1,m_2)=(2,-2)$, $(2,0)$, $(2,2)$  clockwise starting from the top left panel (the same pairs as in Figure \ref{fig:frac_diff}).  The points show $\lambda$ at odd integer multiples of $a/R$ for $\ell=2$ pairs with $n_1=n_2=1000$ (black points) and $n_1\neq n_2$ with $n_{1,2}=1000\pm 2$ (magenta points).
\label{fig:lambda}}
\end{figure*}

\subsection{Value of $\lambda(\omega)$}
\label{sec:lambda}

In Figure \ref{fig:lambda} we show the magnitude of the residual factor $\lambda(\omega)$ as a function of $a/R$.  The red lines show $\lambda$ for mode pairs with $\ell=2, n=1000$, and $(m_1,m_2)=(2,-2)$, $(2,0)$ and $(2,2)$. The blue curve is $\chi_h^{(1)}(R)$, the same curve as in Figure \ref{fig:linear_resonances}, and illustrates the variation of $\gv\chi^{(1)}$ with $a/R$ and the location of linear resonances.  We find $\lambda\sim 0.1 - 1$ in the regions outside the narrow resonance windows (as explained in footnote \ref{footnote:near_resonances}, the regions near resonances where $\lambda$  has large peaks should be ignored).

Although the fractional residual $\sigma_{gg}$ tends to decrease with increasing $a/R$ (Fig. \ref{fig:frac_diff}), we see in Figure \ref{fig:lambda} that the total residual $R_{gg}=2\sigma_{gg}\kappa_{\chi^{(2)}gg} \propto \lambda^2 $ is relatively constant with $a/R$ (modulo $\epsilon^2$ and the resonance-to-resonance fluctuations). This is because the inner turning point of the non-static tide, which occurs at the radius where $N(r)\simeq\omega $, moves inward to smaller $r$ as $\omega=2\Omega=2\pi f_{\rm gw}$ decreases.  The amplitude of the high-order $g$-modes are larger at smaller $r$, and therefore $\kappa_{\chi^{(2)}gg}$ increases significantly as $f_{\rm gw}$ decreases (WAQB discuss this effect in the context of three-mode coupling of $g$-modes to $\gv\chi^{(1)}$).  As a result, although globally $\omega/N$ decreases, the finite frequency corrections remain significant in the region that dominates the coupling (where $\omega \approx N(r)$) and the residual only slowly approaches the static tide limit $R_{gg}\rightarrow \mathcal{O}(\epsilon^2)$ as $f_{\rm gw}\rightarrow 0$.  This slow convergence of the non-static linear tide to the static tide solution in local regions where $N^2$ is small is also noted in \citet{Terquem:98}.

\begin{figure}
\centering
\includegraphics[width=3.45in]{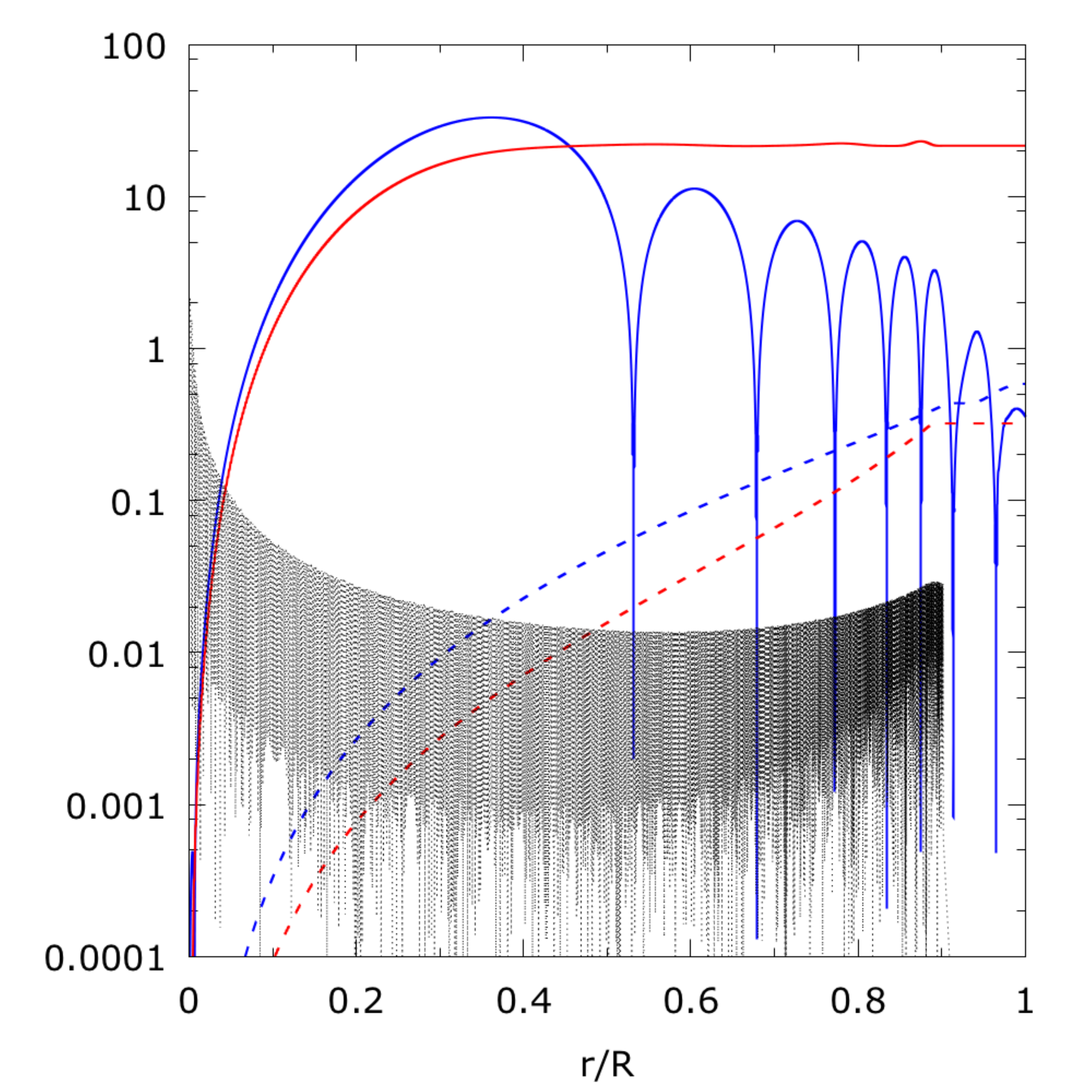} 
\caption{Cumulative coupling integral $\kappa_{\chi^{(2)}gg}(< r)$ (red lines; divided by $10^8\epsilon^2$) and the radial displacement $\chi^{(2)}_r$ of the $(\ell,m)=(4,4)$ harmonic of the nonlinear tide (blue lines; divided by $R$) as a function of radius $r/R$.  The dashed lines are for $a/R=5$ ($f_{\rm gw}\simeq 380\trm{ Hz}$) and the solid lines are for $a/R=34$ ($f_{\rm gw}\simeq 22\trm{ Hz}$),  neither of which are near a linear resonance.  The $g$-mode in the calculation is a self-coupled $g$-mode with $(\ell,m,n)=(2,-2,1000)$ whose frequency $\omega_g\simeq 10^{-4}\omega_0$.  The dotted black line shows its horizontal displacement $g_h$ (divided by $10^6R$).  
\label{fig:kap3_nltide_vs_radius}}
\end{figure}

\begin{figure}
\centering
\includegraphics[width=3.4in]{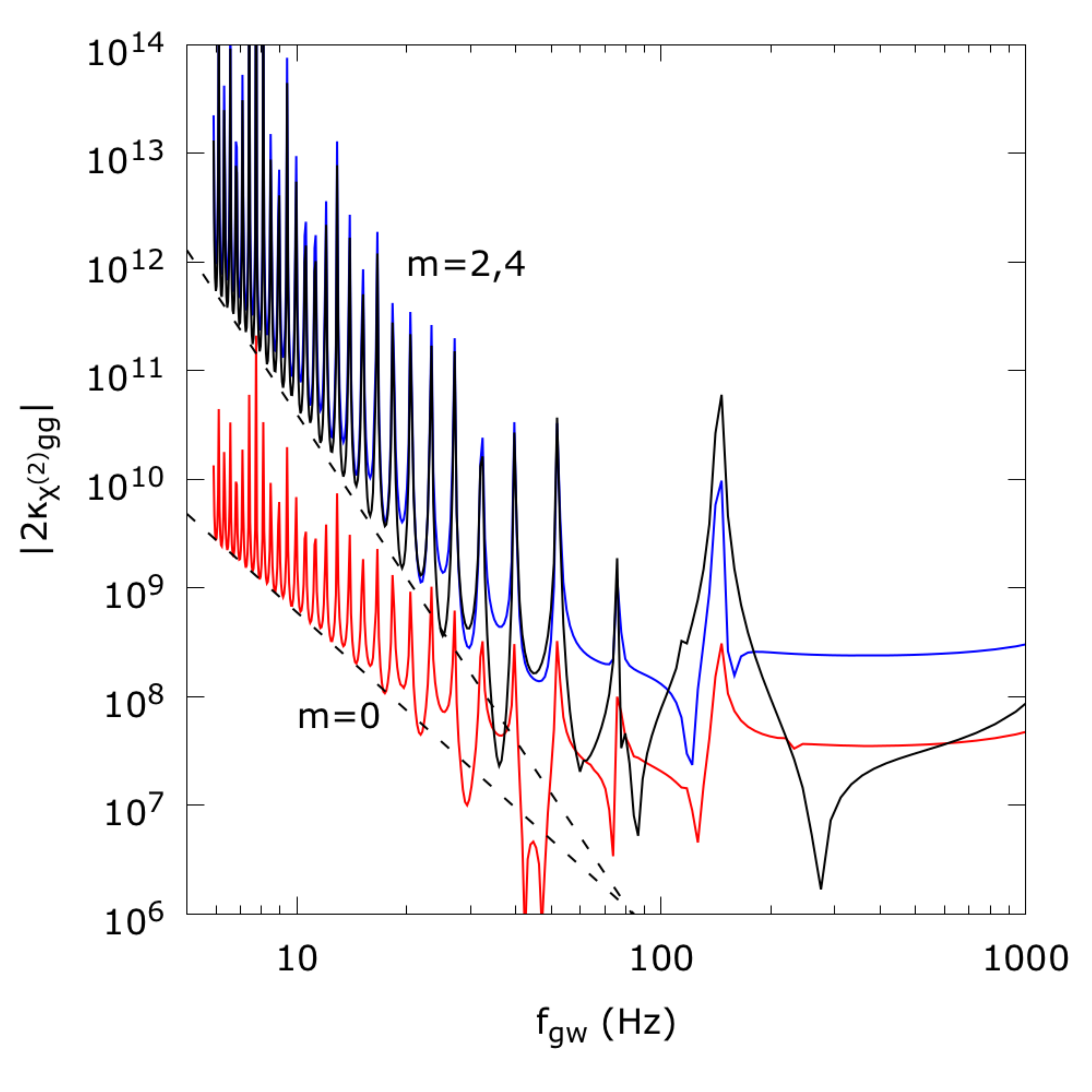} 
\caption{Magnitude of $2\kappa_{\chi^{(2)}gg}=\mathcal{R}_{gg}/\sigma_{gg}$ (divided by $\epsilon^2$) as a function of $f_{\rm gw}$ for $g$-mode pairs with $\ell=2$, $n=1000$ and $m_1+m_2=\{0,2,4\}$ (red, blue, and black solid lines, respectively).   The black dashed lines show $f_{\rm gw}^{-3}$ and $f_{\rm gw}^{-5}$ scalings.
\label{fig:kap3_nltide_vs_fgw}}
\end{figure}

We illustrate this point further in Figures \ref{fig:kap3_nltide_vs_radius} and \ref{fig:kap3_nltide_vs_fgw}.  
In Figure  \ref{fig:kap3_nltide_vs_radius} we show, as a function of $r$, the radial displacement $\chi^{(2)}_r$ of the $(\ell,m)=(4,4)$ harmonic of the nonlinear tide and the cumulative integral $\kappa_{\chi^{(2)}gg}(< r)$ for $a/R=5$ and $a/R=34$, corresponding to$f_{\rm gw}\simeq 380\trm{ Hz}$ and $22 \trm{ Hz}$ (neither of which is near a linear resonance).  We assume an  $(\ell,m,n)=(2,-2,1000)$ self-coupled $g$-mode and also plot the horizontal displacement $g_h$.  At $f_{\rm gw}\simeq 380\trm{ Hz}$, $\chi^{(2)}_r$ has an effective wavelength $\approx R$ and most of the contribution to $\kappa_{\chi^{(2)}gg}$ comes from large radii where $\chi^{(2)}_r$ peaks (the crust at $r\ga0.9R$ does not contribute because we assume it is not buoyant and therefore the $g$-mode displacement rapidly vanishes there).   At $f_{\rm gw}\simeq 22\trm{ Hz}$,  $\chi^{(2)}_r$ has a short wavelength ($\ll R$) and most of the contribution to $\kappa_{\chi^{(2)}gg}$ comes from the region near the inner turning point of  $\chi^{(2)}_r$ (located at $r\simeq 0.2 R$), where the density and displacements of all three modes are large.

In Figure \ref{fig:kap3_nltide_vs_fgw} we show $2\kappa_{\chi^{(2)}gg}$ as a function of $f_{\rm gw}$.  For $g$-mode pairs that couple to the $m=0$ and $m=4$ nonlinear tide we find $\kappa_{\chi^{(2)}gg}\propto f_{\rm gw}^{-5}$ while for pairs that couple to the $m=2$ nonlinear tide $\kappa_{\chi^{(2)}gg}\propto f_{\rm gw}^{-3}$.  The former scale more strongly because the $m=0$ and $m=4$ nonlinear tide is driven by two instances of the non-static linear tide $\{\chi^{(1)}_{m=2}, \chi^{(1)}_{m=-2}\}$ and $\{\chi^{(1)}_{m=2}, \chi^{(1)}_{m=2}\}$, respectively. The finite frequency corrections to the linear tide therefore enter the calculation twice.  The $m=2$ nonlinear tide, by contrast, is only driven by a single instance of the non-static linear tide $\{\chi^{(1)}_{m=2}, \chi^{(1)}_{m=0}\}$.

Figure  \ref{fig:lambda} also shows $\lambda$ for $g$-mode pairs with $n_1\neq n_2$.  As the wavelength of the oscillatory component of $\gv\chi^{(2)}$ decreases with decreasing $f_{\rm gw}$, it begins to couple well to daughter pairs with $|n_1-n_2|\la n_{\chi^{(2)}}$, where $n_{\chi^{(2)}}$ is the effective radial order of $\gv\chi^{(2)}$ (i.e., the number of nodes).  This is a consequence of linear momentum conservation and is discussed in WAQB in the context of three-mode coupling to the linear tide.  At $a/R\la 5$, $n_{\chi^{(2)}}\simeq 0$ and the coupling is much stronger for daughter pairs with $n_1=n_2$ (black points).  At larger $a/R$, $n_{\chi^{(2)}}>0$ and the coupling is also strong for pairs with $n_1\neq n_2$ (magenta points).   As a result, the effective residual $R_{g_1g_2}$ (and thus $\lambda$) in the characteristic Equation (\ref{eq:char_eqn_finite_freq}) is, after summing over the $\approx n_{\chi^{(2)}}$ strongly coupled daughters, larger than the $n_1=n_2$ case by a factor of $\approx n_{\chi^{(2)}}$.  This collective driving of strongly coupled, multi-mode systems is described in WAQB and results in growth rates that are 
$\approx \sqrt{n_{\chi^{(2)}}}$ faster than the single daughter pair estimate.

\subsection{Growth rate and the number of energy $e$-foldings}
\label{sec:efoldings_finite_freq}

\begin{figure}
\centering
\includegraphics[width=3.4in]{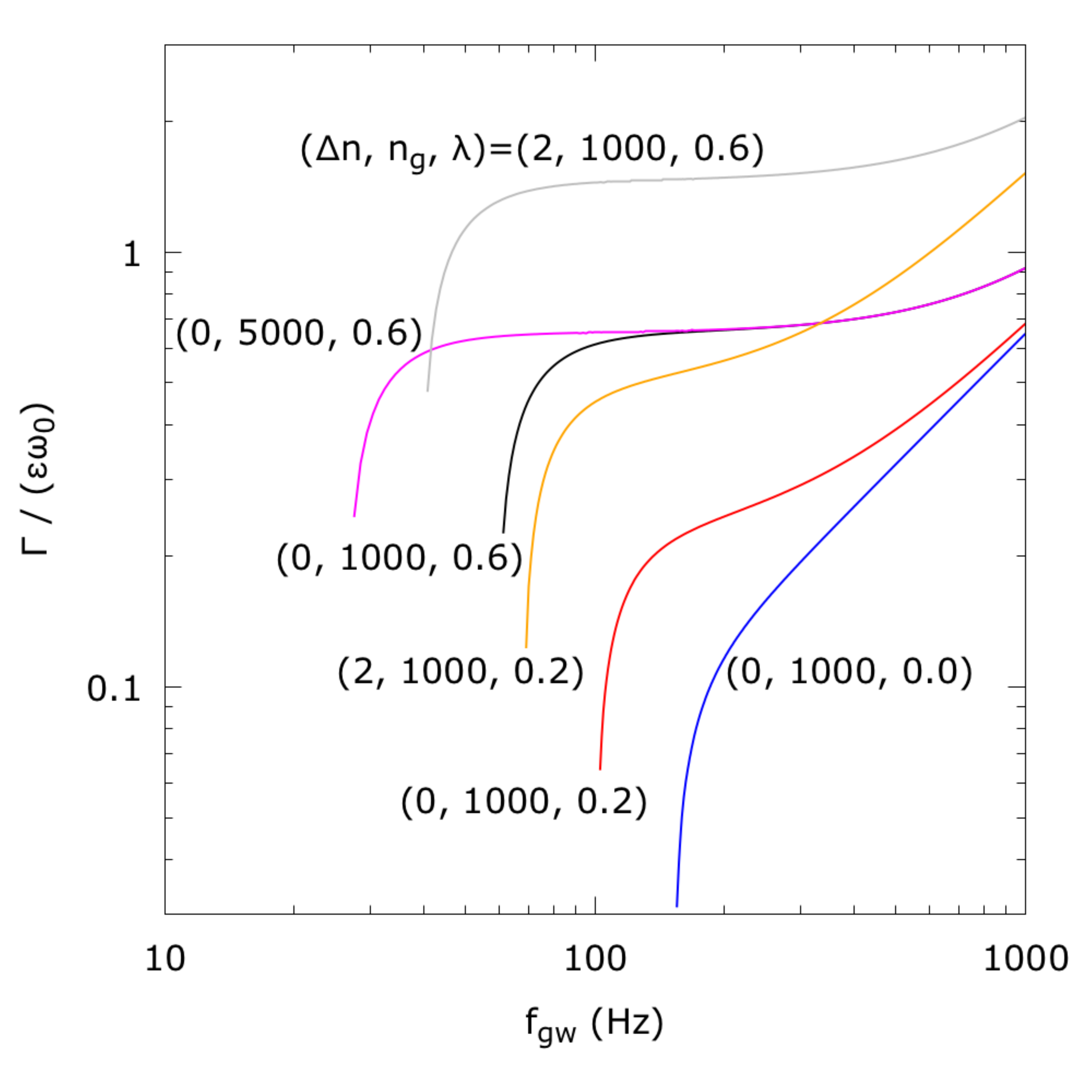} 
\caption{Nonlinear growth rate $\Gamma$ (in units of $\epsilon \omega_0$) of $\ell=2$ $g$-modes (and the $p$-modes to which they couple) as a function of $f_{\rm gw}$ for different values of the parameters $(\Delta n, n_g, \lambda)$. 
\label{fig:growth_rate_finite_freq}}
\end{figure}

In Figure \ref{fig:growth_rate_finite_freq} we plot the growth rate $\Gamma$  as a function of $f_{\rm gw}$, found by numerically solving the eigenvalue problem defined by Equation (\ref{eq:eig2form}). We show results for $g$-modes with $\ell=2$ and a range of parameter values $(\Delta n, n_g, \lambda)$ defined such that $g$-mode pairs in the range $n_g\pm\Delta n$ couple to each other with a residual factor $\lambda$. Based on Figure \ref{fig:lambda}, $\lambda\sim 0.1-1$ when the system is not near a linear resonance.  For $\lambda=0$ we recover the incompressible limit result that $\Gamma\simeq 2\epsilon\Omega$ when $\omega_g \la 2\epsilon\Omega$. For $\lambda\neq 0$ and $\Delta n=0$, we find growth rates $\Gamma\simeq \lambda\epsilon \omega_0$ when $\omega_g \la \lambda\epsilon\omega_0$, consistent with Equation (\ref{eq:low_freq_soln_finite_freq}). And for the reasons given in \S~\ref{sec:lambda}, we find that the effective $\lambda$ increases by a factor of $\approx \sqrt{2\Delta n}$ for $\Delta n\neq 0$. In comparison to the $\lambda=0$ results, the finite frequency corrections yield significantly larger growth rates and the instability is triggered much earlier in the inspiral.

The number of mode energy $e$-foldings is approximately given by
\beq
\label{eq:lnE}
\ln (E/E_i) \simeq 2\int_{t_i}^{t} \Gamma dt = 2\int_{f_{\rm gw,i}}^{f_{\rm gw}} \Gamma \dot{f}^{-1}df,
\eeq
where $f_{\rm gw, i}$ is the gravitational wave frequency when the instability first turns on and  $f/\dot{f}=2t_a/3$ (Equation \ref{eq:ta}) for the gravitational inspiral of two point masses. 

If we ignore finite frequency corrections ($\mathcal{R}_{gg}\simeq0$) and damping, then $\Gamma\approx2\epsilon \Omega$ (Equation \ref{eq:growth_rate_no_finite_freq_no_damping}) and 
\beq
\ln(E/E_i)
=11 \mathcal{M}_{1.2}^{-5/3} 
\left[f_{100}^{1/3}-f_{i,100}^{1/3}\right]
\eeq
for $M'=M$ and a neutron star with dynamical frequency $\omega_0=10^{4}\trm{ rad s}^{-1}$ (where $f_{100}=f_{\rm gw}/100\trm{ Hz}$ and $\mathcal{M}_{1.2}=\mathcal{M}/1.2M_\odot$). In this case,  the instability criterion is $\omega_g \la 2\epsilon\Omega$, and the modes first become unstable at a frequency
\beq
f_{{\rm gw},i} \simeq 149 \left(\frac{\omega_g}{10^{-4}\omega_0}\right)^{1/3}
\trm{ Hz}.
\eeq
If $\omega_g=10^{-4}\omega_0$ (corresponding to $g$-modes with $\ell=2$, $n\simeq1000$) then by $f_{\rm gw}=1000\trm{ Hz}$ the number of $e$-foldings is $\ln(E/E_i)\simeq 11$, i.e., the modes grow by a factor of $\sim 10^5$ in energy.

If we account for finite frequency corrections, $\Gamma\approx\lambda \epsilon \omega_0$ (Equation \ref{eq:Gamma_finite_freq}) and for $M'=M$,
\beq
\ln(E/E_i) \simeq
  44 \mathcal{M}_{1.2}^{-5/3} 
\left(\frac{\lambda}{0.5}\right) 
\left[f_{i,100}^{-2/3}-f_{100}^{-2/3}\right].
\eeq
In this case,  the instability criterion is approximately $\omega_g \la \lambda\epsilon\omega_0$, and the modes first become unstable at a frequency
\beq
f_{{\rm gw,} i} \simeq 63 \left(\frac{\omega_g}{10^{-4}\omega_0}\right)^{1/2}\left(\frac{\lambda}{0.5}\right)^{-1/2} \trm{ Hz}.
\eeq
If $\omega_g=10^{-4}\omega_0$  and $\lambda=0.5$, then  the modes grow by a factor of $\sim 10^{20}$ in energy by $f_{\rm gw}=1000\trm{ Hz}$.   If $\omega_g=10^{-4}\omega_0$ and $\lambda=1$ (corresponding roughly to the effective lambda for $\Delta n=2$; Figure \ref{fig:lambda}), then the modes grow by a factor of $\sim 10^{60}$ in energy by $f_{\rm gw}=1000\trm{ Hz}$.

\begin{figure}
\centering
\includegraphics[width=3.4in]{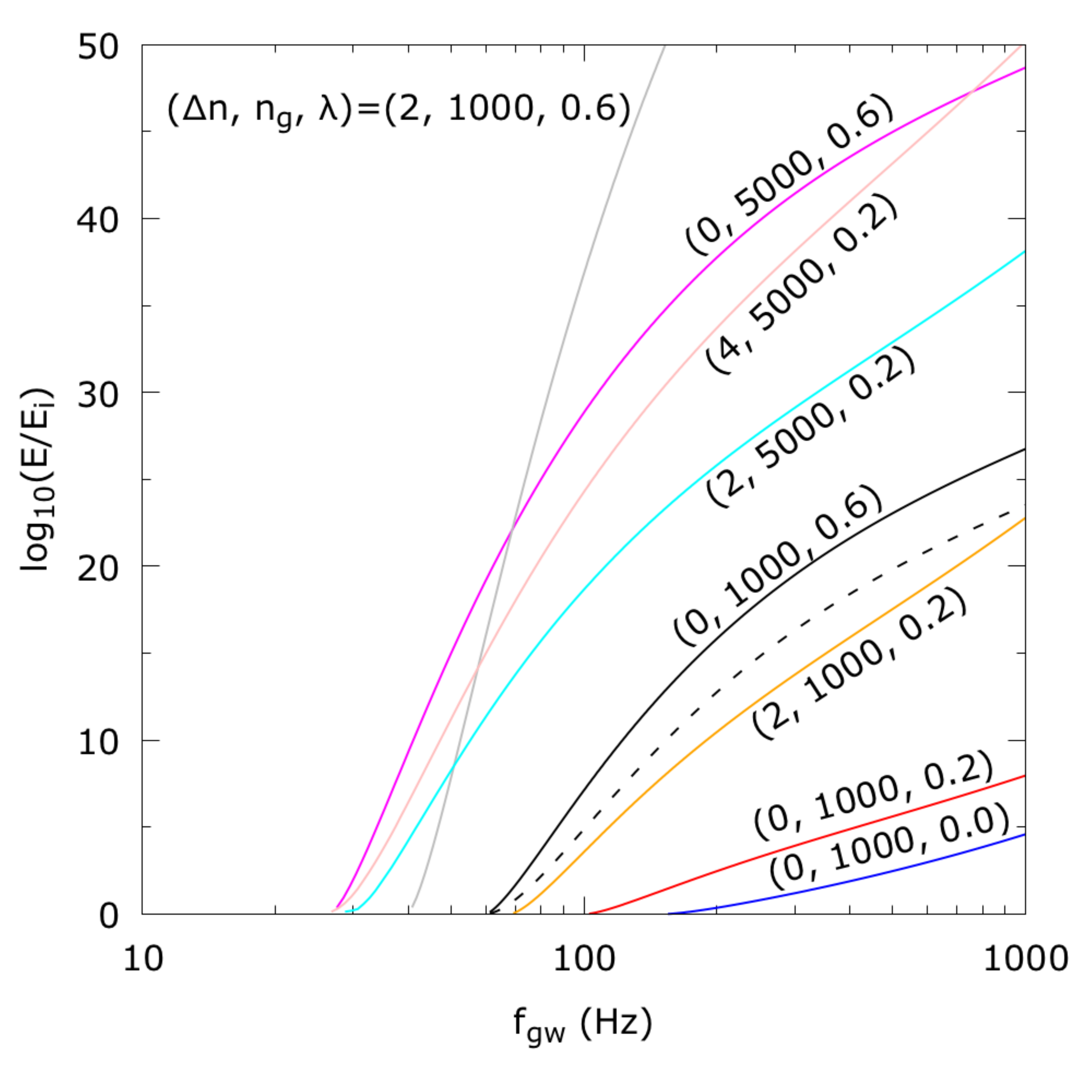} 
\caption{Mode energy growth relative to an initial mode energy $E_i$ as a function of $f_{\rm gw}$ for  $\ell=2$ $g$-modes (and the $p$-modes to which they couple).  The curves are for different values of the parameters $(\Delta n, n_g, \lambda)$.  The solid and dashed and black lines are equivalent except that the latter assumes that the $g$-mode has a linear damping rate $\gamma_g=\omega_g$.
\label{fig:energy_growth_finite_freq}}
\end{figure}

In Figure \ref{fig:energy_growth_finite_freq} we show the growth in mode energy $E$ as a function of $f_{\rm gw}$ for a range of parameter values $(\Delta n, n_g, \lambda)$. We find that depending on the values of $(\Delta n, n_g, \lambda)$,  the modes grow by factors of $\sim10^{20}$ to $\gg 10^{50}$ before the merger.  Moreover, because the growth rates can be large relatively early in the inspiral, the modes can reach large energies already by $f_{\rm gw}\simeq 100\trm{ Hz}$. 

The linear damping rates of $g$-modes increase with decreasing $\omega_g$ and  $\gamma_g\approx \omega_g$ for $\omega_g\approx 10^{-4}\omega_0$ (\S~\ref{sec:pmode_damping_rate}).  We find, however, that even allowing for a linear damping rate $\gamma_g\simeq\omega_g$ does not significantly limit the growth (see solid and dashed black lines in Fig. \ref{fig:energy_growth_finite_freq}). 

\subsubsection{Initial mode energy $E_i$}
\label{sec:initial_energy}

What is the initial mode energy $E_i$ and how many $e$-foldings are needed to reach dynamically significant energies?  A $g$-mode with $\ell=2$ and $m=0$ is linearly driven by the static linear tide to an energy
\beq
\label{eq:Eig}
E_{i,g}\approx (\epsilon I_g)^2E_0 \approx 10^{-24}\left(\frac{\omega_g}{10^{-4}\omega_0}\right)^4\left(\frac{f_{\rm gw}}{100\trm{ Hz}}\right)^4 E_0,
\eeq
where the linear overlap integral $I_g\simeq 0.3 (\omega_g /\omega_0)^2$ (WAB).
A low frequency $g$-mode ($\omega_g \ll \omega$) with $(\ell,m)=(2,\pm2)$ is linearly driven by the non-static linear tide to an energy that is smaller than this by a factor of $\sim (\omega_g/\omega)^4$, where $\omega=2\Omega$ is the tidal frequency.  This yields $E_{i,g}\approx 10^{-35}E_0$ for the values in the equation above. 

The near cancellation of three- and four-wave couplings  $K_{4gg}+\sum K_{3pg}K_{3\bar{p}g}$ implies that in order to prevent the runaway growth of the $g$-mode due to its self-coupling term $K_{4 gg}$, the many $p$-modes to which the linearly driven $g$-mode couples must grow along with it and maintain an energy
\beq
E_p \simeq \left(K_{3pg}\right)^2 E_g \approx \left(\epsilon \frac{\omega_p}{\omega_g}\right)^2 E_g. 
\eeq
For strongly coupled $p$-$g$ pairs, $k_p\simeq k_g$ which for our neutron star model implies 
\beq
\label{eq:wp_vs_wg}
\omega_p \simeq \frac{\Lambda_g Nc_s}{r\omega_g} \simeq 10^3 \Lambda_g \left(\frac{\omega_g}{10^{-4}\omega_0}\right)^{-1}\omega_0.
\eeq
Therefore for $\ell=2$ $g$-modes
\bea
E_p &\approx& 10^8
\left(\frac{\omega_g}{10^{-4}\omega_0}\right)^{-4} \left(\frac{f_{\rm gw}}{100\trm{ Hz}}\right)^4 E_g, 
\eea
i.e., the $p$-modes must maintain a much larger energy than the $g$-mode in order to preserve the three- and four-mode cancellation throughout the \emph{linear} driving of the $\ell=2$ $g$-mode. Thus, by Equation (\ref{eq:Eig})
\beq
E_{p,i}\approx 10^{-16}\left(\frac{f_{\rm gw}}{100\trm{ Hz}}\right)^8 E_0.
\eeq

These estimates suggest that there are $p$-$g$ unstable modes with initial  energies $E_i \sim 10^{-30}-10^{-20} E_0$ at $f_{\rm gw}=100\trm{ Hz}$.  For such modes, growth factors of $E/E_i\sim 10^{20}$ are significant as they correspond to mode energies that are a substantial fraction of $E_0$, the star's binding energy.  Given that a $g$-mode with frequency $\omega_g\approx10^{-4}\omega_0$ breaks at $E\sim 10^{-10}E_0$ (i.e., $|q_g k_g g_r|\sim1$; WAB), even growth factors $\ll 10^{20}$ may be significant.

 We conclude, therefore, that the finite frequency corrections may lead to substantial mode growth prior to merger.  Since each $g$-mode couples to many $p$-modes, there can be a very  large number of excited modes ($\gg 10^3$).  Determining what affect this may have on the inspiral requires an understanding of the saturation of the instability and is left to future work.

\vspace{0.2cm}
\section{Results with linear damping}
\label{sec:results_damping}

In this section we determine how linear damping influences the stability of the tide to $p$-$g$ coupling.  In order to separate the effects of damping from the effects of finite frequency corrections, we will ignore the latter and assume that the cancellation given by Equation (\ref{eq:K4_K3sq_cancellation}) holds.  The characteristic Equation (\ref{eq:char_pg}) is then
\bea
&&\left[-\left(s-m_{g_j}\Omega\right)^2+i\gamma_{g_j}\left(s-m_{g_j}\Omega\right)+\omega_{g_j}^2\right]q_{g_j}
\non &&= -\omega_{g_j}\sum_{g_i}\sum_{p_i}K_{3 p_i \bar{g}_j} K_{3\bar{p}_i g_i} 
\non &&\times
\left[\frac{\left(s-m_{p_i}\Omega\right)^2-i\gamma_{p_i}\left(s-m_{p_i}\Omega\right)}{\left(s-m_{p_i}\Omega\right)^2-i\gamma_{p_i}\left(s-m_{p_i}\Omega\right)-\omega_{p_i}^2}\right]\omega_{g_i} q_{g_i},
\hspace{0.65cm} 
\eea
where we neglect $\mathcal{O}(\epsilon^2)$ terms since they do not alter the stability.  As in \S~\ref{sec:stability_no_finite_freq_no_damping}, we can estimate the stability of this potentially large system of modes by  considering a simple two mode system consisting of a single $p$-$g$ pair.  Since $K_{3pg}\simeq \epsilon \omega_p/\omega_g$, if we let $r=s-m_p \Omega$ and $\alpha=\left(m_p-m_g\right)\Omega$, the characteristic equation becomes (cf. Equation \ref{eq:r_char_no_finite_freq_no_damping})
\bea
\label{eq:pg_time_dep}
&&\left[\left(r+\alpha\right)^2-i\gamma_g\left(r+\alpha\right)-\omega_g^2\right]
\left[r^2-i\gamma_p r-\omega_p^2\right]
\non &&
-\left[r^2-i\gamma_pr\right]\epsilon^2 \omega_p^2\simeq0.
\eea

We consider the solution to this equation for various limits of $\alpha$ and $\gamma_p$.  In \S~\ref{sec:static_damping}, we consider the case of the static tide ($\alpha=0$) with damping and show that all solutions are stable.  In  \S~\ref{sec:nonstatic_damping} we consider the case of the non-static tide with damping.  We show that $p$-modes that are strongly damped with $\gamma_p \gg \Omega$ are unstable and grow, along with the $g$-modes to which they couple, at a rate $\Gamma\sim \epsilon \Omega\left(\gamma_p/\Omega\right)^{1/2}$. In  \S~\ref{sec:pmode_damping_rate} we estimate the damping rate of high-order $p$- and $g$-modes and argue that $\gamma_p\gg \Omega$ if the $p$-modes are above the acoustic cutoff frequency of the neutron star. In \S~\ref{sec:efolding_with_damping} we show that for $p$-$g$ pairs containing such strongly damped $p$-modes, the modes can reach significant amplitudes before the neutron star merges.

\begin{figure}
\centering
\includegraphics[width=3.4in]{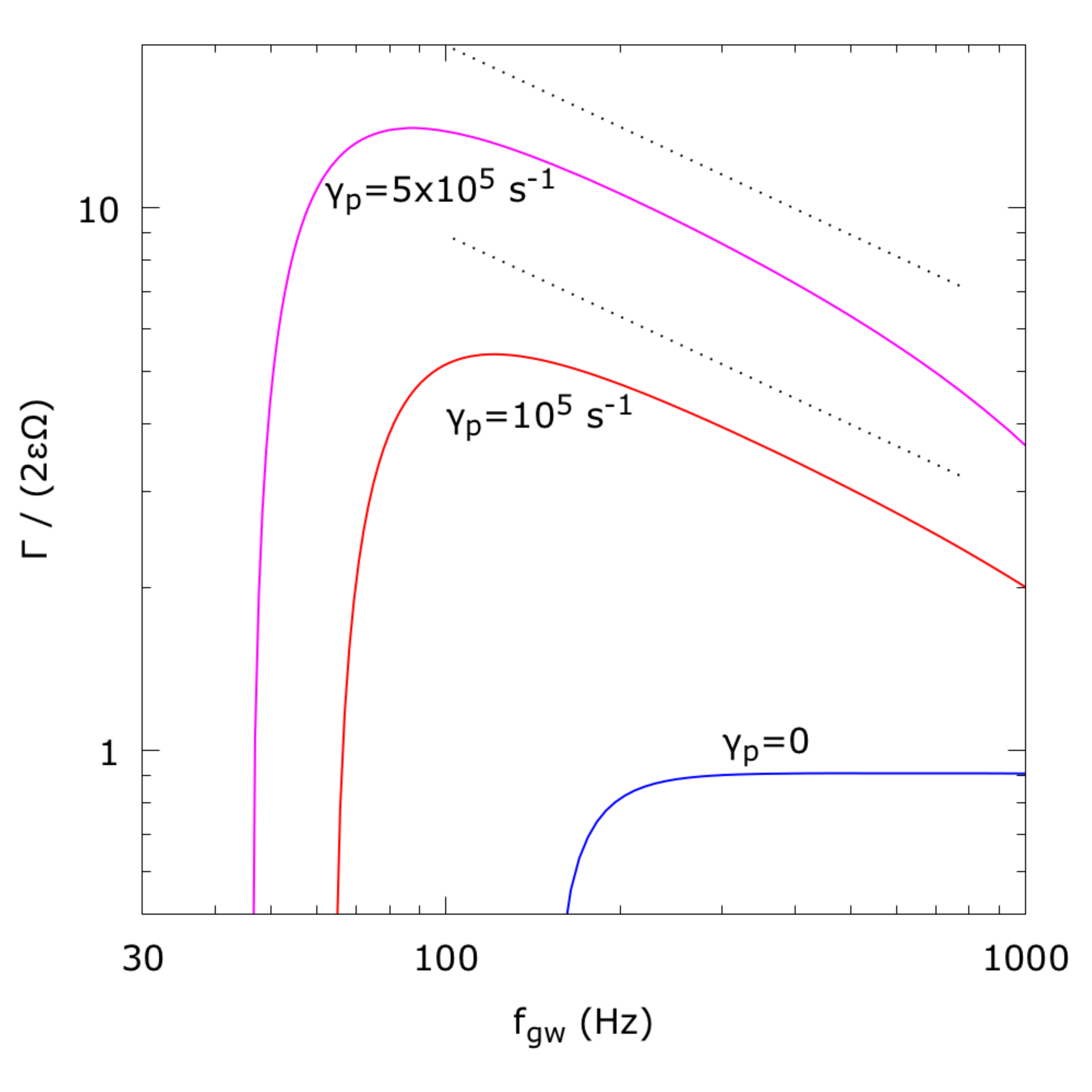} 
\caption{Instability growth rate $\Gamma$ (in units of $2\epsilon \Omega$) including linear mode damping as a function of gravitational wave frequency $f_{\rm gw}$ for an $\ell=2$ $p$-$g$ pair. The mode frequencies are $(\omega_p,\omega_g)\simeq(2\times10^3, 10^{-4}) \omega_0$.  The two upper lines are for $p$-mode damping rates of $\gamma_p=5\times10^5 \trm{ s}^{-1}$ and $10^5\trm{ s}^{-1}$. Both also assume a $g$-mode damping rate $\gamma_g=\omega_g$.  The lower line assumes no damping of either mode. The dotted black lines show $\sqrt{\gamma_p/4\Omega}$.}
\label{fig:growth_rate}
\end{figure}

\subsection{Static tide with linear damping}
\label{sec:static_damping}

By angular momentum conservation (i.e., the coupling coefficient selection rules Equations \ref{eq:selection_rule1}--\ref{eq:selection_rule3}) only $p$-$g$ pairs that satisfy $|m_p-m_g|=m$ couple to harmonic $m$ of the $\ell=2$ linear tide.  Therefore for the case of the static tide $m=\alpha=0$, Equation (\ref{eq:pg_time_dep}) gives, upon substituting $r=-i\beta$,
\bea
&&\beta^4+\beta^3[\gamma_p+\gamma_g]+\beta^2[\omega_p^2(1+\epsilon^2)+\omega_g^2+\gamma_p\gamma_g]
\non &&
+\beta[\omega_p^2(\epsilon^2\gamma_p+\gamma_g)+\gamma_p\omega_g^2]+\omega_p^2\omega_g^2=0.
\eea
The Routh-Hurwitz criterion \citep{Gradshteyn:07} for a quartic equation with real coefficients $a_4x^4+a_3x^3+a_2x^2+a_1x+a_0=0$ shows that the solution is stable since $a_i>0$ for all $i$, $a_3 a_2 > a_4 a_1$, and $a_3a_2a_1 < a_4a_1^2+a_3^2a_0$.

\subsection{Non-static tide with linear damping}
\label{sec:nonstatic_damping}

For the case of the non-static tide $m=\pm 2$,  the low frequency solution to Equation (\ref{eq:pg_time_dep})  is
\bea
\label{eq:eigenvalue_damped}
&& r\simeq -\alpha+\frac{i}{2}\left(\epsilon^2\gamma_p+\gamma_g\right)
\non &&
\pm\left[\omega_g^2-\frac{(\epsilon^2\gamma_p+\gamma_g)^2}{4}
-\epsilon^2\alpha\left[\alpha+i\left(\gamma_p-\gamma_g\right)\right]\right]^{1/2}.\hspace{0.6cm}
\eea
If the last term under the square root dominates, i.e.
\beq
\epsilon \sqrt{\alpha\gamma_p} \gg \omega_g, \epsilon^2\gamma_p, \gamma_g, 
 \trm{ and}\hspace{0.2cm}
\gamma_p\gg |\alpha|, \gamma_g,
\eeq 
then $r \simeq -\alpha \pm \epsilon(1+i)\sqrt{\alpha \gamma_p/2}$ and there is an unstable solution with a growth rate
\beq
\label{eq:Gamma_with_damping}
\Gamma\approx\epsilon\Omega \sqrt{\frac{\gamma_p}{\Omega}}.
\eeq
Thus,  the solution with damping is larger by a factor of $\approx(\gamma_p/4\Omega)^{1/2}$ compared to the solution without damping or finite frequency corrections $\Gamma\approx 2\epsilon\Omega$ (see Equation \ref{eq:growth_rate_no_finite_freq_no_damping}).

To understand why large linear damping can result in a viscous-type instability that enhances the growth rates,  note that the cancellation between three- and four-wave interactions requires that the phases of the $p$- and $g$-mode oscillations be in strict relation.  This relation can only be perfectly satisfied for the static tide (which is why the static tide is always stable).  For the non-static tide \emph{without} linear damping (nor finite frequency corrections), there is a phase offset in the equations of motion due to the Coriolis term $\sim \omega^2$.  As a result, the growth rates scale as $\sim \omega$, as VZH showed (see also Equation \ref{eq:growth_rate_no_finite_freq_no_damping}).  For the non-static tide \emph{with} linear damping, the equations of motion have an additional viscous term $\sim \gamma \omega$ and, in analogy with the effect of the Coriolis term, we find that for large enough $\gamma_p$ it can result in a growth rate $\sim (\gamma_p \omega)^{1/2}$ (Equation \ref{eq:Gamma_with_damping}).

\subsection{Damping rates}
\label{sec:pmode_damping_rate}

First consider the $p$-mode damping rate.  By Equation (\ref{eq:wp_vs_wg}), a $g$-mode with $\omega_g\sim 10^{-4}\omega_0$ couples to a $p$-mode with $\omega_p \sim 10^3\Lambda_g \omega_0$. Such a $p$-mode is well above the  acoustic cutoff frequency of a neutron star ($\omega_{\rm ac}\sim 10^2\omega_0$; WAB).   Because it does not reflect at the stellar surface, it escapes in one group travel time across the star and in a region between $r$ and $r+\Delta r$ it has an effective damping rate 
\beq
\label{eq:gam_p_effective}
\gamma_p \approx \frac{2\pi}{t_{p}(\Delta r)}, 
\eeq
where its group travel time across $\Delta r$ is
\beq
t_{p}(\Delta r) \simeq \int_r^{r+\Delta r} \frac{dr}{c_s}\simeq \frac{\Delta r}{c_s}.
\eeq

The size of the $p$-$g$ coupling region determines $\Delta r$. If the driving rate is slower than the $g$-mode's group travel time across this region, the driving is global and $\Delta r \approx R$ (WAB). In that case $t_p \approx R/c_s \approx 0.7\omega_0^{-1}$ since $c_s\simeq 1.5 R\omega_0$ for a neutron star core.   We therefore find a $p$-mode damping rate
\beq
\gamma_p \approx10^5\trm{ s}^{-1}.
\eeq
This estimate motivates the value for $\gamma_p$ used in the $e$-folding calculation below. In Appendix \ref{app:local_driving_pmode_damping} we show that it maybe more appropriate to treat the $g$-mode  driving as local rather than global.  In that case, the coupling region $\Delta r \ll R$ and we estimate that the $p$-mode damping rate would be $\gg 10^5\trm{ s}^{-1}$.

In WAB we found a $g$-mode damping rate 
\beq
\gamma_g \approx 3\times10^{-9}\Lambda_g^2 T_8^{-2}\left(\frac{\omega_0}{\omega_g}\right)^{2} \trm{ s}^{-1}, 
\eeq
where $T_8$ is the core temperature in units of $10^8 K$. The minimum $g$-mode frequency such that $\omega_g \ga \gamma_g$ is therefore
\beq
\omega_g \approx 0.7 \Lambda_g^{2/3} T_8^{-2/3} \trm{ rad s}^{-1}.
\eeq
In practice, we find that the modes are $p$-$g$ unstable even if $\omega_g<\gamma_g$, although in this regime the $g$-modes are more correctly treated as traveling waves rather than global standing waves.  In any case, this motivates our choice of $\omega_g=10^{-4}\omega_0$ in the estimates below.  

The damping rate calculations above assume that the modes are oscillating at their natural frequencies.  However, $p$-$g$ pairs will oscillate far from their natural frequencies when they are unstable to nonlinear $p$-$g$ driving by the tide (see Equation (\ref{eq:eigenvalue_damped})).  The $p$-mode, in particular, is forced to oscillate at a frequency $\ll \omega_p$.  Although not entirely clear, we might still expect a $p$-mode whose natural frequency is above the acoustic cutoff to have a large damping rate that is given approximately by Equation (\ref{eq:gam_p_effective}).
 This is because once the $p$-mode leaves the coupling region in the core, it is no longer unstable and forced to oscillate at the nonlinear driving frequency. Instead, upon entering the crust, it will begin to oscillate at its natural frequency.  It therefore does not reflect at the stellar surface, and thus its effective damping rate is approximately the inverse sound crossing time. A caveat is that its group velocity within the coupling region might be less than $c_s$. Therefore, perhaps only the outgoing acoustic waves excited near the outer coupling region have damping rates as large as the estimates above.

\subsection{Number of $e$-foldings with damping}
\label{sec:efolding_with_damping}

Figure \ref{fig:growth_rate} shows the $p$-$g$ growth rate accounting for rapid mode damping.  In order to obtain these results, we numerically solve the eigenvalue problem (Equation \ref{eq:eig2form}).  We assume that the $p$-$g$ pair has frequency $(\omega_p,\omega_g)\simeq (2\times10^3,10^{-4})\omega_0$.  Based on the estimates in \S~\ref{sec:pmode_damping_rate}, this suggests $p$-mode damping rates $\gamma_p\approx 10^5\trm{ s}^{-1}$ and $g$-mode damping rates $\gamma_g\approx\omega_g$.   We find that the numerical calculation agrees well with the analytic estimate given by Equation (\ref{eq:Gamma_with_damping}).  The growth rates are significantly larger when damping is included and the instability is triggered much earlier in the inspiral.

By Equation (\ref{eq:lnE}),  the growth rate of Equation (\ref{eq:Gamma_with_damping}) results in an energy $e$-folding 
\beq
\label{eq:efold_damp}
\ln(E/E_i)
\simeq 198 \mathcal{M}_{1.2}^{-5/3}\left(\frac{\gamma_p}{10^5 \trm{ s}^{-1}}\right)^{1/2} 
 \left[f_{i,100}^{-1/6}-f_{100}^{-1/6}\right]
\eeq
for $M'=M$ and $\omega_0=10^4\trm{ rad s}^{-1}$. The instability criterion is  $\omega_g \la \epsilon\sqrt{2\Omega \gamma_p}$, and the modes become unstable at a frequency
\beq
\label{eq:fi_damp}
f_{{\rm gw},i} \simeq 56\left(\frac{\omega_g}{10^{-4}\omega_0}\right)^{2/5}\left(\frac{\gamma_p}{10^5 \trm{ s}^{-1}}\right)^{-1/5}\trm{ Hz}.
\eeq
If the instability begins at $f_{{\rm gw},i}=60\trm{ Hz}$ then  the mode energy grows by a factor of $E/E_i \sim 10^{35}$ by $f_{\rm gw}=1000\trm{ Hz}$. 

 We show in Appendix \ref{app:local_driving_pmode_damping}  that the $g$-mode driving might be in the local regime, in which case $\gamma_p \gg 10^5\trm{ s}^{-1}$.  The modes would then grow by $E/E_i \gg 10^{35}$.

\section{Conclusions}
\label{sec:conclusions}

We analyze the stability of the tide in coalescing binary neutron stars to $p$-$g$ mode coupling.  Previous studies did not account for either the relevant four-wave interactions (WAB) or the compressibility of the linear tide (VZH).  In order to account for both effects, we first solve for the non-static linear tide $\gv\chi^{(1)}$, and thereby obtain the finite frequency corrections to the static tide (such as compressibility).  We then use this solution to evaluate the three- and four-mode coupling coefficients.  We find that the finite frequency corrections are significant and undo the near-exact cancellation between the coefficients found in the incompressible limit.  As a result, the instability begins earlier in the inspiral and the $p$-$g$ growth rates are $\sim \omega_0/\omega$ larger than those of the incompressible limit.  We find that the unstable $p$-$g$ modes can potentially reach significant energies well before the neutron star merges.

In a separate analysis, we investigate the effects of linear damping on the $p$-$g$ instability (ignoring finite frequency corrections in order to disentangle the two effects).  For small linear damping rates, the damping acts to slow the mode growth.  However, if the $p$-mode damping rate is sufficiently large, it induces a viscous-type instability and the $p$-$g$ growth rate increases with increasing $\gamma_p$. We find that $p$-modes above the acoustic cutoff frequency of the neutron star have effective damping rates $\gamma_p\ga \omega_0$ and increase the instability growth rates relative to the inviscid limit by a factor of $\sim (\omega_0/\omega)^{1/2}$. This viscous mechanism can therefore also drive modes to significant energy before the merger.

In this study we focused on the stability and growth rate of $p$-$g$ coupling.  We did not attempt to solve for the instability's saturation and therefore did not determine its influence on the tidal dissipation rate. Thus, we do not know to what extent the instability affects the phase evolution of gravitational waves from coalescing neutron star binaries.  Nonetheless, its early onset ($f_{\rm gw} \approx 50\trm{ Hz}$), rapid growth rates, and large number of excited modes ($\gg 10^3$), suggests the instability's impact could be significant and motivates the study of its nonlinear saturation.

\acknowledgments{
The author thanks Gordon Ogilvie for pointing out the argument for the stability of the static tide given in Appendix A and also thanks Phil Arras and the referee for valuable comments on this manuscript.  This research was supported by NASA grant NNX14AB40G.
}

\begin{appendix}

\section{Stability of a static tidal field}
\label{app:static_tidal_field}

Consider a neutron star that is deformed by a \emph{static} tidal field.  We show here that a short wavelength perturbation to this static, tidally deformed neutron star is stable to all orders in the tidal factor $\epsilon = (M'/M)(R/a)^3$ as long as $\epsilon \la 1$ (we thank G. Ogilvie for bringing this argument to our attention).  

The response of a static, tidally deformed star to a small perturbation $\gv{\xi}$ is given by
\beq
\rho \frac{\partial^2 \gv{\xi}}{\partial t^2}=-\grad\delta p -\delta\rho\grad\phi-\rho\grad\delta\phi,
\eeq
where $\delta p, \delta \rho$, and $\delta \phi$ are Eulerian perturbations and the gravitational potential $\phi$ is that of the star plus the (time independent) external tidal potential. Since the background is taken to be the tidally deformed star (not a spherically symmetric star), the background quantities $\rho, p$, etc. account for the exact static tide response to all orders in the tidal factor $\epsilon$.  If we assume that $\rho$ and $p$ vanish at the surface of the star and the perturbation varies with time as $\gv{\xi}\sim e^{is t}$, the energy equation of the perturbation is 
\bea
s^2\int_V \rho |\gv{\xi}|^2 dV &=& -\frac{1}{4\pi G}\int_\infty |\grad \delta \phi|^2 dV + \int_V \left[\frac{|\delta p|^2}{\Gamma_1 p}+\left(\gv{\xi}^\ast\cdot\grad p\right)\left(\gv{\xi}\cdot\gv{A}\right)\right]dV,
\eea
where $V$ is the volume of the tidally distorted star and $\gv{A}=\grad\ln \rho - (1/\Gamma_1)\grad\ln p$ is the Schwarzschild discriminant.  Since the perturbation is assumed to be small, we only account for its energy to lowest  (i.e., quadratic) order.  

The three contributions to $s^2$ are: the perturbed gravity term (negative and therefore destabilizing), the perturbed pressure term (positive and therefore stabilizing),  and the buoyancy term, whose influence on the stability we now consider.  In a spherical star, $(\gv{\xi}^\ast\cdot \grad p)(\gv{\xi}\cdot\grad \gv{A})=\rho N^2|\xi_r|^2$, and the buoyancy term is stabilizing if $N^2>0$, where $N^2\equiv \grad p\cdot\gv{A}/\rho$ is the Brunt-V\"ais\"al\"a buoyancy frequency.  Consider now a star deformed by a static tide.  Since the hydrostatic equilibria are barotropic, the vectors $\grad p$ and $\gv{A}$ are parallel and  $(\gv{\xi}^\ast\cdot \grad p)(\gv{\xi}\cdot\grad \gv{A})=\rho N^2|\xi_{\hat{g}}|^2$, where now $\rho N^2$ is that of the tidally deformed star and $\xi_{\hat{g}}$ is the component of $\gv{\xi}$ along $\gv{g}=-\grad \phi$.  If we start with a stably stratified spherical star and apply a static tidal deformation, we still have $N^2>0$ as long as $\epsilon \la 1$.  This is because the Eulerian perturbation to $N^2$ due to the static tide is $\delta N^2/N^2 \sim \epsilon$ (see, e.g., \citealt{Burkart:13}).  The buoyancy term in the tidally deformed star is therefore stabilizing if the spherical star is stably stratified, as is the case for neutron stars.  

For short wavelength perturbations, such as those we consider in the main text, the perturbed gravity term $\delta\phi$ in the energy equation should be unimportant (i.e., it is appropriate to work in the Cowling approximation).  We therefore conclude that a short wavelength perturbation to a static, tidally deformed neutron star is stable to all orders in $\epsilon$.

\section{Useful form for the sum rule integral}
\label{app:sum_rule_useful_form}

We derive here an expression of the sum rule integral (Equation \ref{eq:K3sq_sum_rule}) that makes its similarities with $K_{4cd}$ more apparent.  Assuming the Cowling approximation, the linear operator acting on some vector $\gv{\xi}$ can be written as \citep{Reisenegger:94a}
\bea
\frac{\gv{f}_1\left[\gv{\xi}\right]}{\rho}  &=&
-\grad\left(\frac{\delta p}{\rho}\right)+\frac{N^2c_s^2}{g^2}\left(\div{\gv{\xi}}\right)\gv{g}
 =\grad\left[c_s^2 \div{\gv{\xi}} + \gv{g}\cdot\gv{\xi}\right]  +\frac{N^2c_s^2}{ g^2}\left(\div{\gv{\xi}}\right) \gv{g}
\eea
(the two restoring forces are the pressure variation $\delta p$ and the buoyancy associated with $N^2$).
For any two vectors $\gv{\xi}$ and $\gv{\eta}$, we want to evaluate 
\bea
\int d^3x\, \gv{\xi} \cdot \gv{f}_1\left[\gv{\eta}\right] &=& 
\int d^3x \rho \,   \gv{\xi} \cdot\left(\grad\left[c_s^2 \div{\gv{\eta}} + \gv{g}\cdot\gv{\eta}\right]  +\frac{N^2c_s^2 }{g^2} \left(\div{\gv{\eta}}\right) \gv{g}\right).
\eea
We can use integration by parts to simplify the first term 
\bea
\int d^3x \rho \,   \gv{\xi} \cdot \grad\left[c_s^2 \div{\gv{\eta}} + \gv{g}\cdot\gv{\eta}\right]
&=& \int d\gv{S} \cdot \rho \gv{\xi}  \left[c_s^2 \div{\gv{\eta}} + \gv{g}\cdot\gv{\eta}\right] -
\int d^3x \div{\left(\rho \gv{\xi} \right)} \left[c_s^2 \div{\gv{\eta}} + \gv{g}\cdot\gv{\eta}\right]
\non &=& 
-\int d^3x \left\{\rho c_s^2 \left(\div{\gv{\xi}}\right)\left(\div{\gv{\eta}}\right) 
+\rho  \left(\div{\gv{\xi}}\right)  \gv{g}\cdot\gv{\eta}
 +\gv{\xi}\cdot \grad\rho \left[c_s^2 \div{\gv{\eta}} + \gv{g}\cdot\gv{\eta}\right]\right\},
\eea
where the surface term vanishes because we are interested in vectors that satisfy the surface boundary condition
\bea
\left(c_1 \gv{\eta}\cdot\gv{\hat{r}}+c_2\div{\gv{\eta}}\right)_{r=R}=0,
\eea
where $c_1$ and $c_2$ are constants (Equation 5 in Reisenegger 1994).  We then have
\bea
\int d^3x\, \gv{\xi} \cdot \gv{f}_1\left[\gv{\eta}\right] &=& 
-\int d^3x \left\{\Gamma_1 p  \left(\div{\gv{\xi}}\right)\left(\div{\gv{\eta}}\right) 
+\rho  \left(\div{\gv{\xi}}\right)  \gv{g}\cdot\gv{\eta}
- \gv{\xi} \cdot
\left(
\rho \frac{N^2c_s^2 }{g^2} \left(\div{\gv{\eta}}\right) \gv{g} - \grad\rho \left[c_s^2 \div{\gv{\eta}} + \gv{g}\cdot\gv{\eta}\right]
\right)\right\}.\hspace{0.7cm}
\eea
Using the definition of $N^2$, the integrand term within large parentheses simplifies to
\bea
\rho c_s^2\div{\gv{\eta}}\left[-\frac{N^2}{g}-\d{\ln \rho}{r}\right] +g \d{\rho}{r}\eta_r
&=& \rho g \div{\gv{\eta}}+g \d{\rho}{r}\eta_r
\eea
and we find
\beq
\int d^3x\, \gv{\xi} \cdot \gv{f}_1\left[\gv{\eta}\right] =
\int d^3x 
\left[
-\Gamma_1 p  \left(\div{\gv{\xi}}\right)\left(\div{\gv{\eta}}\right)+ 
 \rho g \left(\xi_r \div{\gv{\eta}}+ \eta_r\div{\gv{\xi}}\right)+ g \d{\rho}{r} \xi_r \eta_r
\right].
\eeq
Note that this is symmetric in $\gv{\xi}\leftrightarrow \gv{\eta}$,  as expected given that $\gv{f}_1$ is an Hermitian operator, i.e.,
\beq
\int d^3x\, \gv{\xi} \cdot \gv{f}_1\left[\gv{\eta}\right] =\int d^3x\, \gv{f}_1\left[\gv{\xi}\right] \cdot \gv{\eta}.
\eeq
Equation (\ref{eq:K3sq_sum_rule}) can therefore be written as
\bea
\sum_{b\, \in\, \{p\}} K_{3\bar{b}c}^\ast K_{3\bar{b}d}
&\simeq&
- \frac{1}{E_0}\int d^3x \,\left[\left(1- \frac{2\omega^2}{\langle \omega_b^2\rangle}\right) \gv{\psi}_{ac}^\ast \cdot \gv{f}_1\left[\gv{\psi}_{ad}\right]
-\frac{\omega_0^2}{\langle \omega_b^2\rangle} \left(\gv{\psi}_{ac}^\ast \cdot  \gv{f}_1\left[\gv{\zeta}_{ad}\right]+\gv{\psi}_{ad}\cdot \gv{f}_1\left[\gv{\zeta}_{ac}^\ast\right]\right)\right]
\non&\simeq&
- \frac{1}{E_0}\int d^3x 
\left[
-\Gamma_1 p  \left(\div{\gv{\psi}_{ac}^\ast}\right)\left(\div{\gv{\psi}_{ad}}\right)+ 
 \rho g \left(\psi_{ac,r}^\ast \div{\gv{\psi}_{ad}}+ \psi_{ad,r}\div{\gv{\psi}_{ac}^\ast}\right)+ g \d{\rho}{r} \psi_{ac,r}^\ast \psi_{ad,r}
\right]
\non &&
-\frac{1}{E_0}
\int d^3x \frac{\Gamma_1 p}{\langle \omega_b^2\rangle}
  \left[2\omega^2\left(\div{\gv{\psi}_{ac}^\ast}\right)\left(\div{\gv{\psi}_{ad}}\right)+ 
\omega_0^2\left\{\left(\div{\gv{\psi}_{ac}^\ast}\right)\left(\div{\gv{\zeta}_{ad}}\right)
+\left(\div{\gv{\zeta}_{ac}^\ast}\right)\left(\div{\gv{\psi}_{ad}}\right)\right\}\right]
\non &\simeq&
- \frac{1}{E_0}\int d^3x 
\left[
-\Gamma_1 p  \left(\div{\gv{\psi}_{ac}^\ast}\right)\left(\div{\gv{\psi}_{ad}}\right)+ 
 \rho g \left(\psi_{ac,r}^\ast \div{\gv{\psi}_{ad}}+ \psi_{ad,r}\div{\gv{\psi}_{ac}^\ast}\right)+ g \d{\rho}{r} \psi_{ac,r}^\ast \psi_{ad,r}
\right]
\non &&
-\frac{1}{E_0}
\int d^3x \frac{\Gamma_1 p}{\langle \omega_b^2\rangle}
  \left[2\omega^2\d{\psi_{ac,r}^\ast}{r}\d{\psi_{ad,r}}{r}+ 
\omega_0^2\left\{\d{\psi_{ac,r}^\ast}{r}\d{\zeta_{ad,r}}{r}
+\d{\zeta_{ac,r}^\ast}{r}\d{\psi_{ad,r}}{r}\right\}\right],
\label{eq:K3sq_v1}
\eea
where we dropped terms smaller than $\mathcal{O}(\omega_c^{-2})$ given that $\langle \omega_b^2\rangle \propto \omega_c^{-2}$, $\gv{\psi}_{ac} \sim \mathcal{O}(\omega_c^{-1})$ and 
$\div{\gv{\psi}_{ac}} \sim d\psi_{ac,r}/dr \sim \mathcal{O}(\omega_c^{-2})$. In terms of the  $\mathcal{O}(\omega_c^{-1})$ function
\beq
 z_{ac}(r,\theta,\phi)\equiv
r\left[ \div{\gv{\psi}_{ac}} -\d{\psi_{ac,r}}{r}\right]
= 2\psi_{ac,r} + \frac{1}{\sin\theta}\left[\d{\left(\sin\theta \psi_{ac,\theta}\right)}{\theta}+\d{\psi_{ac,\phi}}{\phi}\right],
\eeq
we have, after integration by parts,
\bea
\int d^3x \rho g \left(\psi_{ac,r}^\ast \div{\gv{\psi}_{ad}}+ \psi_{ad,r}\div{\gv{\psi}_{ac}^\ast}\right)&=&
\int d^3x \rho g \left(\d{(\psi_{ac,r}^\ast \psi_{ad,r})}{r} +\frac{\psi_{ac,r}^\ast z_{ad}+\psi_{ad,r} z_{ac}^\ast}{r}\right)
\non &=&
 -\int d^3x \left(\left[\rho \d{g}{r}+g\d{\rho}{r}+\frac{2\rho g}{r}\right]\psi_{ac,r}^\ast \psi_{ad,r}- \frac{\psi_{ac,r}^\ast z_{ad}+\psi_{ad,r} z_{ac}^\ast}{r}\right).\hspace{0.5cm}
\eea
We therefore have our final result (Equation \ref{eq:k3pg_sq_final})
\bea
\sum_{b\, \in\, \{p\}} K_{3bc} K_{3\bar{b}d}
&\simeq&
\frac{1}{E_0}\int d^3x
\bigg\{\Gamma_1 p \Big[\big(a^i_{;j}c^j_{;i}\big)\left(a^k_{;s}d^s_{;k}\right)+  \big(a^i_{;j}c^j_{;i}\big)\left(d^k a^s_{;ks}\right)+ \big(c^i a^j_{;ij}\big)\left(a^k_{;s}d^s_{;k}\right)
 + \big(c^i a^j_{;ij}\big)\left(d^k a^s_{;ks}\right)\Big]
  \non &&
-\frac{\rho g}{r} \left[ \psi_{ac,r} z_{ad}+\psi_{ad,r} z_{ac} -\left(2+\d{\ln g}{\ln r}\right)\psi_{ac,r} \psi_{ad,r} \right] 
\non &&
 -\frac{\Gamma_1 p}{\langle \omega_b^2\rangle}
  \left[2\omega^2\d{\psi_{ac,r}}{r}\d{\psi_{ad,r}}{r}+ 
\omega_0^2\left(\d{\psi_{ac,r}}{r}\d{\zeta_{ad,r}}{r}
+\d{\zeta_{ac,r}}{r}\d{\psi_{ad,r}}{r}\right)\right]\bigg\}
+\epsilon^2\mathcal{O}(\omega_c^{-1}),
\label{eq:kap_a_ns_sq}
\eea
where we wrote $\div{\gv{\psi}_{ac}}$ in component form
\beq
\div{\gv{\psi}_{ac}} = \div{\left[\left(\gv{c}\cdot\grad\right)\gv{a}\right]}=
\left(c^j a^i_{;j}\right)_{;i} =a^i_{;j}c^j_{;i} + c^j a^i_{;ji} 
\eeq
and used the fact that $\gv{a}=\gv{\chi}^{(1)}=\left(\gv{\chi}^{(1)}\right)^\ast$. 

\section{Four-mode coupling coefficient $\kappa_{\lowercase{abcd}}$}
\label{app:friendly_four_mode_coup_coef}

Here we give the form of the four-mode coupling coefficient $\kappa_{abcd}$ that we use for numerical calculation. We derive it using the same general approach taken by \citet{Wu:01} and WAQB to derive a numerically useful form for the three-mode coupling coefficient $\kappa_{abc}$.  We find that the coefficient can be broken into five pieces
\beq
\kappa_{abcd} = -\frac{I + II + III + IV + V}{6E_0},
\eeq
and can be written in terms of angular integrals, $\xi_r$, $\xi_h$, $\div{\gv{\xi}}$, and
\beq
\label{eq:aov1}
a^i_{;j}b^j_{;i}= h^{(1)}_{ab} f^{(1)}_{ab}+h^{(2)}_{ab} f^{(2)}_{ab}+h^{(3)}_{ab} f^{(3)}_{ab},
\eeq
where
\bea
\label{eq:hab1}
h^{(1)}_{ab}(r)f^{(1)}_{ab}(\theta,\phi) &\equiv&\left[\frac{\partial a_r}{\partial r} \frac{\partial b_r}{\partial r}-\frac{\Lambda_b^2}{r^2}a_rb_h -\frac{\Lambda_a^2}{r^2}a_hb_r+\frac{2}{r^2}a_rb_r\right]Y_a Y_b,\\
h^{(2)}_{ab}(r)f^{(2)}_{ab}(\theta,\phi)&\equiv&\left[\frac{a_r }{r}\frac{\partial b_h}{\partial r}+\frac{b_r}{r}\frac{\partial a_h}{\partial r}-\frac{1}{r}\frac{\partial (a_h b_h)}{\partial r}\right]\grad Y_a\cdot\grad Y_b,\\
 \label{eq:hab3}
h^{(3)}_{ab}(r)f^{(3)}_{ab}(\theta,\phi) &\equiv& \frac{a_h b_h}{r^2}\nabla_i\nabla^jY_a\nabla_j\nabla^i Y_b,
\eea
as follows:
\bea
I &=& T_{abcd}\int dr \, r^2  p\, G_1 \grad\cdot\gv{a}\grad\cdot\gv{b}\grad\cdot\gv{c}\grad\cdot\gv{d}\\
\non
II &=&
\int dr \, r^2 p \, G_2 \left[ 
\grad\cdot\gv{a}\grad\cdot\gv{b} \left(h_{cd}^{(1)} T_{abcd}+h_{cd}^{(2)} F_{ab,cd}^{(2)}+h_{cd}^{(3)} F_{ab,cd}^{(3)}\right)+ 
\{\trm{permutations of abcd}\}
\right]\\
\non
III &=& \int dr\, r^2 \Gamma_1 p \sum_{i=1}^3\sum_{j=1}^3\left[h^{(i)}_{ab} h^{(j)}_{cd}E^{(ij)}_{ab,cd} +h^{(i)}_{ac} h^{(j)}_{bd}E^{(ij)}_{ac,bd}+h^{(i)}_{ad} h^{(j)}_{bc}E^{(ij)}_{ad,bc}  \right]
\\
\non
\label{eq:kap4_termIV}
IV&=&\int dr\, \Gamma_1 p\bigg\{
-4r^2 \grad\cdot\gv{a}\grad\cdot\gv{b}\grad\cdot\gv{c}\grad\cdot\gv{d}T_{abcd}
\non && 
+r^2 \grad\cdot\gv{a}\left[\grad\cdot\gv{b} \left(h_{cd}^{(1)} T_{abcd}+h_{cd}^{(2)} F_{ab,cd}^{(2)}+h_{cd}^{(3)} F_{ab,cd}^{(3)}\right)+\{b\leftrightarrow c \leftrightarrow d\}\right]
\non &&
+\grad\cdot\gv{a}\bigg(\pd{b_r}{r}\left[\left(2c_r d_r - c_r d_h\Lambda_d^2 
- d_r c_h\Lambda_c^2 \right)T_{abcd}
+c_h d_h\left(\Lambda_c^2\Lambda_d^2 T_{abcd}-F_{ab,cd}^{(3)}\right) \right]
\non &&
 \hspace{1.1cm}
+\pd{b_h}{r}
\Big[\left(c_r-c_h\right)\left(d_hS_{a,bc,d}-d_rF_{ad,bc}^{(2)}\right)
+\left(d_r-d_h\right)\left(c_hS_{a,bd,c}-c_rF_{ac,bd}^{(2)}\right)\Big]+
\{b\leftrightarrow c \leftrightarrow d\} \bigg)
\non &&
+\{\trm{permutations of abcd}\}
\bigg\}
\\
V &=& \int dr  \rho\bigg\{
a_r b_r c_r d_r r^2 \frac{\partial^4\phi}{\partial r^4} T_{abcd}
+a_h b_h c_h d_h \left(\pdd{\phi}{r} - \frac{1}{r}\pd{\phi}{r} \right)
\left(E_{ab,cd}^{(22)}+E_{ac,bd}^{(22)}+E_{ad,bc}^{(22)}\right)
\non &&
+\left(r\frac{\partial^3\phi}{\partial r^3} - 2\pdd{\phi}{r}+\frac{2}{r}\pd{\phi}{r}\right)
\bigg(
a_r b_r c_h d_h F_{ab,cd}^{(2)} +
a_r b_h c_r d_h F_{ac,bd}^{(2)} +
a_r b_h c_h d_r F_{ad,bc}^{(2)} 
 \non &&
+a_h b_r c_r d_h F_{bc,ad}^{(2)} +
a_h b_r c_h d_r F_{bd,ac}^{(2)}+
a_h b_h c_r d_r  F_{cd,ab}^{(2)} \bigg)\bigg\}.
\eea
Here the divergences are with respect to the radial functions. 
The angular integrals are defined as
\bea
\label{eq:Tabcd}
T_{abcd}&\equiv& \int d\Omega\, Y_a Y_b Y_c Y_d,\\
\label{eq:Fabcd}
F_{ab,cd}^{(i)} &\equiv& \int d\Omega \, Y_a Y_b  f_{cd}^{(i)},\\
\label{eq:Eabcd}
E^{(ij)}_{ab,cd} &\equiv& \int d\Omega f^{(i)}_{ab}f^{(j)}_{cd},\\
\label{eq:Sabcd}
S_{a,bc,d} &=&
\int d\Omega Y_a \nabla_i Y_b \nabla^j Y_c \nabla_j \nabla^i Y_d+\Lambda_d^2 F_{ad,bc}^{(2)}.
\eea
Paired subscripts not separated by a comma are symmetric in those indices. Note that $F_{ab,cd}^{(1)}=T_{abcd}$. The integrals $T_{abcd}$, $F_{ab,cd}^{(2)}$, $F_{ab,cd}^{(3)}$, and $S_{a,bc,d}$ are, respectively, analogs of the three-mode integrals $T_{abc}$, $F_b$, $V_b$, and $S$ defined in \citet{Wu:01}; the integrands of the former equal $Y_a$ times the integrands of the latter.  We compute these integrals numerically in our calculations.

\subsection{A relation between $S_{a,bc,d}$ and $E_{ab,cd}^{(22)}$}
\label{app:Sabcd_Eabcd_relation}

The differential expression under the integral in the definition of $S_{a,bc,d}$ can be written as
\beq
\nabla_i Y_b \nabla^j Y_c \nabla_j \nabla^i Y_d
 =\nabla_j\left(\nabla^i Y_d \nabla_i Y_b\right)\nabla^j Y_c - \nabla_i Y_d \nabla^j Y_c\nabla_j \nabla^i Y_b. 
\eeq
Therefore 
\bea
S_{a,bc,d}+S_{a,cd,b}&=&\int d\Omega \,Y_a \nabla_j\left(\nabla^i Y_d \nabla_i Y_b\right)\nabla^j Y_c +\Lambda_b^2 F_{ab,cd}^{(2)} + \Lambda_d^2 F_{ad,bc}^{(2)}
\non &=& \int d\Omega \,\nabla^i Y_d \nabla_i Y_b\left[\Lambda_c^2 Y_a Y_c - \nabla_j Y_a \nabla^j Y_c\right] +\Lambda_b^2 F_{ab,cd}^{(2)} + \Lambda_d^2 F_{ad,bc}^{(2)}
\non &=&
\Lambda_b^2 F_{ab,cd}^{(2)}+\Lambda_c^2 F_{ac, bd}^{(2)} + \Lambda_d^2 F_{ad,bc}^{(2)}- E_{ad, bc}^{(22)},
\eea
where we used integration by parts and the relation $\nabla^2Y=-\Lambda^2 Y$. We thus have 
\beq
\label{eq:Sabcd_E22abcd}
E_{ac,bd}^{(22)}-S_{a,bd,c}=E_{ad,bc}^{(22)}-S_{a,bc,d}.
\eeq

\section{Expressions for the $\mathcal{O}(\omega_{\lowercase{g}}^{-2})$ contributions to $K_{\lowercase{4gg}}+\sum K_{\lowercase{3pg}} K_{\lowercase{3\bar{p}g}}$}
\label{app:Komc2}

In this appendix we provide expressions for the $\mathcal{O}(\omega_{\lowercase{g}}^{-2})$ contributions to $K_{\lowercase{4gg}}+\sum K_{\lowercase{3pg}}   K_{\lowercase{3\bar{p}g}}$.  In \S~\ref{app:3kaagg_sum_k3pg_sq} we consider $3\kappa_{\chi^{(1)}\chi^{(1)}gg} + \sum K_{3pg} K_{3\bar{p}g}$.
We then describe two methods for calculating the nonlinear tide $\gv{\chi}^{(2)}$, which we need in order to evaluate $2\kappa_{\chi^{(2)}gg}$: (\S~\ref{app:nltide_sum_over_modes}) as a sum over modes and (\S~\ref{app:nltide_direct}) by directly integrating the inhomogeneous equation of motion.

Once we have $\gv{\chi}^{(2)}$ we can calculate $\kappa_{\chi^{(2)}gg}$ similar to how WAQB compute the three-mode coupling coefficient $\kappa_{abc}$.  Note, however, that we cannot use the final expression for $\kappa_{abc}$ in WAQB (A55--A62) since their analysis assumes that all three modes are non-radial modes ($\ell \neq 0$) whereas $\ell=0$ is one of the harmonics of $\gv\chi^{(2)}$.  Instead, we proceed as in VZH and use their Equations D10--D13. When calculating $\gv\chi^{(2)}$ by direct integration (\S~\ref{app:nltide_direct}) we must account for the inhomogeneous term in the equation of motion, which yields, in addition to VZH's Equation D10--D13, an inhomogeneous coupling term
\beq
\kappa_{\chi^{(2)}cd}^{(I)}=\frac{1}{2E_0}\int dr r \left(c_r d_h + c_h d_r-c_h d_h\right) f_{r, \chi^{(2)}} F_{\chi^{(2)}},
\eeq
where $c$ and $d$ label the $g$-modes, $F_{\chi^{(2)}}$ is the three-mode angular integral and $f_{r, \chi^{(2)}}$ is the radial component of the nonlinear tide driving force (given by Equation \ref{eq:fr_nltide} below).

\subsection{$3\kappa_{\chi^{(1)}\chi^{(1)}gg} + \sum K_{3pg} K_{3\bar{p}g}$}
\label{app:3kaagg_sum_k3pg_sq}

Starting from Equations (\ref{eq:k3pg_sq_final}) and (\ref{eq:kaacd}), we decompose the terms that enter at $\mathcal{O}(\omega_g^{-2})$ into eight separate pieces
\beq
\label{eq:Komg2}
3\kappa_{aacd} + \sum_b K_{3bc} K_{3\bar{b}d}
=I+II+III+IV+V+VI+VII+VIII,
\eeq
where $a$ labels the linear tide (whose frequency is $\omega$), $c$ and $d$ label the $g$-modes, and
\beq
I=\frac{1}{E_0}\int d^3x 
\Gamma_1 p \left[\big(c^i_{;j}a^j_{;i}\big) \left(d^k a^s_{;sk}\right)+\big(c^i a^j_{;ij}\big) \left(d^k_{;s} a^s_{;k}\right)
-\div{\gv{a}}\det\left|(a+c +d)^i_{;j}\right|\right]
\eeq

\beq
II=\frac{1}{E_0}\int d^3x 
\Gamma_1 p  \big(c^i a^j_{;ij}\big) \left(d^k a^s_{;sk}\right)
\eeq

\beq
III=
-\frac{1}{E_0}\int d^3x\frac{\rho g}{r} \left( \psi_{ac,r} z_{ad}+\psi_{ad,r} z_{ac} -\left(2+\d{\ln g}{\ln r}\right)\psi_{ac,r} \psi_{ad,r} \right) 
\eeq

\beq
IV=-\frac{1}{E_0}
\int d^3x \frac{\Gamma_1 p}{\langle \omega_b^2\rangle}
  \left[2\omega^2\d{\psi_{ac,r}}{r}\d{\psi_{ad,r}}{r}+ 
\omega_0^2\left\{\d{\psi_{ac,r}}{r}\d{\eta_{ad,r}}{r}
+\d{\eta_{ac,r}}{r}\d{\psi_{ad,r}}{r}\right\}\right]
\eeq

\beq
V=
-\frac{1}{2E_0}\int d^3x \Gamma_1 p\left(\frac{G_2}{\Gamma_1} +2\right)\left(\left(\div{\gv{a}}\right)^2 c^i_{;j}d^j_{;i} + 2\left(\div{\gv{a}}\right) \left(\div{\gv{c}}\right) a^i_{;j}d^j_{;i}+ 2\left(\div{\gv{a}}\right) \left(\div{\gv{d}}\right) a^i_{;j}c^j_{;i} \right)
\eeq

\beq
VI=-\frac{1}{2E_0}\int d^3x \Gamma_1 p\big(a^i_{;j}a^j_{;i}\big)\left(c^k_{;s}d^s_{;k}\right)
\eeq

\beq
VII= -\frac{1}{2E_0}\int d^3x \Gamma_1 p\left(\div{\gv{c}}\det\left|\left(a+a+d\right)^i_{;j}\right|
+\div{\gv{d}}\det\left|\left(a+a+c\right)^i_{;j}\right|\right)
\eeq

\beq
VIII=  -\frac{1}{2E_0}\int d^3x \rho  a^i a^j c^k d^s \phi_{;ijks}
\eeq
In \S~\ref{sec:corrections_wc3&4} we show that term $I$ is $\mathcal{O}(\omega_c^{-2})$ due to cancellations between its individual  $\mathcal{O}(\omega_c^{-3})$ terms.  From the definitions of $\gv\psi_{ac}$ and $\gv\zeta_{ac}$ (Equations \ref{eq:psi1} and \ref{eq:zeta1}) and the four-mode angular integrals defined in Appendix \ref{app:friendly_four_mode_coup_coef}, we can reduce the $\mathcal{O}(\omega_c^{-2})$  portion of these terms to radial integrals. With the double sum over all harmonics of the linear tide mode $a$ implicit, we find:
\bea
 \label{eq:t1}
I &=& 
-\frac{1}{E_0}\int dr\Bigg\{
\d{}{r}\left(\Gamma_1 p \div{\gv{a}}\left[a_hS_{a,cd,a}-a_rF_{aa,cd}^{(2)}+\left(a_r - a_h\right)\left(E_{ac,ad}^{(22)}-S_{a,ad,c}\right)\right]\right)
\non &&
+ \Gamma_1 p 
\d{\left(\div{\gv{a}}\right)}{r}\left(a_r-a_h\right)\left(\Lambda_c^2 F_{ac,ad}^{(2)}+\Lambda_d^2 F_{ad,ac}^{(2)}\right)
\non &&+
  \Gamma_1 p \div{\gv{a}}
\bigg(\pd{a_r}{r}\left[\Lambda_c^2  F_{ac,ad}^{(2)} +\Lambda_d^2 F_{ad,ac}^{(2)}+\Lambda_c^2\Lambda_d^2 T_{abcd}-F_{aa,cd}^{(3)}\right]
+\pd{a_h}{r}\left[E_{ac,ad}^{(22)} +E_{ad,ac}^{(22)}-S_{a,ac,d}-S_{a,ad,c}\right]
\non &&
+\frac{\left(a_r-a_h\right)}{r}
\left[\Lambda_c^2 F_{ac,ad}^{(2)}+\Lambda_d^2 F_{ad,ac}^{(2)}
 \right]
+\frac{a_r}{r} \left(\Lambda_c^2\left[F_{aa,cd}^{(2)}-F_{ac,ad}^{(2)}\right]+\Lambda_d^2\left[F_{aa,cd}^{(2)}-F_{ad,ac}^{(2)}\right]-2\Lambda_c^2\Lambda_d^2 T_{aacd}\right)
\non &&
+\frac{a_h}{r}\left[2\Lambda_a^2\Lambda_c^2\Lambda_d^2 T_{aacd}+E_{ac,ad}^{(32)} +E_{ad,ac}^{(32)}-\Lambda_c^2 \left(S_{a,cd,a}+F_{ac,ad}^{(3)}\right)-\Lambda_d^2 \left(S_{a,cd,a}+F_{ad,ac}^{(3)}\right)\right]\bigg)
\Bigg\}\, c_h d_h
\eea

\beq
\label{eq:t2}
II= \frac{1}{E_0}\int dr \Gamma_1 p \left(\div{\gv{a}}\right)^2 E_{ac,ad}^{(22)} \, c_h d_h
\eeq

\bea
\label{eq:t3}
III&=&-\frac{1}{E_0} \int dr \rho g  \frac{\left(a_r-a_h\right)}{r}\Bigg[
 a_r\left(\left[4-\d{\ln g}{\ln r}\right]E_{ac,ad}^{(22)}-\Lambda_c^2 F_{ac,ad}^{(2)}-\Lambda_d^2 F_{ad,ac}^{(2)}\right) 
 \non &&
 +a_h\left(E_{ac,ad}^{(32)}+E_{ad,ac}^{(32)}-\left[2\Lambda_a^2-\d{\ln g}{\ln r}\right] E_{ac,ad}^{(22)}\right) \Bigg]\,c_h d_h
\eea

\bea
\label{eq:t4}
IV &=&
-\frac{1}{E_0}
\int dr \frac{\Gamma_1 p}{\langle \omega_p^2\rangle} \left(a_r-a_h\right)\bigg[\left(\omega^2+\omega'^2\right)\left(a'_r-a'_h\right)E_{ac,a'd}^{(22)}
\non &&
+ g\left[\Gamma_1 \d{\ln \Gamma_1}{\ln p}-\left(\pd{\ln \Gamma_1}{\ln \rho}\right)_S\right] \div{\gv{a}}
\left(\Lambda_c^2F_{ac,ad}^{(2)}+\Lambda_d^2 F_{ad,ac}^{(2)}\right)\bigg]\, \d{c_h}{r}\d{d_h}{r} 
\eea

\bea
\label{eq:t5}
V&=& -\frac{1}{2E_0} \int dr \bigg\{
\left(\frac{G_2}{\Gamma_1} +2\right)\div{\gv{a}}
\bigg[\Gamma_1 p\div\gv{a} \left(\Lambda_c^2 \Lambda_d^2  T_{aacd}+
F_{aa,cd}^{(3)}-\Lambda_c^2F_{aa,cd}^{(2)} -\Lambda_d^2F_{aa,cd}^{(2)}\right)
\non &&
 -2\rho g\left(a_r - a_h\right)
\left(\Lambda_c^2F_{ac,ad}^{(2)}+\Lambda_d^2F_{ad,ac}^{(2)}\right) \bigg]
+\d{}{r}\left[r \Gamma_1 p\left(\frac{G_2}{\Gamma_1} +2\right)\left(\div{\gv{a}}\right)^2\right] F_{aa,cd}^{(2)}
\bigg\}\, c_h d_h
\eea

\bea
\label{eq:t6}
VI &=&
-\frac{1}{2E_0}\int dr \bigg\{
 \Gamma_1 p \Bigg[
\left(h^{(1)}_{aa}T_{aacd}+h^{(2)}_{aa} F_{cd,aa}^{(2)}+h^{(3)}_{aa} F_{cd,aa}^{(3)}\right)\Lambda_c^2 \Lambda_d^2 
+\left(h^{(1)}_{aa}E_{aa,cd}^{(13)}+h^{(2)}_{aa} E_{aa,cd}^{(23)}+h^{(3)}_{aa} E_{aa,cd}^{(33)}\right) 
\non &&
- \left(h^{(1)}_{aa}E_{aa,cd}^{(12)}+h^{(2)}_{aa} E_{aa,cd}^{(22)}+h^{(3)}_{aa} E_{aa,cd}^{(32)}\right)\left(\Lambda_c^2  + \Lambda_d^2\right)
\Bigg]
+\d{}{r}\left[r\Gamma_1 p \left(h^{(1)}_{aa}E_{aa,cd}^{(12)}+h^{(2)}_{aa} E_{aa,cd}^{(22)}+h^{(3)}_{aa} E_{aa,cd}^{(32)}\right)\right]
 \bigg\} c_h d_h
 \non
\eea

\beq
\label{eq:t7}
VII= 
 \frac{1}{E_0}\int  dr \rho g\frac{\left(a_r-a_h\right)}{r}\left[a_h\left(\Lambda_c^2 S_{c,da,a}+\Lambda_d^2 S_{d,ca,a}\right)-a_r\left(\Lambda_c^2F_{ac,ad}^{(2)}+\Lambda_d^2F_{ad,ac}^{(2)}\right)
\right] c_h d_h
\eeq

\beq
 \label{eq:t8}
VIII= 
 -\frac{1}{2E_0} \int dr  \rho\Bigg[
a_h^2 \left(\pdd{\phi}{r} - \frac{1}{r}\pd{\phi}{r} \right)
\left(E_{aa,cd}^{(22)}+E_{ac,ad}^{(22)}+E_{ad,ac}^{(22)}\right)
+a_r^2 \left(r\frac{\partial^3\phi}{\partial r^3} - 2\pdd{\phi}{r}+\frac{2}{r}\pd{\phi}{r}\right)
 F_{aa,cd}^{(2)} \Bigg] c_h d_h
\eeq

\subsection{Calculating $\kappa_{\chi^{(2)}cd}$ with $\gv{\chi}^{(2)}$ evaluated as a sum over modes}
\label{app:nltide_sum_over_modes}

Expanding the nonlinear tide as a sum over modes
\beq
\label{eq:nltide_expand}
\gv{\chi}^{(2)}(\gv{x},t)=\sum \chi_a^{(2)}(t)  \gv{\xi}_a(\gv{x}),
\eeq
Equation (\ref{eq:chi2}) becomes
\beq
\ddot{\chi}_a^{(2)}+\omega_a^2 \chi_a^{(2)}=\omega_a^2 \left[V_a(t) + K_a(t)\right]^\ast
= \omega_a^2 \left[\sum_b\sum_m U_{ab}^{(m)} \chi_b^{(1)}+ \sum_{bc} \kappa_{abc} \chi_b^{(1)}\chi_c^{(1)}\right] e^{-im_a \Omega t},
\label{eq:nltide_eom_expand}
\eeq
where (cf., Equations 20--22 in WAQB)
\beq
V_a(t)= -\frac{1}{E_0}\int d^3x \rho\, \gv{\xi}_a \cdot\left(\gv \chi^{(1)}\cdot \grad\right)\grad U,
\eeq
\beq
K_{a}(t)= \frac{1}{E_0} \int d^3x \, \gv{\xi}_a \cdot\gv{f}_2\left[\gv \chi^{(1)} ,\gv \chi^{(1)}\right].
\eeq
 The amplitudes of the nonlinear tide expansion are therefore given by the steady state solution to Equation \ref{eq:nltide_eom_expand} (cf., Equation 59 in WAQB)
\beq
\label{eq:chi2_expansion}
\chi_a^{(2)}(t)=\frac{\omega_a^2\left(V_a + K_a\right)}{\omega_a^2 - (m_a \Omega)^2} e^{-im_a \Omega t}.
\eeq
For the static tide ($m_a=m_b=m_c=m=0$), 
\beq
\chi_a^{(1)} = U_a = -\frac{1}{E_0}\int d^3x \rho\, \gv{\xi}_a^\ast \cdot \grad U
\eeq
and the solution is (cf. VZH Equation 23)
\beq
\chi_a^{(2)}=V_a + K_a =  \sum_b \left(U_{ab} U_b + \sum_{c} \kappa_{abc} U_b U_c\right).
\eeq
In the more general case that includes the non-static linear tide ($\ell=2$, $m=\{-2,0,2\}$), we first write
\beq
\gv{\chi}^{(1)}(\gv{x},t)
=\gv\chi_{-2}(\gv{x},t)+\gv\chi_{0}(\gv{x},t)+\gv\chi_{2}(\gv{x},t)
=\epsilon \left[W_{20} \tilde{\gv{\chi}}_{0}(\gv{x})+2 W_{22}\tilde{\gv{\chi}}_{2}(\gv{x})
\cos(2\Omega t)\right].
\eeq
The harmonics that contribute to $\gv{\chi}^{(2)}$ are $\ell_a=\{0, 2, 4\}$ and $m_a=\{0,\pm2,\pm4\}$.  The coupling coefficients for the different values of $m_a$ are then
\bea
\left(V_a+K_a\right)_{m_a=0}&=&
\epsilon^2\left[W_{20}W_{20} \left(\left[\hat{J}_{a\tilde{\chi}_0} +2 \hat{\kappa}^{(I)}_{a\tilde{\chi}_0}\right]+ \hat{\kappa}^{(H)}_{a\tilde{\chi}_0\tilde{\chi}_0}\right)T_{[0,0,0]}+
2W_{22}W_{22}\left(\left[\hat{J}_{a\tilde{\chi}_2}+2 \hat{\kappa}^{(I)}_{a\tilde{\chi}_2}\right]+\hat{\kappa}^{(H)}_{a\tilde{\chi}_2\tilde{\chi}_2}\right)T_{[0,2,-2]}\right],
\non\\
\left(V_a+K_a\right)_{m_a=\pm 2}&=&
 \epsilon^2 W_{20}W_{22}\left(\left[\hat{J}_{a\tilde{\chi}_0}+2 \hat{\kappa}^{(I)}_{a\tilde{\chi}_0}\right]
 +\left[\hat{J}_{a\tilde{\chi}_2}+2 \hat{\kappa}^{(I)}_{a\tilde{\chi}_2}\right]+2\hat{\kappa}^{(H)}_{a\tilde{\chi}_0\tilde{\chi}_2}\right)T_{[2,0,-2]},
 \\ \non
\left(V_a+K_a\right)_{m_a=\pm 4}&=&
 \epsilon^2 W_{22}W_{22}\left(\left[\hat{J}_{a\tilde{\chi}_2}+2 \hat{\kappa}^{(I)}_{a\tilde{\chi}_2}\right]+\hat{\kappa}^{(H)}_{a\tilde{\chi}_2\tilde{\chi}_2} \right)T_{[4,-2,-2]},
\eea
where $T$ is the three-mode angular integral, which we label here by the values of $m$  (the three $\ell$ values are $\ell_a$ for the nonlinear tide and $\ell=2$ for each linear tide).  The coefficients   $\hat{J}_{ab}$, $\hat{\kappa}_{ab}^{(I)}$, and $\hat{\kappa}_{abc}^{(H)}$ are as defined in Appendix A of WAQB (the hat symbol indicates that the angular integral $T$ is factored out).
 
\subsection{Calculating $\kappa_{\chi^{(2)}cd}$ with $\gv{\chi}^{(2)}$ determined directly from the equation of motion}
\label{app:nltide_direct}

Rather than expand $\gv{\chi}^{(2)}$ as a sum over modes as in \S~\ref{app:nltide_sum_over_modes}, we find that it is numerically more accurate to instead directly solve the inhomogeneous equation of motion
\beq
\label{eq:chi2_direct}
\rho \ddot{\gv \chi}^{(2)}
=\gv{f}_1[\gv \chi^{(2)}]
+\gv{f}_2[\gv \chi^{(1)}, \gv \chi^{(1)}]  - \rho \left(\gv{\chi}^{(1)}\cdot \grad\right) \grad U.
\eeq
We do this by first solving Equation (\ref{eq:chi1}) for the linear tide $\gv{\chi}^{(1)}$.  By angular momentum conservation (in the form of three-mode coupling between the linear and nonlinear tide),  we have that the inhomogeneous driving term and $\gv{\chi}^{(2)}$ both oscillate as $e^{i\omega t}$, where $\omega = m\Omega$ and $m=\{0,\pm2,\pm4\}$ is a harmonic of the nonlinear tide. The time dependence therefore cancels out and the equations reduce to a boundary value problem involving a pair of coupled linear ODEs in the radial direction with inhomogeneous driving terms. We find (cf., \citealt{Pfahl:08})
\bea
\label{eq:dy1}
\d{y_1}{\ln r} &=& \left(\frac{gr}{c_s^2}-3\right)y_1+\left(\frac{g k_h^2 r}{\omega^2}-\frac{gr}{c_s^2}\right)y_2-\frac{f_h}{\rho r\omega^2} \\
\label{eq:dy2}
\d{y_2}{\ln r}&=&\left(\frac{\omega^2-N^2}{g/r}\right)y_1 + \left(1-\d{\ln m}{\ln r} + \frac{N^2}{g/r}\right) y_2 + \frac{f_r}{\rho g},
\eea
where, letting $\gv{a}$ and $\gv{b}$ represent harmonics of $\gv{\chi}^{(1)}$ and $\gv{c}$ represent a harmonic of $\gv\chi^{(2)}$,
\bea
y_1 &=& \frac{c_r}{r},\hspace{0.5cm}
y_2 =\frac{\psi}{gr},\hspace{0.5cm}
\psi=  g c_r -c_s^2\div{\gv{c}},
\eea
and the radial and horizontal driving forces are (see also Equations \ref{eq:fr_full} and \ref{eq:fh_full} below)
\bea
\label{eq:fr_nltide}
f_r &=&\sum_{ab}\int d\Omega\, Y_c^\ast \left\{\gv{f}_2\left[\gv{a},\gv{b}\right] - \rho \left(\gv{a}\cdot\grad\right)\grad U_b \right\}\cdot\gv{\hat{r}},\\
f_h&=&\sum_{ab}\int d\Omega \, \grad_\perp Y_c^\ast \cdot \left\{\gv{f}_2\left[\gv{a},\gv{b}\right] - \rho \left(\gv{a}\cdot\grad\right)\grad U_b \right\}_\perp.
\eea
Here we use the notation $U_b$ to indicate harmonic $b$ of the tidal potential (and not the linear driving coefficient found in WAQB). Equations (\ref{eq:dy1}, \ref{eq:dy2}) assume that $\omega=m_c\Omega\neq 0$. If $\omega = m_c\Omega =0$ (i.e., the static nonlinear tide) and $\ell_c\neq0$, then the nonlinear tide is instead given by the solution to
\bea
\psi &=&  \frac{r f_h}{\rho \Lambda_c^2}\\
\label{eq:dpsi_om_zero}
\d{\psi}{r}&=&\frac{N^2}{g}\psi -N^2c_r + \frac{f_r}{\rho}.
\eea
The first equation gives $\psi(r)$ and we can use that result to get $c_r$ from the second equation.   If $\ell_c=0$ then Equation (\ref{eq:dpsi_om_zero}) still holds and the equation for $dc_r/dr$ is, by the definition of the divergence,
\beq
\d{c_r}{r}=-\frac{2}{r}c_r+\div{\gv{c}}
= \left(\frac{g}{c_s^2}-\frac{2}{r}\right)c_r-\frac{\psi}{c_s^2}.
\eeq

Starting from Equation (I37) in Schenk et al. 2002, the radial and horizontal driving forces can be written in terms of the linear tide displacement $\gv a$ and $\gv b$ as
\bea
\label{eq:fr_full}
f_r
&=&\sum_{ab}\Bigg\{-\frac{1}{2}\rho\left[a_r b_r \dd{g}{r} T +\d{}{r}\left(\frac{g}{r}\right)a_h b_hF_c \right]
+\frac{1}{2}\rho g\left[\div{\gv{a}}\div{\gv{b}} T- 
\left(h_{ab}^{(1)} T + h_{ab}^{(2)} F_c + h_{ab}^{(3)} V_c\right)\right]
\non &&
+\rho g\left[ \pd{a_r}{r}\bigg\{\frac{\Lambda_b^2}{r} b_h -\frac{2}{r}b_r\bigg\}T
 + \frac{\left(a_r-a_h\right)}{r}\pd{b_h}{r} F_c \right]
+ \left(\frac{2}{r}a_r -\frac{\Lambda_a^2}{r}a_h \right)\pd{\left[\Gamma_1 p \div{\gv{b}}\right]}{r}T
-\pd{a_h}{r} \frac{\Gamma_1 p \div{\gv{b}}}{r}F_c 
\non &&
-\frac{1}{2}\pd{}{r}\left[\Gamma_1 p\left(h_{ab}^{(1)} T + h_{ab}^{(2)} F_c + h_{ab}^{(3)} V_c\right) \right]
-\frac{1}{2}\pd{}{r}\left[\Gamma_1 p\left(\Gamma_1+\pd{\ln \Gamma_1}{\ln \rho}\right)\div{\gv{a}}\div{\gv{b}} T\right]
\non && 
-\rho\left[a_r \pdd{U_{b}}{r}T+\frac{a_h U_{b}\left(\ell_b -1\right)}{r^2}F_c \right]\Bigg\},\\
\label{eq:fh_full}
f_h&=&
\sum_{ab}\Bigg\{-\frac{1}{2}\rho\d{}{r}\left(\frac{g}{r}\right)\left[a_r b_h F_a + a_h b_r F_b\right]
+\rho g\bigg[
\frac{\left(a_r-a_h\right)}{r}\left(\frac{b_r}{r}-\div{\gv{b}}\right)F_b
 +\pd{a_r}{r}\frac{\left(b_r-b_h\right)}{r}F_a
+\frac{\left(a_r-a_h\right)}{r} \frac{b_h}{r}G_b\bigg]
\non &&
-\frac{\left(a_r-a_h \right)}{r} \pd{\left[\Gamma_1 p \div{\gv{b}}\right]}{r}F_b 
- \frac{\Gamma_1 p \div{\gv{b}}}{r}\left[\left(\frac{a_r}{r}-\div{\gv{a}}\right)F_a
+\frac{a_h}{r}G_a\right]
\non &&
- \frac{1}{2}\Lambda_c^2 \frac{\Gamma_1 p}{r} \left(h_{ab}^{(1)}T + h_{ab}^{(2)}F_c+\frac{a_h b_h}{r^2} V_c\right)
-\frac{1}{2} \frac{p}{r}\left(\Gamma_1^2+\pd{\Gamma_1}{\ln \rho}\right)\div{\gv{a}}\div{\gv{b}} \Lambda_c^2 T
\non &&
-\rho\bigg[a_r \pd{(U_b/r)}{r} F_a+\frac{a_h \ell_b U_b}{r^2}F_b\bigg]
-\rho\frac{a_h U_b}{r^2}G_b\Bigg\}.
\eea
At the center we impose the regularity conditions $c_r=\ell_c c_h$ and at the surface we require the fluid to be hydrostatic by imposing $\Delta p = 0$.  We solve the equations by shooting from these boundaries to a fitting point at $r\simeq R/2$ (see, e.g., \citealt{Press:92}).

\section{Damping rate of \lowercase{$p$}-modes in the case of local driving}
\label{app:local_driving_pmode_damping}

In \S~\ref{sec:pmode_damping_rate} we assume that the $g$-mode driving is global.  However, if the driving is local then the coupling region $\Delta r \ll R$.  In the calculation below, we show that in that case the $p$-mode damping rate is much larger than that in the global case (i.e., $\gamma_p\gg 10^5\trm{ s}^{-1}$).   The $g$-mode driving is local if 
\beq
\Gamma t_g(R) \ga 1,
\eeq
where $\Gamma\simeq \epsilon \sqrt{\Omega \gamma_p}$ is the nonlinear growth rate and $t_g(R)\simeq k_r R/\omega_g$ is the $g$-mode's group travel time across the star.  For an equal wavelength $p$-$g$ pair, $k_r \simeq\omega_p / c_s \simeq \omega_p /1.5 R\omega_0$, and  the condition for local driving becomes
\beq
\frac{\omega_p}{\omega_g} \ga \frac{1.5\omega_0}{\epsilon\sqrt{\Omega \gamma_p}}\simeq 6\times10^3 f_{100}^{-5/2}\left(\frac{\gamma_p}{10^5\trm{ s}^{-1}}\right)^{-1/2}.
\eeq
For $\omega_p > \omega_{\rm ac}\sim10^2\omega_0$, wavelength matching implies $\omega_p/\omega_g \ga 10^5\Lambda_g^{-1}$ and driving is local if $f_{\rm gw} \ga 30\trm{ Hz}$.

To estimate $\gamma_p$ in the local driving regime, assume $\Delta r$ is determined by the distance the $g$-mode travels in a nonlinear growth time, i.e., equate the $g$-mode's group travel time across $\Delta r$ to its inverse growth rate
\beq
\label{eq:tcGamma}
t_{g}(\Delta r) \approx \Gamma^{-1}.
\eeq
Solving this equation for $\Delta r$ assuming an equal wavelength $p$-$g$ pair yields a damping rate
\beq
\gamma_p \approx \frac{2\pi c_s}{\Delta r} 
\approx
 \left(2\pi \epsilon \frac{\omega_p}{\omega_g}\right)^2 \Omega 
\approx 
3\times10^7 \Lambda_g^2 f_{100}^5 \left(\frac{10^{-3}\omega_0}{\omega_g}\right)^4 \trm{ s}^{-1}.
\label{eq:gamb_local}
\eeq
We thus see that $\gamma_p\gg 10^{5}\trm{ s}^{-1}$ in the local driving regime.

\end{appendix}

%\vspace{0.5cm}
\bibliographystyle{apj}

\bibliography{ref}

\begin{thebibliography}{43}
\expandafter\ifx\csname natexlab\endcsname\relax\def\natexlab#1{#1}\fi

\bibitem[{{Accadia} {et~al.}(2012){Accadia}, {Acernese}, {Alshourbagy},
  {Amico}, {Antonucci}, {Aoudia}, {Arnaud}, {Arnault}, {Arun}, {Astone}, \&
  et~al.}]{Accadia:12}
{Accadia}, T., {et~al.} 2012, Journal of Instrumentation, 7, 3012

\bibitem[{{Arras} {et~al.}(2003){Arras}, {Flanagan}, {Morsink}, {Schenk},
  {Teukolsky}, \& {Wasserman}}]{Arras:03}
{Arras}, P., {Flanagan}, E.~E., {Morsink}, S.~M., {Schenk}, A.~K., {Teukolsky},
  S.~A., \& {Wasserman}, I. 2003, \apj, 591, 1129

\bibitem[{{Arras} \& {Socrates}(2010)}]{Arras:10}
{Arras}, P., \& {Socrates}, A. 2010, \apj, 714, 1

\bibitem[{{Barker} \& {Ogilvie}(2010)}]{Barker:10}
{Barker}, A.~J., \& {Ogilvie}, G.~I. 2010, \mnras, 404, 1849

\bibitem[{{Bildsten} \& {Cutler}(1992)}]{Bildsten:92}
{Bildsten}, L., \& {Cutler}, C. 1992, \apj, 400, 175

\bibitem[{{Burkart} {et~al.}(2013){Burkart}, {Quataert}, {Arras}, \&
  {Weinberg}}]{Burkart:13}
{Burkart}, J., {Quataert}, E., {Arras}, P., \& {Weinberg}, N.~N. 2013, \mnras,
  433, 332

\bibitem[{{Chabanat} {et~al.}(1998){Chabanat}, {Bonche}, {Haensel}, {Meyer}, \&
  {Schaeffer}}]{Chabanat:98}
{Chabanat}, E., {Bonche}, P., {Haensel}, P., {Meyer}, J., \& {Schaeffer}, R.
  1998, Nuclear Physics A, 635, 231

\bibitem[{{Christensen-Dalsgaard}(2008)}]{Dalsgaard:08}
{Christensen-Dalsgaard}, J. 2008, \apss, 316, 113

\bibitem[{{Damour} {et~al.}(2012){Damour}, {Nagar}, \& {Villain}}]{Damour:12}
{Damour}, T., {Nagar}, A., \& {Villain}, L. 2012, \prd, 85, 123007

\bibitem[{{Essick} \& {Weinberg}(2016)}]{Essick:16}
{Essick}, R., \& {Weinberg}, N.~N. 2016, \apj, 816, 18

\bibitem[{{Flanagan} \& {Hinderer}(2008)}]{Flanagan:08}
{Flanagan}, {\'E}.~{\'E}., \& {Hinderer}, T. 2008, \prd, 77, 021502

\bibitem[{{Flanagan} \& {Racine}(2007)}]{Flanagan:07}
{Flanagan}, {\'E}.~{\'E}., \& {Racine}, {\'E}. 2007, \prd, 75, 044001

\bibitem[{{Goodman} \& {Dickson}(1998)}]{Goodman:98}
{Goodman}, J., \& {Dickson}, E.~S. 1998, \apj, 507, 938

\bibitem[{{Gradshteyn} {et~al.}(2007){Gradshteyn}, {Ryzhik}, {Jeffrey}, \&
  {Zwillinger}}]{Gradshteyn:07}
{Gradshteyn}, I.~S., {Ryzhik}, I.~M., {Jeffrey}, A., \& {Zwillinger}, D. 2007,
  {Table of Integrals, Series, and Products}

\bibitem[{{Harry}(2010)}]{Harry:10}
{Harry}, G.~M. 2010, Classical and Quantum Gravity, 27, 084006

\bibitem[{{Hinderer} {et~al.}(2010){Hinderer}, {Lackey}, {Lang}, \&
  {Read}}]{Hinderer:10}
{Hinderer}, T., {Lackey}, B.~D., {Lang}, R.~N., \& {Read}, J.~S. 2010, \prd,
  81, 123016

\bibitem[{{Ho} \& {Lai}(1999)}]{Ho:99}
{Ho}, W.~C.~G., \& {Lai}, D. 1999, \mnras, 308, 153

\bibitem[{{Kochanek}(1992)}]{Kochanek:92}
{Kochanek}, C.~S. 1992, \apj, 398, 234

\bibitem[{{Kumar} \& {Goodman}(1996)}]{Kumar:96}
{Kumar}, P., \& {Goodman}, J. 1996, \apj, 466, 946

\bibitem[{{Lackey} {et~al.}(2012){Lackey}, {Kyutoku}, {Shibata}, {Brady}, \&
  {Friedman}}]{Lackey:12}
{Lackey}, B.~D., {Kyutoku}, K., {Shibata}, M., {Brady}, P.~R., \& {Friedman},
  J.~L. 2012, \prd, 85, 044061

\bibitem[{{Lackey} \& {Wade}(2015)}]{Lackey:15}
{Lackey}, B.~D., \& {Wade}, L. 2015, \prd, 91, 043002

\bibitem[{{Lai}(1994)}]{Lai:94}
{Lai}, D. 1994, \mnras, 270, 611

\bibitem[{{Lai} \& {Wu}(2006)}]{Lai:06}
{Lai}, D., \& {Wu}, Y. 2006, \prd, 74, 024007

\bibitem[{{Ogilvie}(2014)}]{Ogilvie:14}
{Ogilvie}, G.~I. 2014, \araa, 52, 171

\bibitem[{{Peters} \& {Mathews}(1963)}]{Peters:63}
{Peters}, P.~C., \& {Mathews}, J. 1963, Physical Review, 131, 435

\bibitem[{{Pfahl} {et~al.}(2008){Pfahl}, {Arras}, \& {Paxton}}]{Pfahl:08}
{Pfahl}, E., {Arras}, P., \& {Paxton}, B. 2008, \apj, 679, 783

\bibitem[{{Press} {et~al.}(1992){Press}, {Teukolsky}, {Vetterling}, \&
  {Flannery}}]{Press:92}
{Press}, W.~H., {Teukolsky}, S.~A., {Vetterling}, W.~T., \& {Flannery}, B.~P.
  1992, {Numerical recipes in FORTRAN. The art of scientific computing}

\bibitem[{{Read} {et~al.}(2013){Read}, {Baiotti}, {Creighton}, {Friedman},
  {Giacomazzo}, {Kyutoku}, {Markakis}, {Rezzolla}, {Shibata}, \&
  {Taniguchi}}]{Read:13}
{Read}, J.~S., {et~al.} 2013, \prd, 88, 044042

\bibitem[{{Read} {et~al.}(2009){Read}, {Markakis}, {Shibata}, {Ury$\bar{\rm
  u}$}, {Creighton}, \& {Friedman}}]{Read:09b}
{Read}, J.~S., {Markakis}, C., {Shibata}, M., {Ury$\bar{\rm u}$}, K.,
  {Creighton}, J.~D.~E., \& {Friedman}, J.~L. 2009, \prd, 79, 124033

\bibitem[{{Reisenegger}(1994)}]{Reisenegger:94a}
{Reisenegger}, A. 1994, \apj, 432, 296

\bibitem[{{Reisenegger} \& {Goldreich}(1994)}]{Reisenegger:94b}
{Reisenegger}, A., \& {Goldreich}, P. 1994, \apj, 426, 688

\bibitem[{{Schenk} {et~al.}(2002){Schenk}, {Arras}, {Flanagan}, {Teukolsky}, \&
  {Wasserman}}]{Schenk:02}
{Schenk}, A.~K., {Arras}, P., {Flanagan}, {\'E}.~{\'E}., {Teukolsky}, S.~A., \&
  {Wasserman}, I. 2002, \prd, 65, 024001

\bibitem[{{Somiya}(2012)}]{Somiya:12}
{Somiya}, K. 2012, Classical and Quantum Gravity, 29, 124007

\bibitem[{{Terquem} {et~al.}(1998){Terquem}, {Papaloizou}, {Nelson}, \&
  {Lin}}]{Terquem:98}
{Terquem}, C., {Papaloizou}, J.~C.~B., {Nelson}, R.~P., \& {Lin}, D.~N.~C.
  1998, \apj, 502, 788

\bibitem[{{Unno} {et~al.}(1989){Unno}, {Osaki}, {Ando}, {Saio}, \&
  {Shibahashi}}]{Unno:89}
{Unno}, W., {Osaki}, Y., {Ando}, H., {Saio}, H., \& {Shibahashi}, H. 1989,
  {Nonradial oscillations of stars} (Tokyo: University of Tokyo Press)

\bibitem[{{Van Hoolst}(1994)}]{VanHoolst:94}
{Van Hoolst}, T. 1994, \aap, 286, 879

\bibitem[{{Van Hoolst} \& {Smeyers}(1993)}]{VanHoolst:93}
{Van Hoolst}, T., \& {Smeyers}, P. 1993, \aap, 279, 417

\bibitem[{{Venumadhav} {et~al.}(2014){Venumadhav}, {Zimmerman}, \&
  {Hirata}}]{Venumadhav:14}
{Venumadhav}, T., {Zimmerman}, A., \& {Hirata}, C.~M. 2014, \apj, 781, 23

\bibitem[{{Weinberg} {et~al.}(2013){Weinberg}, {Arras}, \&
  {Burkart}}]{Weinberg:13}
{Weinberg}, N.~N., {Arras}, P., \& {Burkart}, J. 2013, \apj, 769, 121

\bibitem[{{Weinberg} {et~al.}(2012){Weinberg}, {Arras}, {Quataert}, \&
  {Burkart}}]{Weinberg:12}
{Weinberg}, N.~N., {Arras}, P., {Quataert}, E., \& {Burkart}, J. 2012, \apj,
  751, 136

\bibitem[{{Weinberg} \& {Quataert}(2008)}]{Weinberg:08}
{Weinberg}, N.~N., \& {Quataert}, E. 2008, \mnras, 387, L64

\bibitem[{{Wu} \& {Goldreich}(2001)}]{Wu:01}
{Wu}, Y., \& {Goldreich}, P. 2001, \apj, 546, 469

\bibitem[{{Zahn}(1970)}]{Zahn:70}
{Zahn}, J.~P. 1970, \aap, 4, 452

\end{thebibliography}

\end{document}